\documentclass[useAMS,usenatbib,usegraphicx]{mn2e}
\usepackage{color}
\usepackage{longtable}
\usepackage[none,bottom]{}
\usepackage{epsfig}
\usepackage{graphics}
\usepackage{rotating}
\usepackage{subfigure}
\title[Triggered Star Formation and  Evolution of T-Tauri stars]{Triggered Star Formation and  Evolution of T-Tauri stars in and around 
Bright-Rimmed Clouds}
\author[Chauhan et al.]
{Neelam Chauhan$^{1}$, A. K. Pandey$^{1}$\thanks{E-mail: pandey@aries.ernet.in}, 
K. Ogura$^{2}$, D. K. Ojha$^3$, B. C. Bhatt$^4$,  
\newauthor S. K. Ghosh$^3$, P. S. Rawat$^5$ \\
$^1$Aryabhatta Research Institute of Observational Sciences (ARIES), Nainital, 263 129, India\\
$^2$Kokugakuin University, Higashi, Shibuya-ku, Tokyo 150-8440, Japan\\
$^3$Tata Institute of Fundamental Research, Mumbai (Bombay) - 400 005, India\\
$^4$CREST, Indian Institute of Astrophysics, Hosakote 562 114, India\\
$^5$Physics Department, D.S.B. Campus, Kumaun University, Nainital, India}
\begin{document}

\date{}
\pubyear{2008}

\maketitle

\label{firstpage}

\begin{abstract}
The aim of this paper is to quantitatively testify the ``{\it small-scale sequential star formation}" 
hypothesis in and around bright-rimmed clouds (BRCs). As a continuation of the recent attempt by Ogura 
et al. (2007, Paper I), we have carried out $BVI_{c}$ photometry of four more BRC aggregates along with 
deeper re-observations of 2 previously observed BRCs. Again quantitative age gradients are found in almost 
all the BRCs studied in the present work. Archival Spitzer/IRAC data also support this result. The global 
distribution of NIR excess stars in each HII region studied here clearly shows evidence that a series of 
radiation driven implosion (RDI) processes proceeded in the past from near the central O star(s) towards the peripheries of the HII 
region. We found that in general weak-line T-Tauri stars (WTTSs) are somewhat older than classical T-Tauri 
stars (CTTSs). Also the fraction of CTTSs among the T-Tauri stars (TTSs) associated with the BRCs is found 
to decrease with age. These facts are in accordance with the recent conclusion by Bertout et al. (2007) 
that CTTSs evolve into WTTSs. It seems that in general the EW of H$\alpha$ emission in TTSs associated 
with the BRCs decreases with age. The mass function (MF) of the aggregates associated with the BRCs of 
the morphological type `A' seems to follow that found in young open clusters, whereas `B/C' type BRCs 
show significantly steeper MF.
\end{abstract}
\section{Introduction}
It is believed that majority of the stars in the Galaxy form in clusters that may 
contain massive (M $\ga$ 10M$_\odot$) as well as low mass stars. A massive star 
has strong impact on the evolution of its parental molecular cloud. As soon as 
O stars form their strong ultra-violet (UV) radiation photo-ionizes the surrounding gas 
and develops an expanding HII region, thus dispersing the remaining molecular cloud. 
However, the UV radiation can also induce triggering of the next generation star 
formation. This phenomenon is known as  `sequential star formation'.
Observational evidence for this process is often inferred from the spatial 
distribution of young stars and subgroups of OB associations and their age 
distribution (see e.g. Sharma et al. 2007,  Jose et al. 2008, Pandey et al. 2008, 
Samal et al. 2007).

One of the triggered star formation processes is known as the `collect and collapse 
process', which was proposed by Elmegreen $\&$ Lada (1977). 
As an HII region expands the surrounding neutral material is collected between the 
ionization front and the shock front which precedes the former. With time the layer 
gets massive and consequently becomes gravitationally unstable and collapses to 
form stars of the second generation, including massive stars. So this process can 
repeat itself. Recent simulations of this process include Hosokawa $\&$ Inutsuka 
(2005, 2006) and Dale et al. (2007). An observational signature of the process is the 
presence of a dense layer and massive condensations adjacent to  an HII region (e.g.,
Deharveng et al. 2003).
 
Another process which has been frequently supported by numerical simulations 
as well as by observations is radiation driven implosion (RDI) of a molecular cloud 
condensation. In this process a pre-existing dense clump is exposed to the ionizing 
radiation from massive stars of the previous generation. The head part of the clump 
collapses due to  the high pressure of the ionized gas and the self-gravity, 
which consequently leads to the formation of next generation stars. Detailed model 
calculations of the RDI process have been carried out by several authors (e.g.,
 Bertoldi 1989, Lefloch $\&$ Lazareff 1995, Lefloch et al. 1997, De Vries et al. 
2002, Kessel-Deynet $\&$ Burkert 2003, Miao et al. 2006). The signature of the RDI 
process is the anisotropic density distribution in a relatively small molecular cloud 
surrounded by a curved ionization/shock front (bright rim).  
 
Bright-rimmed clouds (BRCs) are small molecular clouds located near the edges of 
evolved HII regions and show the above signature. So they are considered to be 
good laboratories to study the physical processes involved in the
RDI process. Actually  a Submillimeter Common User Bolometer Array (SCUBA) 
imaging survey of the
submillimeter continuum emission from BRCs has revealed the presence of embedded
cores (Morgan et al. 2008, Thompson et al. 2004). Morgan et al. (2004) have shown 
the presence of a ionized boundary layer at the interface between the HII region 
and the BRC molecular cloud. They have also shown that many BRCs may be in a 
post-shocked state and ongoing star formation, which may be due to the interaction 
with the external ionizing radiation. Further 
many BRCs are associated with the signposts of recent/ongoing 
star formation such as Herbig-Haro objects and Infrared Astronomical Satellite (IRAS) 
point sources of low temperature that meet the criteria of young stellar objects (YSOs). 
Sugitani, Fukui $\&$ Ogura (1991) (hereafter SFO91) and Sugitani $\&$ Ogura 
(1994) compiled catalogues of altogether 89 BRCs, associated with 
IRAS point sources for the northern and southern hemispheres, respectively. Subsequently 
Sugitani et al. (1995) carried out near-infrared (NIR) imaging of 44 BRCs and revealed 
that an elongated, small cluster or aggregate of YSOs which are aligned along the direction 
toward the ionizing star is often associated with them. These aggregates showed a tendency 
that `redder' (presumably younger) stars tend to be located inside the BRCs, whereas 
relatively `bluer' (presumably older) stars are found outside the clouds, suggesting an 
age gradient. Thus they advocated a hypothesis called `{\it small-scale sequential star 
formation'} ({\it S$^4$F}), i.e. the propagation of star formation along the axis of the BRCs 
as the ionization/shock front advances further and further into the molecular cloud. 
The H$\alpha$  grism survey of 24 BRCs by Ogura et al. (2002) detected 460 H$\alpha$ 
emission stars (possibly, T-Tauri stars or Herbig Ae/Be stars) and 12 Herbig-Haro 
objects in their vicinities. Again these H$\alpha$ emission stars are found concentrated 
toward the head or just outside of the BRCs and aligned toward the exciting star(s) 
direction. Deep NIR photometry of BRC 14 by Matsuyanagi et al. (2006) revealed that 
three indicators of star formation, i.e., the fraction of YSOs among the sources, the 
amount of extinction and the near IR excesses of the YSOs, show a clear trend from 
outside to the inside of the rim indicating that the YSOs located near the rim are 
relatively younger than those located away from the rim. This result further 
strengthens the $S^4F$ hypothesis.

The best way to quantitatively testify the hypothesis is to estimate the ages of the 
aggregate members and to compare them between different regions with respect to 
the bright rim. Ogura et al. (2007, hereafter referred to as Paper I) undertook 
{\it BVI$_{c}$} photometry of four BRC aggregates (BRCs 11NE, 12, 14 and 37) 
and showed that the stars inside or on the bright rim tend to have younger ages than those 
outside it, which is exactly what is expected from the $S^4F$ hypothesis. The main aim 
of the present study is to further confirm it and to investigate the star formation scenario 
in/around the BRCs. We have extended {\it BVI$_{c}$} photometry to four more BRCs, 
namely BRCs 2, 13, 27 and 38. In addition to them we have re-observed BRCs 11NE and 14 
to obtain deeper data. 

The information about the observations and archival data is given in Sec. 2 and 3, 
respectively. Sec. 4 describes the BRCs studied in the present work. The procedure to 
estimate the membership, age and mass of the YSOs is described in Sec. 5. The star 
formation scenario, evolution of disk of T-Tauri stars and mass functions in the BRC 
regions are studied in Sec. 6, 7 and 8, respectively. In sec. 9 the conclusions 
of the present study are summarized.

\section{Observations and Data reductions}
\begin{table*}
\caption{Log of optical observations}
\begin{tabular}{|p{.8in}|p{.8in}|p{2.2in}|p{1.1in}|}
\hline
Region &Telescope  &filter; exposure time(sec)$\times$No. of frames & date of observations\\
\hline
BRC 2  &HCT, Hanle   &B : 600$\times$4;   V : 300$\times$4;  I$_c$ : 180$\times$4 & 2006.10.27 \\
BRC 11 &HCT, Hanle   &B : 600$\times$4;   V : 300$\times$4;  I$_c$ : 180$\times$4 & 2006.10.28 \\
BRC 13 &HCT, Hanle   &B : 600$\times$4;   V : 300$\times$4;  I$_c$ : 180$\times$4 & 2006.10.27 \\
BRC 14 &HCT, Hanle   &B : 600$\times$4;   V : 300$\times$4;  I$_c$ : 180$\times$4 & 2006.10.27 \\
BRC 27 &HCT, Hanle   &B : 600$\times$4;  V : 300$\times$4;   I$_c$ : 180$\times$4 & 2006.10.28 \\
BRC 38 &ST, Nainital   &B :1800$\times$4;   V : 300$\times$8;  I$_c$ : 600$\times$3 & 2006.10.26 \\

\hline
\end{tabular}
\end{table*}
 $BVI_{c}$ CCD observations of BRCs 2, 11NE, 13, 14 and 27 were carried out using the 
2048 $\times$ 2048 pixel$^2$ CCD camera mounted on 2.0-m {\it Himalayan Chandra 
Telescope}  (HCT) of the Indian Astronomical Observatory (IAO), Hanle, India on 
2006 October 27 and 28. The instrument Himalaya Faint Object Spectrograph Camera 
(HFOSC) was used in the imaging mode. The details of the site, HCT and HFOSC can be found 
at the HCT website (http://www.crest.ernet.in). The sky at the time of 
observations was photometric with a seeing size (FWHM) of $\sim$ $1^{\prime\prime}.5$. The 
observations of the BRCs 2, 13, 27 were standardized on  same night by observing 
standard stars in the SA113 field (Landolt 1992). The observations of BRCs 11NE and 14 
were transformed to the standard system by using the {\it BVI$_{c}$} magnitudes given 
in Paper I. 

The {\it BVI$_{c}$} observations of BRC 38 were obtained by 
using 2048 $\times$ 2048 pixel$^2$ CCD camera mounted at {f/13} Cassegrain focus of 
the 1.04-m Sampurnanand Telescope (ST) at Aryabhatta Research Institute of Observational 
Sciences (ARIES), Nainital, India. The details of the CCD camera can be found in our 
earlier paper (e.g. Jose et al. 2008, Pandey et al. 2008). 
To improve the signal to noise ratio, the 
observations were carried out in a binning mode of 2 $\times$ 2 pixels. During the 
observations the seeing was about $2^{\prime\prime}.1$. SA98 field of Landolt (1992) 
was observed on 2006, October 26 to standardize the observations. The log of the HCT and ST 
observations is tabulated in Table 1. A number of bias and twilight flat frames 
were also taken during the observing runs.

The data analysis was carried out at ARIES, Nainital, India. The initial processing of 
the data frames was done using various tasks available under the IRAF data reduction 
software package. The photometric measurements of the stars were performed using 
DAOPHOT II software package (Stetson 1987). The point spread function 
(PSF) was obtained for each frame using several uncontaminated stars. Aperture 
photometry was carried out for the standard stars to estimate 
the atmospheric extinction and to calibrate the observations. The following transformation
equations were used to calibrate the observations:

{\it $(B-V) = m_1 (b - v) + c_1 $}

{\it $(V-I_c) = m_2 (v - i) + c_2 $}

\hspace{0.4cm}{\it     $V= v + m_3 (v - i) + c_3 $}

where {\it b, v, i} are the instrumental magnitudes corrected for the atmospheric 
extinctions, and {{\it B, V, I$_c$} are the standard magnitudes; {\it c$_1$, c$_2$, c$_3$} 
and {\it m$_1$, m$_2$, m$_3$} are zero-point constants and colour-coefficients, 
respectively. The values of the zero-point constants and the colour-coefficients are given 
in Table 2.
\begin{table}
\caption{The zero-point constants, colour-coefficients and extinction-coefficients}
\begin{tabular}{|lll|}
\hline
Parameters &  HCT  &ST \\
\hline
Zero-point constants & &\\
c1 &-0.344 $\pm$ 0.024 &-0.305 $\pm$ 0.011\\
c2 &0.101 $\pm$ 0.005 &0.541 $\pm$ 0.009\\
c3 &-0.799 $\pm$ 0.017 &-3.394 $\pm$ 0.010\\

colour-coefficients & &\\
m1 &0.855 $\pm$ 0.017 &0.981 $\pm$ 0.008\\
m2 &1.063 $\pm$ 0.005 &0.990 $\pm$ 0.011\\
m3 &0.078 $\pm$ 0.015 &0.031 $\pm$ 0.009\\

Extinction-coefficients& &\\
K$_b$&0.219 $\pm$ 0.009 &0.301 $\pm$ 0.010\\
K$_v$&0.122 $\pm$ 0.007 &0.199 $\pm$ 0.009\\
K$_i$&0.056 $\pm$ 0.008 &0.088 $\pm$ 0.010 \\
\hline
\end{tabular}
\end{table}

The standard deviations of the standardization residuals, $\Delta$, between the standard 
and transformed magnitudes and colours of the standard stars, are found to be $\Delta$V = 
0.006, $\Delta$(B - V) = 0.007 and $\Delta$(V - I$_c$) = 0.007 for the HCT data, whereas 
for the ST observations these values are 0.001, 0.010 and 0.002 respectively. 
The photometric accuracies depend on the brightness of the stars, and the typical 
DAOPHOT errors in B, V and I$_c$ bands at V $\sim$ 18 are smaller than 0.01 mag. Near the 
limiting magnitude of V $\sim$ 21, which is practically the same for HCT and ST, the 
DAOPHOT errors increase to 0.11, 0.05, 0.02 
mag in the {\it B, V} and {\it I$_c$} bands, respectively.
The {\it B, V} and {\it I$_c$} photometric data for the stars along with their 
positions, equivalent widths (EWs) and corresponding 2MASS data are given in Table 3.
\begin{table*}
\tiny
\centering
\rotcaption{{\it B, V} and {\it I$_c$} photometric data for the stars along with their positions, EWs and corresponding 2MASS data.}
\begin{sideways}
\begin{minipage}{245mm}
\begin{tabular}{|rllrrrrrrrrrrr|}
\hline
{\bf S.No.}&  {\bf   RA }    & {\bf  DEC }    &{\bf B $\pm$ eB} &{\bf V $\pm$ eV } & {\bf I$_c$ $\pm$ eI$_c$} &{\bf EW [H$\alpha$]}&{\bf 2MASS Name}    &{\bf J $\pm$ eJ} & {\bf H $\pm$ eH }& {\bf K $\pm$ eK } & {\bf Q flag}&{\bf C flag}& {\bf ID(Ogura}\\
   &  {\bf  (2000) } & {\bf (2000)} &{\bf (mag)} &{\bf (mag)} & {\bf (mag)} &{(\AA)} &&{\bf (mag)}&{\bf (mag)}&{\bf (mag)}&& &{\bf et al. 2002)}\\
\hline
{\bf  BRC 2}&             &             &         &        &        &       &       &      \\
   1 &  00  03 57.1& +68 33 46.4 &   20.087 $\pm$  0.009 & 18.049 $\pm$ 0.003 &  15.136 $\pm$ 0.004& 16.3 &00035705+6833465&13.067 $\pm$ 0.026&11.906 $\pm$ 0.031&11.220 $\pm$ 0.021&AAA&000  & 5\\
   2 &  00  03 57.3& +68 33 23.0 &                       & 22.450 $\pm$ 0.044 &  17.832 $\pm$ 0.004& 274.4&00035728+6833229&14.863 $\pm$ 0.036&13.768 $\pm$ 0.035&13.174 $\pm$ 0.033&AAA&000  & 6\\
   3 &  00  03 59.1& +68 32 47.4 &                       & 21.134 $\pm$ 0.014 &  17.133 $\pm$ 0.005& 28.1 &00035905+6832472&14.681 $\pm$ 0.035&13.804 $\pm$ 0.042&13.315 $\pm$ 0.037&AAA&000  & 8\\
   4 &  00  04 01.6& +68 34 14.2 &                       & 17.975 $\pm$ 0.037 &  15.649 $\pm$ 0.014&  2.7 &00040165+6834137&13.737 $\pm$ 0.040&12.834 $\pm$ 0.037&12.447 $\pm$ 0.029&AAA&ccc  & 9\\
   5 &  00  04 01.8& +68 34 00.1 &                       & 22.786 $\pm$ 0.074 &  18.087 $\pm$ 0.007& 21.7 &00040176+6833599&15.423 $\pm$ 0.048&13.756 $\pm$ 0.035&12.639 $\pm$ 0.033&AAA&000  &10\\
   6 &  00  04 01.8& +68 34 34.3 &   18.906 $\pm$  0.009 & 16.950 $\pm$ 0.005 &  13.991 $\pm$ 0.004& 20.9 &00040183+6834344&11.359 $\pm$ 0.049&10.059 $\pm$ 0.051& 9.099 $\pm$ 0.039&EEE&000  &12\\
   7 &  00  04 02.6& +68 34 26.0 &                       & 19.489 $\pm$ 0.036 &  16.795 $\pm$ 0.011& 19.4 &00040261+6834263&14.644 $\pm$ 0.045&13.355 $\pm$ 0.040&12.617 $\pm$ 0.033&AAA&ccc  &14\\
   8 &  00  04 07.6& +68 33 24.8 &   22.113 $\pm$  0.046 & 19.673 $\pm$ 0.006 &  16.363 $\pm$ 0.002& 18.2 &00040758+6833250&14.158 $\pm$ 0.026&12.990 $\pm$ 0.032&12.539 $\pm$ 0.028&AAA&000  &21\\
   9 &  00  04 11.7& +68 33 25.2 &                       & 20.455 $\pm$ 0.008 &  16.470 $\pm$ 0.003& 18.2 &00041165+6833253&14.104 $\pm$ 0.034&12.978 $\pm$ 0.030&12.461 $\pm$ 0.021&AAA&000  &22\\
  10 &  00  04 15.2& +68 33 01.8 &   18.424 $\pm$   0.01 & 16.617 $\pm$ 0.004 &  14.148 $\pm$ 0.002& 18.2 &00041520+6833019&12.126 $\pm$ 0.032&11.019 $\pm$ 0.032&10.324 $\pm$ 0.023&AAA&000  &25\\
  11 &  00  03 58.4& +68 34 06.6 &                       & 20.695 $\pm$ 0.040 &  17.521 $\pm$ 0.012& 6.8  &00035828+6834062&14.661 $\pm$ 0.034&13.175 $\pm$ 0.029&12.276 $\pm$ 0.023&AAA&000  &7 \\
  12 &  00  04 04.6& +68 34 52.0 &   21.649 $\pm$  0.033 & 19.356 $\pm$ 0.004 &  16.025 $\pm$ 0.004& 23.2 &00040454+6834519&13.599 $\pm$ 0.029&12.364 $\pm$ 0.029&11.625 $\pm$ 0.019&AAA&000  &16\\
  13 &  00  04 05.6& +68 33 44.3 &                       & 17.352 $\pm$ 0.013 &  14.926 $\pm$ 0.003& 804.5&00040563+6833442&12.867 $\pm$ 0.035&11.660 $\pm$ 0.032&10.787 $\pm$ 0.023&AAA&000  &19\\
  14 &  00  03 38.0& +68 34 55.6 &                       & 21.268 $\pm$ 0.017 &  17.758 $\pm$ 0.006&      &00033798+6834554&15.120 $\pm$ 0.047&14.215 $\pm$ 0.048&13.660 $\pm$ 0.045&AAA&000  &  \\
  15 &  00  03 54.5& +68 33 43.2 &                       & 23.591 $\pm$ 0.127 &  18.730 $\pm$ 0.007&      &00035445+6833444&14.951 $\pm$ 0.047&13.762 $\pm$ 0.050&13.027 $\pm$ 0.044&AAA&000  &  \\
  16 &  00  04 14.0& +68 32 21.5 &    21.74 $\pm$  0.029 & 19.904 $\pm$ 0.005 &  17.020 $\pm$ 0.003& 238.6&00041398+6832215&14.026 $\pm$ 0.031&12.709 $\pm$ 0.051&11.810 $\pm$ 0.033&AAA&000  &23\\
  17 &  00  04 14.7& +68 32 48.8 &                       & 21.905 $\pm$ 0.025 &  18.924 $\pm$ 0.007&      &00041473+6832490&13.585 $\pm$ 0.026&12.480 $\pm$ 0.032&12.052 $\pm$ 0.021&AAA&000  &24 \\
&            &             &         &        &        &       &       &      & \\
{\bf BRC 11NE}&&&&&&&& \\
  18 &  02  51 37.4& +60 06 26.6 &20.088 $\pm$ 0.007 & 18.481 $\pm$ 0.002 & 16.268 $\pm$ 0.002&27.4  &02513737+6006267 & 14.550 $\pm$ 0.042 & 13.527 $\pm$  0.043 & 12.940 $\pm$ 0.033 & AAA &000&1  \\
  19 &  02  51 54.5& +60 08 26.6 &20.497 $\pm$ 0.010 & 18.837 $\pm$ 0.004 & 16.530 $\pm$ 0.003&50.9  &02515451+6008266 & 14.582 $\pm$ 0.056 & 13.567 $\pm$  0.043 & 12.825 $\pm$ 0.038 & AAA &c0c&4  \\
  20 &  02  51 58.7& +60 08 05.8 &21.109 $\pm$ 0.018 & 19.503 $\pm$ 0.004 & 16.996 $\pm$ 0.003&6.8   &02515869+6008060 & 14.918 $\pm$ 0.029 & 14.016 $\pm$  0.043 & 13.466 $\pm$ 0.040 & AAA &000&5  \\
  21 &  02  52 11.1& +60 07 15.2 &21.649 $\pm$ 0.029 & 19.860 $\pm$ 0.005 & 17.332 $\pm$ 0.002&25.5  &02521113+6007154 & 15.634 $\pm$ 0.053 & 14.509 $\pm$  0.058 & 13.988 $\pm$ 0.050 & AAA &000&7  \\
  22 &  02  52 15.1& +60 05 18.5 &21.220 $\pm$ 0.023 & 19.797 $\pm$ 0.008 & 17.153 $\pm$ 0.005&17.9  &02521503+6005188 & 15.113 $\pm$ 0.047 & 14.089 $\pm$  0.042 & 13.640 $\pm$ 0.044 & AAA &000&8  \\
  23 &  02  51 54.2& +60 07 43.5 &20.456 $\pm$ 0.012 & 18.598 $\pm$ 0.006 & 15.919 $\pm$ 0.003&      &02515419+6007437 & 14.116 $\pm$ 0.034 & 13.144 $\pm$  0.037 & 12.791 $\pm$ 0.033 & AAA &000&3  \\
  24 &  02  51 59.7& +60 06 39.3 &20.803 $\pm$ 0.012 & 19.202 $\pm$ 0.004 & 16.893 $\pm$ 0.002&49.4  &02515975+6006394 & 15.306 $\pm$ 0.048 & 14.236 $\pm$  0.042 & 13.515 $\pm$ 0.038 & AAA &000&6  \\
  25 &  02  51 52.1& +60 07 10.0 &19.964 $\pm$ 0.009 & 18.334 $\pm$ 0.003 & 15.988 $\pm$ 0.002&      &02515212+6007102 & 14.131 $\pm$ 0.032 & 12.975 $\pm$  0.033 & 12.136 $\pm$ 0.026 & AAA &000&   \\
  26 &  02  52 01.3& +60 06 15.3 &                   & 21.882 $\pm$ 0.028 & 18.491 $\pm$ 0.004&      &02520131+6006154 & 15.629 $\pm$ 0.053 & 14.406 $\pm$  0.056 & 13.627 $\pm$ 0.042 & AAA &000&   \\
  27 &  02  51 59.9& +60 05 32.0 &                   & 21.713 $\pm$ 0.021 & 18.551 $\pm$ 0.006&      &02515993+6005323 & 16.155 $\pm$ 0.091 & 14.914 $\pm$  0.081 & 14.142 $\pm$ 0.068 & AAA &000&   \\
&            &             &         &        &        &       &       &        &   \\ 	      					 
{\bf BRC 11}&            &             &         &        &        &       & &    & \\        				 
  28 &  02  51 32.8& +60 03 54.3 &                   & 19.871 $\pm$ 0.013 & 16.965 $\pm$ 0.012&7.2  &02513283+6003542 & 13.005 $\pm$ 0.026 &	11.523 $\pm$  0.032 & 10.447 $\pm$ 0.022	& AAA &000&1  \\
  29 &  02  51 25.6& +60 06 04.8 &19.816 $\pm$ 0.006 & 18.318 $\pm$ 0.002 & 15.967 $\pm$ 0.001&     &02512557+6006048 & 14.609 $\pm$ 0.038 &	13.142 $\pm$  0.033 & 12.095 $\pm$ 0.019	& AAA &000&  \\
&            &             &         &        &        &       &       &    &   \\
{\bf BRC 11E}&&&&&&&& \\
  30 &  02  52 13.6& +60 03 26.2 &                    & 20.991 $\pm$ 0.014 & 18.151 $\pm$ 0.015&136.4    &02521362+6003262 & 15.592 $\pm$ 0.074 & 14.685 $\pm$ 0.064 & 14.157 $\pm$ 0.072 & AAA & ccc&1 \\
  31 &  02  52 14.2& +60 03 11.7 &20.687 $\pm$ 0.016  & 19.129 $\pm$ 0.005 & 16.647 $\pm$ 0.008&         &02521422+6003114 & 14.311 $\pm$ 0.036 & 13.278 $\pm$ 0.038 & 12.532 $\pm$ 0.032 & AAA & 000&   \\
&            &             &         &        &        &       &       &       \\ &
{\bf BRC 13}&            &             &         &        &        &       &       &\\     
  32 &  03  00 51.1& +60 39 36.3 &20.129 $\pm$   0.014 &   18.477 $\pm$   0.022  &                   &99.2 & 03005107+6039360 & 13.508 $\pm$ 0.044& 12.421 $\pm$ 0.043 & 11.804 $\pm$ 0.038& AAA& ccc & 6 \\
  33 &  03  00 51.6& +60 39 48.9 &                     &   21.683 $\pm$   0.031  & 17.783 $\pm$ 0.003&20.5 & 03005161+6039489 & 15.270 $\pm$ 0.055& 14.176 $\pm$ 0.044 & 13.478 $\pm$ 0.068& AAA& c00 & 7 \\
  34 &  03  00 52.7& +60 39 31.6 &21.216 $\pm$   0.018 &   19.667 $\pm$   0.006  & 17.147 $\pm$ 0.003&602.3& 03005265+6039317 & 15.043 $\pm$ 0.050& 14.025 $\pm$ 0.050 & 13.291 $\pm$ 0.042& AAA& 000 &10 \\
  35 &  03  00 53.6& +60 40 24.9 &21.955 $\pm$   0.042 &   19.702 $\pm$   0.013  & 16.893 $\pm$ 0.009&     & 03005350+6040252 & 14.376 $\pm$ 0.042& 13.157 $\pm$ 0.045 & 12.751 $\pm$ 0.038& AAA& ccc &11 \\
  36 &  03  00 55.4& +60 39 42.7 &                     &   20.841 $\pm$   0.015  & 17.869 $\pm$ 0.004&180.5& 03005542+6039427 & 15.789 $\pm$ 0.075& 14.496 $\pm$ 0.059 & 13.935 $\pm$ 0.056& AAA& 000 &12 \\
  37 &  03  00 56.0& +60 40 26.3 &                     &   20.713 $\pm$   0.022  & 17.244 $\pm$ 0.008& 8.0 & 03005601+6040265 & 14.695 $\pm$ 0.053& 13.591 $\pm$ 0.057 & 12.945 $\pm$ 0.054& AAA& cc0 &13 \\
  38 &  03  00 44.8& +60 40 09.1 &21.874 $\pm$   0.038 &   19.923 $\pm$   0.009  & 17.283 $\pm$ 0.015& 16.7& 03004476+6040092 & 14.683 $\pm$ 0.039& 13.756 $\pm$ 0.039 & 13.008 $\pm$ 0.036& AAA& 000 & 2   \\
  39 &  03  00 45.3& +60 40 39.5 &20.536 $\pm$    0.01 &   18.722 $\pm$   0.005  & 16.399 $\pm$ 0.004& 14.8& 03004529+6040395 & 14.517 $\pm$ 0.038& 13.672 $\pm$ 0.038 & 13.327 $\pm$ 0.040& AAA& 000 & 3 \\
&            &             &         &        &        &       &       &       \\
{\bf BRC 14}&            &             &   &        &        &       &       &\\     
  40 &  03  01 24.0& +60 30 42.2 &                   &   21.397 $\pm$ 0.029  & 17.971 $\pm$ 0.005&125.4 &03012400+6030423&15.940 $\pm$ 0.010&  14.870 $\pm$ 0.010 & 14.410 $\pm$ 0.010&&&29\\
  41 &  03  01 24.7& +60 30 09.6 &                   &   21.998 $\pm$ 0.045  & 18.197 $\pm$ 0.006&      &                &15.680 $\pm$ 0.010&  14.360 $\pm$ 0.010 & 13.780 $\pm$ 0.010&&&30\\
  42 &  03  01 25.6& +60 29 39.0 &                   &   19.644 $\pm$ 0.006  & 16.857 $\pm$ 0.003& 9.5  &03012556+6029392&14.730 $\pm$ 0.010&  13.610 $\pm$ 0.010 & 13.080 $\pm$ 0.010&&&31\\
  43 &  03  01 26.4& +60 30 53.9 & 20.374$\pm$ 0.012 &   18.351 $\pm$ 0.003  & 15.816 $\pm$ 0.004&10.6  &03012638+6030539&14.050 $\pm$ 0.010&  13.030 $\pm$ 0.010 & 12.570 $\pm$ 0.010&&&32\\
  44 &  03  01 27.2& +60 30 56.9 &                   &   20.927 $\pm$ 0.018  & 18.063 $\pm$ 0.006&58.5  &03012722+6030569&16.150 $\pm$ 0.020&  15.090 $\pm$ 0.020 & 14.520 $\pm$ 0.020&&&33\\
  45 &  03  01 27.4& +60 30 39.7 & 22.791$\pm$ 0.092 &   20.661 $\pm$ 0.016  & 17.794 $\pm$ 0.006&21.3  &                &15.510 $\pm$ 0.010&  14.370 $\pm$ 0.010 & 13.820 $\pm$ 0.010&&&34\\
  46 &  03  01 29.3& +60 31 13.6 & 20.097$\pm$ 0.009 &   18.277 $\pm$ 0.003  & 15.866 $\pm$ 0.002&49.4  &03012930+6031136&14.720 $\pm$ 0.010&  13.420 $\pm$ 0.010 & 12.420 $\pm$ 0.010&&&35\\
  47 &  03  01 34.0& +60 27 45.6 & 22.614$\pm$ 0.076 &   20.349 $\pm$ 0.011  & 17.343 $\pm$ 0.003&11.4  &                &15.200 $\pm$ 0.010&  14.190 $\pm$ 0.010 & 13.700 $\pm$ 0.010&&&39\\
  48 &  03  01 34.4& +60 30 08.5 &                   &   20.462 $\pm$ 0.012  & 17.100 $\pm$ 0.003&19.4  &                &14.750 $\pm$ 0.010&  13.410 $\pm$ 0.010 & 12.680 $\pm$ 0.010&&&40\\
  49 &  03  01 36.4& +60 29 06.1 &                   &   21.481 $\pm$ 0.034  & 17.928 $\pm$ 0.005&54.7  &03013640+6029061&15.660 $\pm$ 0.010&  14.180 $\pm$ 0.010 & 13.120 $\pm$ 0.010&&&41\\
  50 &  03  01 37.0& +60 31 00.2 &                   &   20.347 $\pm$ 0.012  & 17.175 $\pm$ 0.017&17.1  &03013695+603100 &14.920 $\pm$ 0.010&  13.880 $\pm$ 0.010 & 13.360 $\pm$ 0.010&&&42\\
  51 &  03  01 37.1& +60 29 41.2 &                   &   20.355 $\pm$ 0.010  & 17.228 $\pm$ 0.004& 6.5  &                &15.770 $\pm$ 0.010&  15.160 $\pm$ 0.020 & 14.870 $\pm$ 0.020&&&43\\
  52 &  03  01 43.3& +60 28 51.5 &                   &   22.110 $\pm$ 0.051  & 18.337 $\pm$ 0.012&13.3  &                &15.530 $\pm$ 0.010&  14.030 $\pm$ 0.010 & 13.240 $\pm$ 0.010&&&46\\ 
  53 &  03  01 50.0& +60 28 50.5 &                   &   21.694  $\pm$   0.032  & 18.183 $\pm$0.006  &     &                 &15.650$\pm$ 0.010&  14.310$\pm$ 0.010&  13.800$\pm$  0.010&&&47 \\
  54 &  03  01 04.2& +60 31 25.3 &                     &20.579 $\pm$ 0.016 &   17.760  $\pm$ 0.004 &44.8 & 03010418+6031252& 15.640 $\pm$ 0.010&  14.460 $\pm$ 0.010&  13.800 $\pm$ 0.010&&&1 \\
  55 &  03  01 06.2& +60 30 17.6 & 22.387 $\pm$ 0.062  &20.709 $\pm$ 0.017 &   17.481  $\pm$ 0.008 &79.8 & 03010623+6030176& 15.670 $\pm$ 0.010&  14.610 $\pm$ 0.010&  14.070 $\pm$ 0.010&&&3 \\
  56 &  03  01 06.6& +60 30 36.0 &                     &22.287 $\pm$ 0.067 &   18.596  $\pm$ 0.006 &     &                 & 16.420 $\pm$ 0.020&  15.420 $\pm$ 0.020&  14.900 $\pm$ 0.020&&&4 \\
  57 &  03  01 07.7& +60 29 21.8 & 20.223 $\pm$ 0.011  &18.335 $\pm$ 0.002 &   15.968  $\pm$ 0.003 &31.5 & 03010774+6029218& 14.300 $\pm$ 0.010&  13.150 $\pm$ 0.010&  12.340 $\pm$ 0.010&&&5 \\
  58 &  03  01 11.5& +60 30 46.3 &                     &20.875 $\pm$ 0.024 &   17.987  $\pm$ 0.018 &86.6 &  03011150+6030464&16.100 $\pm$ 0.020&  15.070 $\pm$ 0.020&  14.490 $\pm$ 0.010&&&6 \\
  59 &  03  01 13.4& +60 29 31.9 &                     &21.807 $\pm$ 0.040 &   18.383  $\pm$ 0.006 &13.7 &                 & 15.550 $\pm$ 0.010&  14.350 $\pm$ 0.010&  13.690 $\pm$ 0.010&&&8 \\
  60 &  03  01 16.1& +60 29 47.1 &                     &21.138 $\pm$ 0.023 &   17.779  $\pm$ 0.004 &25.8 & 03011610+6029470& 15.820 $\pm$ 0.010&  14.770 $\pm$ 0.010&  14.280 $\pm$ 0.010&&&10 \\
\hline                                                                                                                     
\end{tabular}
\end{minipage}
\end{sideways}
\end{table*}
\begin{table*}                                                                                                                
\tiny
\begin{sideways}

\begin{tabular}{|rllrrrrrrrrrrr|}
&\multicolumn{12}{c}{\bf Table 3 cont.}& \\

\hline
  61 &  03  01 17.0& +60 29 23.2 & 22.261 $\pm$ 0.058  &19.904 $\pm$ 0.008 &   17.179  $\pm$ 0.003 &16.0 & 03011705+6029232& 15.350 $\pm$ 0.010&  14.360 $\pm$ 0.010&  13.970 $\pm$ 0.010&&&12 \\
  62 &  03  01 20.3& +60 30 02.3 &                     &20.338 $\pm$ 0.012 &   17.666  $\pm$ 0.003 &38.4 & 03012024+6030024& 15.580 $\pm$ 0.010&  14.310 $\pm$ 0.010&  13.330 $\pm$ 0.010&&&18 \\
  63 &  03  01 20.6& +60 29 31.7 & 22.594 $\pm$ 0.079  &20.767 $\pm$ 0.019 &   17.990  $\pm$ 0.004 & 9.1 &                 & 15.750 $\pm$ 0.010&  14.690 $\pm$ 0.010&  14.150 $\pm$ 0.010&&&20 \\
  64 &  03  01 21.2& +60 29 44.3 &                     &20.297 $\pm$ 0.017 &   17.608  $\pm$ 0.005 &     &                 & 15.790 $\pm$ 0.010&  14.670 $\pm$ 0.010&  14.040 $\pm$ 0.010&&&23 \\
  65 &  03  01 21.2& +60 30 10.5 &                     &20.969 $\pm$ 0.018 &   17.774  $\pm$ 0.004 &24.3 &                 & 15.880 $\pm$ 0.010&  14.750 $\pm$ 0.010&  13.970 $\pm$ 0.010&&&24 \\
  66 &  03  01 32.0& +60 29 36.3 &                     &21.907 $\pm$ 0.046 &   19.193  $\pm$ 0.015 &     &                 & 17.600 $\pm$ 0.020&  16.550 $\pm$ 0.030&  15.720 $\pm$ 0.030&&&\\
  67 &  03  01 21.9& +60 29 29.5 &                     &20.588 $\pm$ 0.013 &   17.660  $\pm$ 0.004 &     &03012186+6029296 & 15.630 $\pm$ 0.070&  14.700 $\pm$ 0.070&  14.150 $\pm$ 0.070&AAA & 000&\\
  68 &  03  01 51.4& +60 27 22.7 &                     &22.305 $\pm$ 0.059&    18.613  $\pm$ 0.008 &     &03015137+6027224 & 15.590 $\pm$ 0.073&  14.560 $\pm$ 0.070&  13.900 $\pm$ 0.050&AAA & 000&\\
  69 &  03  01 19.4& +60 29 38.9 &                     &21.924 $\pm$ 0.042& 19.621 $\pm$ 0.016 &&                 & 17.730 $\pm$ 0.020& 16.71 $\pm$ 0.010 &15.930 $\pm$ 0.010 &    & & \\
  70 &  03  00 47.1& +60 28 53.6 &                     &20.298 $\pm$ 0.011& 17.664 $\pm$ 0.019 &&03004713+6028535 & 15.030 $\pm$ 0.050& 14.05 $\pm$ 0.060 &13.386 $\pm$ 0.053 &AAA &ccc & \\
  71 &  03  01 20.3& +60 29 49.3 &15.932 $\pm$0.008    &14.746 $\pm$ 0.015& 13.256 $\pm$ 0.026 &&03012029+6029493 & 11.910 $\pm$ 0.028& 10.97 $\pm$ 0.030 &10.171 $\pm$ 0.023 &AAA & 000& \\
  72 &  03  01 23.5& +60 31 50.6 &                     &20.853 $\pm$ 0.021& 18.096 $\pm$ 0.005 &&03012352+6031507 & 16.070 $\pm$ 0.100& 15.02 $\pm$ 0.090 &14.330 $\pm$ 0.072 &AAA &000&\\
  73 &  03  01 14.1& +60 29 37.4 &                     &21.553 $\pm$ 0.034& 19.196 $\pm$ 0.017 &&                 & 17.370 $\pm$ 0.010& 16.39 $\pm$ 0.010 &15.590 $\pm$ 0.010 &    &&\\
  74 &  03  01 01.1& +60 30 45.2 &                     &21.045 $\pm$ 0.024& 19.013 $\pm$ 0.020 &&                 & 16.770 $\pm$ 0.010& 15.64 $\pm$ 0.010 &14.840 $\pm$ 0.010 &    &&\\
  75 &  03  00 58.0& +60 30 13.4 &                     &19.776 $\pm$ 0.012& 17.100 $\pm$ 0.021 &&03005792+6030133 & 14.930 $\pm$ 0.046& 13.89 $\pm$ 0.050 &13.158 $\pm$ 0.039 &AAA &000&\\
  76 &  03  01 00.9& +60 33 26.7 &                     &20.708 $\pm$ 0.021& 17.603 $\pm$ 0.005 &&03010092+6033265 & 15.680 $\pm$ 0.076& 14.85 $\pm$ 0.090 &14.324 $\pm$ 0.098 &AAA &000&\\ 
  77 &  03  01 02.9& +60 31 22.4 &                     &21.023 $\pm$ 0.025& 18.045 $\pm$ 0.005 &&03010291+6031223 & 15.880 $\pm$ 0.093& 15.02 $\pm$ 0.104 &14.438 $\pm$ 0.097 &AAA &000&\\
  78 &  03  00 57.9& +60 31 21.7 &                     &20.848 $\pm$ 0.021& 17.721 $\pm$ 0.004 &&03005798+6031217 & 15.970 $\pm$ 0.090& 15.10 $\pm$ 0.092 &14.550 $\pm$ 0.099 &AAA &000&\\
  79 &  03  00 51.8& +60 32 10.8 &                     &20.733 $\pm$ 0.019& 17.931 $\pm$ 0.004 &&03005180+6032106 & 15.830 $\pm$ 0.097& 14.89 $\pm$ 0.101 &14.270 $\pm$ 0.089 &AAA &ccc &\\
  80 &  03  01 05.2& +60 31 55.4 &17.823 $\pm$ 0.002   &16.269 $\pm$ 0.001& 14.261 $\pm$ 0.005 &&03010520+6031552 & 12.780 $\pm$ 0.020& 11.88 $\pm$ 0.030 &11.312 $\pm$ 0.022 &AAA &000& \\
&             &                   &                 &             &                 &                &                 &   && &   \\			       	     
  {\bf BRC 27}      &             &                   &                 &                   &                 &                &                 &   && &   \\ 	     
  81 &  07  03 52.8& -11 23 13.2 &  18.954$\pm$0.018&  17.465 $\pm$ 0.015&   15.313 $\pm$ 0.028 & 4.6 &07035271-1123132 &13.801 $\pm$  0.047& 13.026  $\pm$ 0.050& 12.848 $\pm$ 0.039 &AAA &000& 2  \\
  82 &  07  03 53.8& -11 24 28.4 &                  &  20.018 $\pm$ 0.009&   16.761 $\pm$ 0.002 & 27.7&07035372-1124285 &15.008 $\pm$  0.043& 14.211  $\pm$ 0.051& 13.960 $\pm$ 0.057 &AAA &000& 4  \\
  83 &  07  03 57.1& -11 24 32.8 &  20.764$\pm$0.069&  19.139 $\pm$ 0.005&   16.476 $\pm$ 0.004 & 6.1 &07035712-1124327 &14.789 $\pm$  0.033& 13.968  $\pm$ 0.021& 13.756 $\pm$ 0.053 &AAA &000& 7  \\
  84 &  07  04 02.9& -11 23 37.3 &  20.678$\pm$0.085&  19.011 $\pm$ 0.014&   16.327 $\pm$ 0.015 & 8.4 &07040290-1123375 &13.489 $\pm$  0.043& 12.400  $\pm$ 0.049& 11.875 $\pm$ 0.033 &AAA &000& 8  \\  
  85 &  07  04 03.1& -11 23 50.6 &                  &  20.176 $\pm$ 0.011&   17.398 $\pm$ 0.003 &72.6 &07040308-1123504 &15.583 $\pm$  0.060& 14.303  $\pm$ 0.043& 13.567 $\pm$ 0.040 &AAA &000&10\\  
  86 &  07  04 04.3& -11 23 55.7 &  20.895$\pm$0.074&  19.616 $\pm$ 0.011&   16.722 $\pm$ 0.004 &168.3&07040426-1123556 &14.949 $\pm$  0.044& 13.995  $\pm$ 0.042& 13.559 $\pm$ 0.047 &AAA &000&12  \\
  87 &  07  04 04.8& -11 23 39.8 &  20.026$\pm$0.036&  18.318 $\pm$ 0.003&   15.970 $\pm$ 0.003 &26.6 &07040470-1123397 &14.089 $\pm$  0.040& 13.060  $\pm$ 0.043& 12.527 $\pm$ 0.037 &AAA &000&14  \\
  88 &  07  04 05.3& -11 23 13.2 &  20.508$\pm$0.049&  19.095 $\pm$ 0.004&   16.546 $\pm$ 0.003 &38.8 &07040519-1123132 &14.393 $\pm$  0.071& 13.226  $\pm$ 0.073& 12.472 $\pm$ 0.040 &AAA &000&15  \\
  89 &  07  04 06.0& -11 23 58.9 &  19.815$\pm$0.030&  18.224 $\pm$ 0.006&   15.925 $\pm$ 0.004 &22.0 &07040593-1123587 &14.360 $\pm$  0.033& 13.444  $\pm$ 0.026& 12.951 $\pm$ 0.031 &AAA &000&16  \\
  90 &  07  04 06.0& -11 23 15.7 &                  &  20.053 $\pm$ 0.008&   17.314 $\pm$ 0.003 & 4.2 &07040603-1123156 &15.030 $\pm$  0.062& 13.933  $\pm$ 0.044& 13.264 $\pm$ 0.035 &AAA &000&17  \\
  91 &  07  04 06.5& -11 23 36.2 &                  &  20.585 $\pm$ 0.013&   16.839 $\pm$ 0.003 &318.1&07040644-1123360 &14.652 $\pm$  0.049& 13.788  $\pm$ 0.050& 13.381 $\pm$ 0.072 &AAA &ccc&18  \\
  92 &  07  04 06.5& -11 23 16.4 &   19.70$\pm$0.026&  18.083 $\pm$ 0.003&   15.744 $\pm$ 0.002 &33.8 &07040656-1123163 &13.851 $\pm$  0.062& 12.932  $\pm$ 0.043& 12.543 $\pm$ 0.031 &AAA &c00&19  \\
  93 &  07  03 52.6& -11 26 16.8 & 18.311 $\pm$  0.010 &  16.865 $\pm$ 0.002 &  15.109 $\pm$ 0.002&19.8  &07035249-1126168 &13.657  $\pm$ 0.027 & 12.855  $\pm$  0.030 & 12.588$\pm$ 0.029&AAA &000 & 1 \\
  94 &  07  03 55.0& -11 25 14.5 & 20.349 $\pm$  0.045 &  18.769 $\pm$ 0.004 &  16.153 $\pm$ 0.009&4.6   &07035499-1125145 &14.593  $\pm$ 0.030 & 13.817  $\pm$  0.040 & 13.600$\pm$ 0.047&AAA &000 & 5 \\
  95 &  07  03 56.4& -11 25 41.5 &                     &  20.435 $\pm$ 0.019 &  17.396 $\pm$ 0.008&11.0  &07035638-1125413 &15.671  $\pm$ 0.073 & 14.931  $\pm$  0.070 & 14.439$\pm$ 0.088&AAA &000 & 6 \\
  96 &  07  04 04.1& -11 26 35.5 &                     &  20.515 $\pm$ 0.091 &  17.267 $\pm$ 0.021&36.1  &07040408-1126354 &15.349  $\pm$ 0.048 & 14.595  $\pm$  0.070 & 14.146$\pm$ 0.062&AAA &000 &11 \\
  97 &  07  04 08.2& -11 23 54.6 &  17.168$\pm$   0.006&  15.949 $\pm$ 0.003 &  14.346 $\pm$ 0.002& 39.1 &07040803-1123547 &13.094  $\pm$ 0.033 & 12.430  $\pm$  0.037 & 12.216$\pm$ 0.030&AAA &000 &22 \\
  98 &  07  04 08.2& -11 23 09.6 &  21.783$\pm$   0.144&  20.338 $\pm$ 0.008 &  17.413 $\pm$ 0.003&926.8 &07040816-1123097 &15.411  $\pm$ 0.111 & 14.568  $\pm$  0.055 & 14.205$\pm$ 0.075&EAA &ccc &23 \\
  99 &  07  04 09.4& -11 24 38.1 &                     &  21.053 $\pm$ 0.013 &  17.261 $\pm$ 0.003&137.2 &07040925-1124381 &15.003  $\pm$ 0.039 & 14.222  $\pm$  0.054 & 13.729$\pm$ 0.053&AAA &ccc &24 \\
 100 &  07  04 09.8& -11 23 16.4 &  16.234$\pm$   0.004&  15.106 $\pm$ 0.002 &  13.525 $\pm$ 0.003&53.2  &07040995-1123164 &11.698  $\pm$ 0.024 & 10.663  $\pm$  0.021 & 9.849 $\pm$ 0.021&AAA &ccc &25 \\
 101 &  07  04 12.0& -11 24 23.0 &                     &  20.330 $\pm$ 0.014 &  16.850 $\pm$ 0.004&69.5  &07041195-1124227 &14.658  $\pm$ 0.047 & 13.866  $\pm$  0.054 & 13.473$\pm$ 0.047&AAA &ccc &27 \\
 102 &  07  04 13.0& -11 24 03.2 &  19.017$\pm$  0.016 &  17.570 $\pm$ 0.002 &  15.695 $\pm$ 0.003&293.7 &07041292-1124031 &15.317  $\pm$ 0.060 & 14.369  $\pm$  0.047 & 13.931$\pm$ 0.047&AAA &000 &28 \\
 103 &  07  04 13.4& -11 24 55.8 &  16.822$\pm$   0.005&  15.519 $\pm$ 0.003 &  13.742 $\pm$ 0.004&17.5  &07041352-1124557 &12.135  $\pm$ 0.028 & 11.269  $\pm$  0.024 & 10.795$\pm$ 0.023&AAA &000 &29 \\
 104 &  07  04 14.2& -11 23 17.2 &  18.986$\pm$  0.016 &  17.596 $\pm$ 0.002 &  15.548 $\pm$ 0.002&33.1  &07041424-1123169 &13.833  $\pm$ 0.028 & 12.949  $\pm$  0.022 & 12.358$\pm$ 0.026&AAA &000 &31 \\
 105 &  07  04 14.2& -11 23 37.3 &                     &  20.843 $\pm$ 0.025 &  17.507 $\pm$ 0.004&&07041427-1123371 &15.435 $\pm$ 0.064 & 14.551 $\pm$ 0.060 & 14.034 $\pm$ 0.059&AAA &000 &32 \\
 106 &  07  04 08.4& -11 20 05.3 &  13.529$\pm$  0.028 &  12.533 $\pm$ 0.026 &  11.936 $\pm$ 0.024&&07040831-1120052 &13.600 $\pm$ 0.028 & 12.564 $\pm$ 0.026 & 11.919 $\pm$ 0.024&AAA &000 & \\
 107 &  07  04 03.1& -11 23 27.6 &  12.919$\pm$  0.038 &  11.533 $\pm$ 0.037 &  10.710 $\pm$ 0.026&&07040314-1123275 &13.033 $\pm$ 0.038 & 11.573 $\pm$ 0.037 & 10.694 $\pm$ 0.026&AAA &000 & \\
 108 &  07  03 54.7& -11 20 11.0 &  15.869$\pm$  0.074 &   14.946$\pm$ 0.076 &  14.385 $\pm$ 0.074&&07035465-1120110 &15.933 $\pm$ 0.074 & 14.976 $\pm$ 0.076 & 14.368 $\pm$ 0.074&AAA &000 & \\
 109 &  07  03 52.3& -11 21 01.1 &  15.618$\pm$  0.065 &  14.546 $\pm$ 0.068 &  13.794 $\pm$ 0.050&&07035228-1121009 &15.705 $\pm$ 0.065 & 14.586 $\pm$ 0.068 & 13.777 $\pm$ 0.050&AAA &000 & \\
 110 &  07  04 12.2& -11 20 20.8 &  15.618$\pm$  0.065 &  14.546 $\pm$ 0.068 &  13.794 $\pm$ 0.050&&07041215-1120205 &15.848 $\pm$ 0.072 & 14.463 $\pm$ 0.049 & 13.640 $\pm$ 0.040&AAA &000 & \\
 111 &  07  04 05.8& -11 20 03.8 &  15.743$\pm$  0.072 &  14.426 $\pm$ 0.049 &  13.656 $\pm$ 0.040&&07040576-1120038 &14.827 $\pm$ 0.048 & 13.459 $\pm$ 0.037 & 12.631 $\pm$ 0.030&AAA &000 & \\
 112 &  07  04 16.8& -11 24 32.4 &  14.723$\pm$  0.048 &  13.421 $\pm$ 0.037 &  12.647 $\pm$ 0.030&&07041680-1124324 &14.123 $\pm$ 0.026 & 13.213 $\pm$ 0.028 & 12.595 $\pm$ 0.024&AAA &000 & \\
 113 &  07  04 15.1& -11 26 22.6 &  14.062$\pm$  0.026 &  13.182 $\pm$ 0.028 &  12.612 $\pm$ 0.024&&07041508-1126224 &14.151 $\pm$ 0.033 & 13.094 $\pm$ 0.032 & 12.441 $\pm$ 0.027&AAA &000 & \\
 114 &  07  04 19.9& -11 22 22.4 &  14.078$\pm$  0.033 &  13.063 $\pm$ 0.032 &  12.458 $\pm$ 0.027&&07041999-1122224 &14.352 $\pm$ 0.029 & 13.340 $\pm$ 0.024 & 12.666 $\pm$ 0.029&AAA &000 & \\
 115 &  07  04 15.1& -11 23 39.8 &  16.111$\pm$  0.086 &  15.008 $\pm$ 0.076 &  14.249 $\pm$ 0.075&&07041500-1123398 &16.200 $\pm$ 0.086 & 15.049 $\pm$ 0.076 & 14.232 $\pm$ 0.075&AAA &c00 & \\
&            &             &         &        &        &       &                 &      &    &            &             &     \\			      

{\bf BRC 38}&            &             &         &        &         &       & &  &            &             &   &              \\
 116 &  21  40 26.2& +58 14 24.7 &22.249 $\pm$ 0.032  &  20.232 $\pm$ 0.014 &   17.812 $\pm$  0.012&      &21402612+5814243&15.182 $\pm$ 0.076&	14.262 $\pm$ 0.051& 14.004 $\pm$ 0.062&	AAA& 000  &1   \\
 117 &  21  40 28.1& +58 15 14.4 &     0 $\pm$     0  &    20.2 $\pm$ 0.011 &   17.296 $\pm$  0.013&23.94 &21402800+5815142&14.506 $\pm$ 0.038&	13.411 $\pm$ 0.035& 12.939 $\pm$ 0.035&	AAA& 000  &3   \\
 118 &  21  40 31.7& +58 17 55.3 &22.632 $\pm$ 0.047  &  20.275 $\pm$ 0.012 &   17.054 $\pm$  0.008& 3.04 &21403159+5817551&14.028 $\pm$ 0.032&	12.889 $\pm$ 0.031& 12.393 $\pm$ 0.028&	AAA& 000  &4   \\
 119 &  21  40 37.0& +58 14 38.0 &18.448 $\pm$ 0.012  &  16.664 $\pm$ 0.004 &   14.441 $\pm$  0.015&55.86 &21403691+5814378&11.902 $\pm$ 0.024&	10.886 $\pm$ 0.030& 10.234 $\pm$ 0.018&	AAA& 000  &6   \\
 120 &  21  40 37.0& +58 15 03.2 &21.957 $\pm$ 0.027  &  20.142 $\pm$ 0.011 &   17.069 $\pm$   0.01&25.84 &21403704+5815029&14.269 $\pm$ 0.029&	13.284 $\pm$ 0.041& 12.821 $\pm$ 0.029&	AAA& 000  &7   \\
 121 &  21  40 41.3& +58 15 11.5 &19.971 $\pm$ 0.011  &  18.055 $\pm$ 0.005 &   15.403 $\pm$  0.011&25.08 &21404116+5815112&12.968 $\pm$ 0.031&	11.614 $\pm$ 0.035& 10.676 $\pm$ 0.019&	AAA& 000  &9   \\
 122 &  21  40 41.5& +58 14 25.8 & 22.43 $\pm$ 0.046  &  20.582 $\pm$ 0.016 &   17.642 $\pm$  0.015&14.06 &21404156+5814255&13.650 $\pm$ 0.029&	12.618 $\pm$ 0.032& 12.166 $\pm$ 0.028&	AAA& 000  &10   \\
 123 &  21  40 44.9& +58 15 03.6 &     0              &  21.232 $\pm$ 0.023 &    17.95 $\pm$  0.015&113.24&21404485+5815033&14.617 $\pm$ 0.038&	13.347 $\pm$ 0.040& 12.658 $\pm$ 0.030&	AAA& 000  &11   \\
 124 &  21  40 48.0& +58 15 37.8 &22.524 $\pm$ 0.037  &  20.546 $\pm$ 0.015 &   16.845 $\pm$   0.01&16.34 &21404803+5815376&13.894 $\pm$ 0.026&	12.954 $\pm$ 0.033& 12.667 $\pm$ 0.028&	AAA& 000  &12   \\
 125 &  21  40 49.0& +58 17 09.6 &     0              &  21.541 $\pm$ 0.033 &   17.932 $\pm$  0.012&59.66 &21404908+5817093&14.141 $\pm$ 0.031&	12.859 $\pm$ 0.038& 12.133 $\pm$ 0.018&	AAA& 000  &15   \\
 126 &  21  40 27.4& +58 14 21.5 &21.205 $\pm$  0.02  &  19.702 $\pm$ 0.013 &   17.041 $\pm$  0.017&57.00 &21402732+5814212&14.303 $\pm$ 0.042&	13.303 $\pm$ 0.040& 12.878 $\pm$ 0.039&	AAA& 000  &2  \\
 127 &  21  40 36.5& +58 13 46.2 &21.377 $\pm$ 0.016  &  19.213 $\pm$ 0.007 &   16.289 $\pm$  0.024 &4.18 &21403655+5813458&13.514 $\pm$ 0.024&	12.582 $\pm$ 0.032& 12.245 $\pm$ 0.026&	AAA& 000  &5  \\
 128 &  21  40 42.7& +58 19 37.6 &   0.               &  21.135 $\pm$ 0.021 &   17.456 $\pm$  0.014 &     &21404282+5819373&13.935 $\pm$ 0.032&	12.545 $\pm$ 0.036& 11.640 $\pm$ 0.024&	AAA& 000  &\\
 129 &  21  41 12.0& +58 20 33.7 &   0.               &  21.726 $\pm$ 0.037 &   18.972 $\pm$  0.021 &     &21411208+5820336&16.171 $\pm$ 0.098&	15.152 $\pm$ 0.089& 14.523 $\pm$ 0.090&	AAA& 000  & \\
 130 &  21  40 45.1& +58 19 50.2 &0.                  &  22.364 $\pm$ 0.093 &   18.643 $\pm$  0.011&      &21404517+5819506&14.668 $\pm$ 0.026&	13.121 $\pm$ 0.030& 12.214 $\pm$ 0.019&	AAA& 000  &\\
 131 &  21  39 49.2& +58 14 37.0 &23.277 $\pm$ 0.176  &  21.125 $\pm$ 0.029 &   17.511 $\pm$  0.024&     &21394918+5814365&14.592 $\pm$ 0.026&	13.607 $\pm$ 0.037& 12.934 $\pm$ 0.023&	AAA& 000  &   \\
 132 &  21  39 56.4& +58 13 47.7 &20.981 $\pm$ 0.021  &  19.037 $\pm$ 0.014 &   16.289 $\pm$  0.024&     &21395635+5813475&13.338 $\pm$ 0.028&	12.273 $\pm$ 0.036& 11.528 $\pm$ 0.023&	AAA& 000  &   \\
 133 &  21  40 21.8& +58 14 45.6 &23.012 $\pm$ 0.068  &  21.092 $\pm$ 0.023 &   17.536 $\pm$  0.007&     &21402176+5814454&14.784 $\pm$ 0.036&	13.775 $\pm$ 0.047& 13.043 $\pm$ 0.028&	AAA& 000  &   \\
\hline																		       
\end{tabular}                                                    							      			       
\end{sideways}
\end{table*}

\section {Archive Data}
\subsection { Near-infrared data from 2MASS}
NIR $JHK_s$ data for the stars in the BRC regions have been obtained from the Two 
Micron All Sky Survey (2MASS) Point Source Catalog (PSC) (Cutri et al. 2003). Sources 
having uncertainty $\leq$ 0.1 mag (S/N $\geq$ 10) in all the three bands were selected 
to ensure high quality data. The $JHK_s$ data were transformed from the 2MASS system to 
the California Institute of Technology (CIT) system using the relations given in the 2MASS 
website. For BRC 14 we have adopted the $JHK_s$ data by Matsuyanagi et al. (2006), 
which were obtained with the infrared camera SIRIUS mounted on the University 
of Hawaii 2.2-m telescope.
\subsection { Mid Infrared data from Spitzer-IRAC}
We have also used archived mid-infrared (MIR) data from Infrared Array Camera (IRAC) 
of the {\it Spitzer} telescope. We obtained basic calibrated data (BCD) using the software 
Leopard. Mosaicking was performed using the MOPEX (Mosaicker and Point Source Extractor) 
software provided by {\it Spitzer Science Center} (SSC). All of our mosaics were built at 
the native instrument resolution of 1$^{\prime\prime}$.2/pixel with the standard BCDs. 
We used the standard IRAF photometry routines in the {\it apphot} package to detect sources 
and perform aperture photometry in each IRAC band. The FWHM
of every detection is measured and all detections with  FWHM$>$ 3$^{\prime\prime}$.6 are 
considered resolved and removed. The detections are also examined visually in each band to 
remove non-stellar objects and false detections. 
The sources with photometric uncertainties $\leq$ 0.2 mag in each band were considered as good detections.
The photometry was done using an aperture radius of 3$^{\prime\prime}$.6 and the background 
estimation was done within a concentric sky annulus of the inner and outer radii of 3$^{\prime\prime}$.6 and 
8$^{\prime\prime}$.4,  respectively. We adopted the zero-point magnitudes for 
the standard aperture radius (12$^{\prime\prime}$) and background annulus of 
(12$^{\prime\prime}$-22$^{\prime\prime}$.4) of 19.67, 18.93, 16.85 and 17.39 in the 3.6, 
4.5, 5.8 and 8.0 $\micron$ bands, respectively. Aperture corrections were made using the 
values described in the IRAC Data Handbook (Reach et al. 2006).
\section{Description of the BRCs studied}
A brief description of BRCs  studied is given below.

{\bf BRC 2 : } Sharpless 171 (= NGC 7822) is a large HII region associated with the 
Cepheus OB4 association (Yang \& Fukui 1992). This region contains three BRCs, 
BRCs 1-3 (SFO91). A star cluster Be 59, containing nine O7-B3 stars, is located at the 
centre of the HII region. {Recently, Pandey et al. (2008) have made photometric studies
of Be 59 and its surrounding region in detail. The distance to the cluster was estimated to be 1.0 Kpc}. 
The age of these massive stars is found 
to be about 1-4 Myr with an average of $\sim$ 2 Myr. It was also found that the stars
around BRC 1, which is located about 3 pc towards west of Be 59, are younger 
than those in the cluster. This seems to support triggered star 
formation in the BRC 1 region due to the massive stars in Be 59.

BRC 2 is located about 17 pc north of Be 59. On the basis of MIR observations by 
IRAC of the {\it Spitzer Space Telescope}, 
Megeath et al. (2004) have reported a cluster of young stars near the edge of BRC 2. 
The distribution of YSOs suggests that their formation is triggered by 
a photo-evaporation driven shock propagating into the BRC 2 cloud. 

{\bf BRCs 11NE, 13 and 14 : } The large H II region IC 1848 = S199, associated with the 
radio source W5, is located in the Perseus arm at the distance of about 1.9 kpc 
(SFO91). In fact it is composed of two adjacent HII regions namely, IC 1848W and IC 1848E 
Vallee et al. 1979, Karr \& Martin 2003, Koenig et al. 2008).
IC 1848W is ionized by HD 17505 (O6V) and 
HD 17520 (O9V), whereas IC 1848E is ionized by HD 18326 (O7V). The former harbours a 
young cluster (age $\sim$ 1 Myr; Feinstein et al. 1986). Carpenter et al. (2000) 
reported several deeply embedded star forming sites 
in the W3/W4/W5 region and put forward the notion of triggered star formation in this 
complex. Based on a multi-wavelength study of the W5 star forming region, Karr $\&$ 
Martin (2003) investigated the star formation scenario and supported triggered star 
formation in this region. 

SFO91 lists four BRCs, BRCs 11 - 14 around IC 1848. BRC 11 is situated near the 
southern edge of IC 1848W, BRC 12 near its northern edge and BRCs 13 and 14 at the
 eastern edge of IC 1848E.  
There are two more BRCs in the vicinity of BRC11, which are designated as BRC 
11NE and BRC 11E, respectively, by Ogura et at. (2002). They are not listed in 
SFO91 because of the lack of associated IRAS point sources. However, Ogura et al. 
(2002) found several H$\alpha$ emission stars in the vicinity of BRC 11NE in 
contrast to one or two in and around BRC 11 and BRC 11E. Moreover BRC 11NE 
appears to be associated with a more or less clear aggregate of young stars just 
outside its tip.  So BRC 11NE was selected as one of the target BRCs 
in Paper I to show an age gradient. In the present study we have aimed to increase 
the sample stars for age determination by reaching a deeper limiting magnitude. 

BRC 14 is associated with the molecular cloud IC 1848A to its east, which harbours 
a bright infrared young cluster AFGL 4029 (Deharveng et al. 1997). The optical and NIR 
study by these authors revealed that AFGL 4029 is an active star formation site. A 
deeper NIR survey of the BRC 14 region by Matsuyanagi et al. (2006)  
supports sequential star formation in this region propagating from the west. Paper I determined the ages of the 
stars associated with 
BRC 14 and found a quantitative evidence for the {\it S$^4$F} hypothesis.  
We are repeating the study with deeper data for this BRC too.

{\bf BRC 27 : } BRC 27 is located at the outer edge of S296 at a distance of 1.15
Kpc (SFO91) and associated with the active star forming region 
Canis Major R1 (CMa R1). The location of S296 coincides with 
the boundary of an expanding neutral hydrogen shell. 
Shevchenko et al. (1999) have estimated the ages of the stellar contents of CMa R1
ranging from $<$ 1 Myr to 8 Myr.
Herbst $\&$ Assousa (1977) suggested that the star formation in the CMa R1 region could 
have been triggered by a supernova explosion.

{\bf BRC 38 : } Cepheus OB2, located at a distance of $\sim$870 pc 
(Contreras et al. 2002), is a complex of a stellar aggregate and a bubble-shaped structure 
of atomic and molecular gas (Patel et al. 1994, 1998). The clusters NGC 7160 and Tr 37
are located near the centre of the bubble and near its edge, respectively. There is evidence  that the star 
formation at the edge of the bubble was triggered by a supernova explosion which took place 
near the centre of the bubble (Sicilia-Aguilar et al. 2004, 2005). Tr 37 harbours an O6 
star  HD 206267, which excites the relatively evolved HII region IC 1396. The age of 
Tr 37 is 
estimated as $\sim$3-5 Myr (Contreras et al. 2002). IC 1396 has a rich population of BRCs 
including BRCs 32-42 (SFO91), among which BRCs 37 and 38 have been studied 
extensively (see, e.g., Getman et al 2007 and Ikeda et al. 2008). In particular, 
Paper I reported quantitative evidence for $S^4F$ in BRC 37, and Ikeda et al. (2008) 
confirmed 
sequential star formation in this region. Getman (2007) provided detailed qualitative 
discussion on $S^4F$ based on the {\it Chandra} X-ray data for BRC 38.
\section{Membership and Age determination of member stars}\label{rd}
 The aggregates associated with BRCs are very loose and are composed of 
a small number of stars. Since BRCs are found at low galactic latitudes, the fields can be 
significantly contaminated by foreground/background stars. To understand  
star formation in BRCs it is necessary to identify stars 
directly related to them. We selected probable members associated with the BRCs 
using the following criteria. 

The spectra of some  pre-main sequence (PMS) stars, specifically   
CTTSs, show emission lines, among which usually H$\alpha$ is the 
strongest. Therefore, H$\alpha$ emission 
stars can be considered as good candidates for PMS stars associated with BRCs. In 
the present study we use H$\alpha$ emission stars found by Ogura et al. (2002)
in the vicinity of BRCs. However, some of them may not be directly associated with the BRCs
(see Sec 6.3).
\begin{figure}
\centering
\includegraphics[scale = .42, trim = 5 5 5  5, clip]{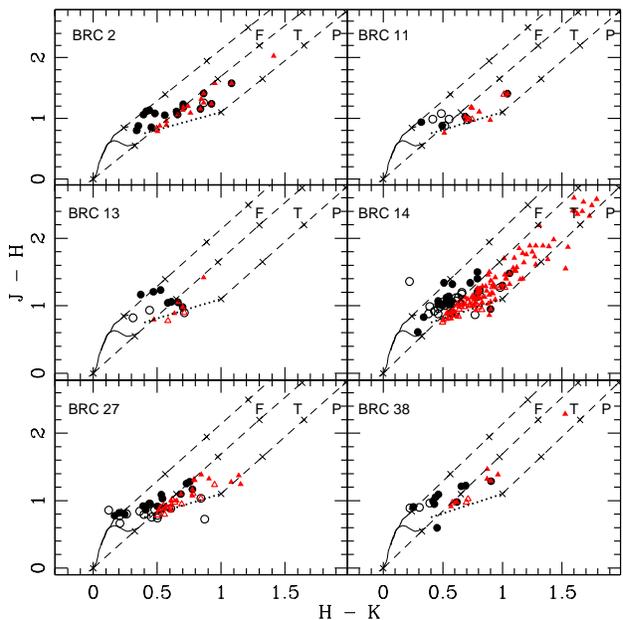}
\caption{ (J-H)/(H-K) colour-colour diagrams for BRCs 2, 11NE, 13, 14, 27 and 38.
The sequences for dwarfs (thin solid curve) and giants (thick solid curve) are from
Bessel $\&$ Brett (1988). The dotted line represents the intrinsic locus of CTTSs (Meyer
et al. 1997).
The three parallel dashed lines represent the reddening vectors.
The crosses on the dashed lines are separated by $A_V$ = 5 mag.
The open and filled circles are H$\alpha$ emission stars lying in outside and
on/inside the bright rims (see Fig. A1) respectively. The open and filled triangles are
NIR excess stars lying in  outside and on/inside the bright rims respectively.}
\label{fig1}
\end{figure}

Since many PMS stars also show NIR excesses caused by circumstellar disks, NIR 
photometric surveys have also emerged as a powerful tool to detect low-mass PMS stars. 
To identify NIR excess stars from the 2MASS PSC, we used NIR ${(J - H)/(H - K)}$ 
colour-colour (NIR-CC) diagrams. Figure 1 shows NIR-CC diagrams for the studied BRCs. 
The thin and thick solid curves represent the unreddened main-sequence and giant 
branches (Bessell $\&$ Brett 1988), respectively. The dotted line indicates the locus 
of intrinsic CTTSs (Meyer et al. 1997). The curves are also in the CIT system. 
The parallel dashed lines are the reddening vectors drawn from the tip 
(spectral type M4) of the giant branch (``upper reddening line''), 
from the base (spectral type A0) of the main-sequence branch (``middle 
reddening line'') and from the tip of the intrinsic CTTS line (``lower reddening 
line''). The extinction ratios $A_J/A_V = 0.265, 
A_H/A_V = 0.155$ and $A_K/A_V=0.090$ have been adopted from Cohen et al. 
(1981). We classified sources into three regions in the NIR-CC diagrams 
(cf. Ojha et al. 2004a). `F' sources are 
located between the upper and middle reddening lines and are considered to be either field
stars (main-sequence stars, giants) or Class III and Class II sources with
small NIR excesses. `T' sources are located between the middle and lower reddening lines.
 These sources are considered to be mostly CTTSs (Class II objects).
There may be an overlap in NIR colours of Herbig Ae/Be stars 
and CTTSs in the `T' region (Hillenbrand et al. 1992). `P' sources are
those located in the region redward of the `T' region and are most likely
Class I objects (protostar-like objects; Ojha et al. 2004b). So,  
objects falling in the `T' and `P'  regions of NIR-CC diagrams are considered 
as NIR excess stars and probable members of the BRC aggregates. These are 
included in the analysis of the  present study in addition to H$\alpha$ emission stars. 
However, we selected only those H$\alpha$ emission stars, as probable members associated
with the BRCs, that lie rightward of the upper reddening line. 
 It is worthwhile, however, to mention that Robitaille et al. (2006) 
have recently shown that there is a significant 
overlap between protostars and CTTSs in the NIR-CC space.

The spatial distribution of the probable YSOs (i.e., H$\alpha$ emission and NIR excess 
stars) for each BRC is shown in Fig. A1, which is available in electronic form only.
In Fig. A1 we have also demarcated the two 
regions for each BRC, i.e., on/inside and outside the bright rim. The NIR-CC
diagrams (Fig. 1) were used to estimate $A_V$ for each of these stars by tracing back to 
the intrinsic CTTS
line of Meyer et al. (1997) along the reddening vector (for details see, Paper I). 
The  $A_V$ for stars lying in the `F' region is estimated by tracing them 
back to the extension of the intrinsic CTTS line.
Fig. 2 shows dereddened $V_0, (V-I_c)_0$ colour-magnitude (CM) diagrams for those stars.

\begin{figure}
\includegraphics[scale = .45, trim = 15 12 10 10, clip]{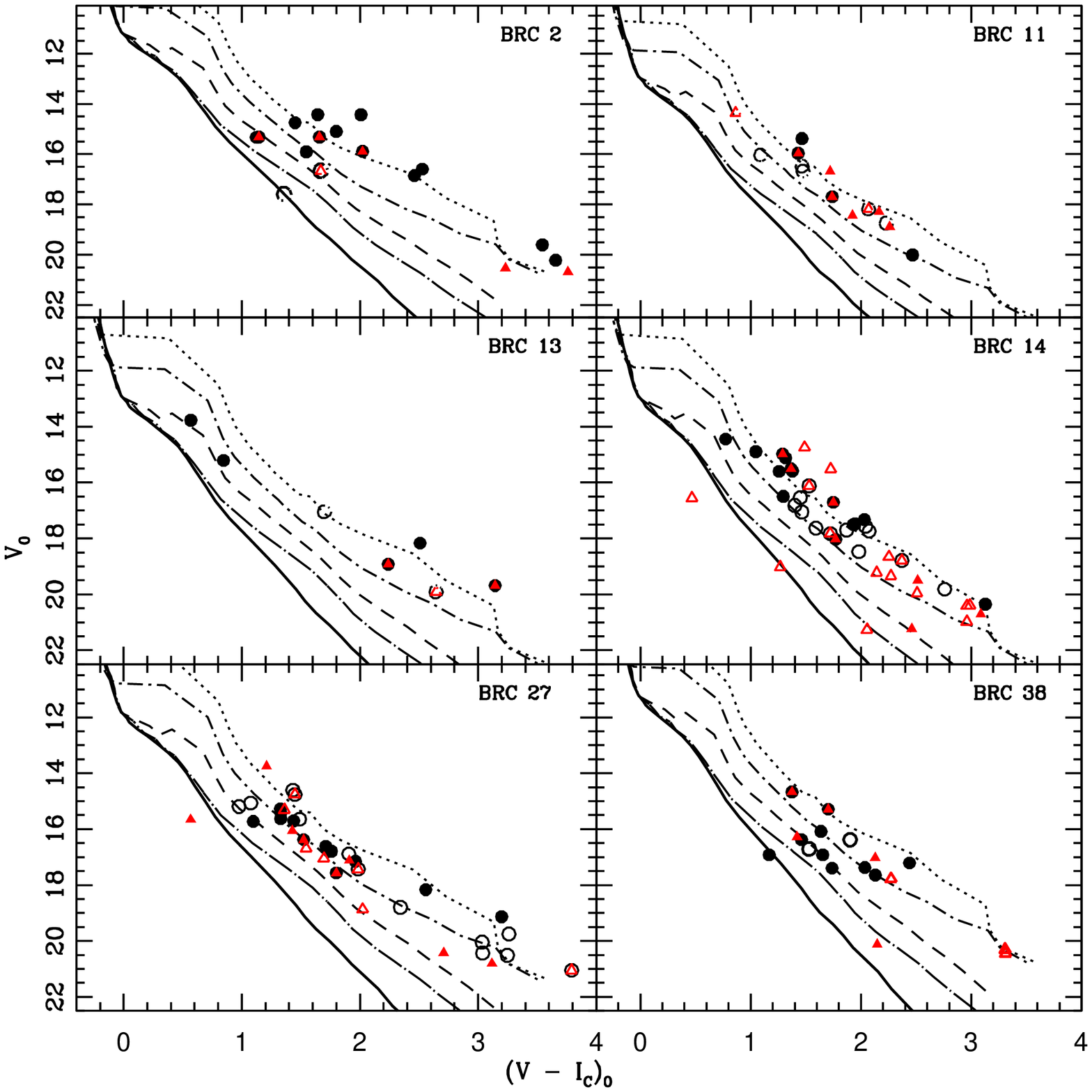}
\noindent{\footnotesize {\footnotesize \hspace{6.0cm} \bf Figure 2.} $V_0$/($V-I_c)_0$ 
colour-magnitude diagrams for probable YSOs in BRCs 2, 
11NE,13, 14, 27 and 38. The 2 Myr isochrone (thick curve) by Girardi et al. (2002) and PMS 
isochrones of 1 (dotted), 3 (dashed-dotted), 10 (dashed), 30 (large dashed-dotted)
Myr by Siess et al. (2000) are also shown. All
the isochrones are corrected for the distances of the respective BRCs. The symbols are same as in Fig. 1.}
\label{fig2}
\end{figure}
In Fig. 2 the post-main sequence isochrone for 2 Myr by Girardi et al. 
(2002), which is practically a ZAMS line, and PMS isochrones for 1, 3, 10, 
30 Myr for the solar metallicity by Siess et al. (2000) are also plotted. The distances are
 taken from SFO 91 barring for BRC 38. In the case of BRC 38 a distance of 
870 pc has been adopted from Contreras et al. (2002). The age of each YSO was 
estimated by referring to the isochrones. The mass of the YSOs was estimated using
the $V_0/(V-I_c)_0$ colour-magnitude diagram as discussed in Pandey at al. (2008).
The resultant A$_V$ values, ages and masses are given in Table 4.
\begin{table*}
\caption{Dereddened magnitude, colours, age, and mass of the YSOs associated with the BRCs.}
\begin{tabular}{|p{.7in}|p{0.5in}|p{0.6in}|p{.5in}|p{.5in}|p{.4in}|p{.6in}|p{.6in}|p{.6in}|p{.7in}|}
\hline
{\bf S.No.}&  {\bf   RA }    & {\bf  DEC }    &{\bf V$_0$} &{\bf (B-V)$_0$} & {\bf (V-I)$_0$} &{\bf A$_V$ $\pm$ $\sigma$}    &{\bf Age $\pm$ $\sigma$} & {\bf Mass $\pm$ $\sigma$} & {\bf ID(Ogura}\\
   &  {\bf  (2000) } & {\bf (2000)} &{\bf (mag)} &{\bf (mag)} & {\bf (mag)} &{\bf (mag)} &{\bf (Myr)}&{\bf (M$_\odot$)} &{\bf et al. 2002)}\\
\hline
{\bf  BRC 2}&             &             &         &        &        &       &       &      \\
   1 &  00  03 57.1& +68 33 46.4 &  15.101 &  1.149 &  1.800 & 3.0 $\pm$0.4 &  0.7 $\pm$ 0.0& 0.52 $\pm$ 0.01& 5  \\
   2 &  00  03 57.3& +68 33 23.0 &  19.604 &      - &  3.543 & 2.9 $\pm$0.5 &  0.3 $\pm$ 0.0& 0.14 $\pm$ 0.01& 6  \\
   3 &  00  03 59.1& +68 32 47.4 &  20.219 &      - &  3.656 & 0.9 $\pm$0.5 &  0.3 $\pm$ 0.0& 0.11 $\pm$ 0.01& 8  \\
   4 &  00  04 01.6& +68 34 14.2 &  15.911 &      - &  1.546 & 2.1 $\pm$0.5 &  4.0 $\pm$ 0.9& 0.78 $\pm$ 0.06& 9  \\
   5 &  00  04 01.8& +68 34 00.1 &  16.857 &      - &  2.460 & 5.9 $\pm$0.5 &  1.1 $\pm$ 0.1& 0.30 $\pm$ 0.02&10  \\
   6 &  00  04 01.8& +68 34 34.3 &  14.431 &  1.246 &  2.008 & 2.5 $\pm$0.6 &  0.1 $\pm$ 0.0& 0.47 $\pm$ 0.01&12  \\
   7 &  00  04 02.6& +68 34 26.0 &  15.329 &      - &  1.123 & 4.2 $\pm$0.6 & 16.1 $\pm$ 3.0& 1.11 $\pm$ 0.04&14  \\
   8 &  00  04 07.6& +68 33 24.8 &  14.750 &  0.995 &  1.451 & 4.9 $\pm$0.4 &  1.2 $\pm$ 0.1& 0.85 $\pm$ 0.03&21  \\
   9 &  00  04 11.7& +68 33 25.2 &  16.596 &      - &  2.527 & 3.9 $\pm$0.4 &  0.2 $\pm$ 0.1& 0.29 $\pm$ 0.00&22  \\
  10 &  00  04 15.2& +68 33 01.8 &  14.428 &  1.170 &  1.643 & 2.2 $\pm$0.4 &  0.6 $\pm$ 0.0& 0.63 $\pm$ 0.01&25  \\
  11 &  00  03 58.4& +68 34 06.6 &  15.320 &      - &  1.144 & 5.4 $\pm$0.4 & 12.3           &                & 7  \\
  12 &  00  04 04.6& +68 34 52.0 &  15.891 &  1.298 &  2.022 & 3.5 $\pm$0.4 &  0.9 $\pm$ 0.0& 0.40 $\pm$ 0.01&16  \\
  13 &  00  04 05.6& +68 33 44.3 &  15.319 &      - &  1.658 & 2.0 $\pm$0.4 &  1.2 $\pm$ 0.1& 0.63 $\pm$ 0.02&19  \\
  14 &  00  03 38.0& +68 34 55.6 &  20.526 &      - &  3.231 & 0.7 $\pm$0.6 &  4.5 $\pm$ 0.2& 0.14 $\pm$ 0.02&\\
  15 &  00  03 54.5& +68 33 43.2 &  20.680 &      - &  3.761 & 2.2 $\pm$0.4 &  1.0           & 0.09 $\pm$ 0.01&\\
  16 &  00  04 14.0& +68 32 21.5 &  16.680 &  0.921 &  1.665 & 3.2 $\pm$0.5 &  6.7 $\pm$ 1.1& 0.68 $\pm$ 0.03&23  \\
  17 &  00  04 14.7& +68 32 48.8 &  17.600 &      - &  1.355 & 4.3 $\pm$0.4 &  $>$30       &                  &24  \\
&            &             &         &        &        &       &       &      & \\
{\bf BRC 11NE}&&&&&&&& \\
  18 &  02  51 37.4& +60 06 26.6 &  16.495 &  1.028 &  1.463 & 2.0 $\pm$0.6 &  1.5 $\pm$ 0.2 & 0.91 $\pm$ 0.01 &1  \\
  19 &  02  51 54.5& +60 08 26.6 &  18.196 &  1.485 &  2.065 & 0.6 $\pm$0.6 &  1.3 $\pm$ 0.1 & 0.44 $\pm$ 0.01 &4  \\
  20 &  02  51 58.7& +60 08 05.8 &  18.760 &  1.421 &  2.226 & 0.7 $\pm$0.5 &  1.5 $\pm$ 0.1 & 0.35 $\pm$ 0.01 &5  \\
  21 &  02  52 11.1& +60 07 15.2 &  16.047 &  0.674 &  1.087 & 3.8 $\pm$0.7 &  4.5 $\pm$ 0.8 & 1.45 $\pm$ 0.05 &7  \\
  22 &  02  52 15.1& +60 05 18.5 &  16.692 &  0.557 &  1.471 & 3.1 $\pm$0.6 &  1.5 $\pm$ 0.2 & 0.83 $\pm$ 0.04 &8  \\
  23 &  02  51 54.2& +60 07 43.5 &  15.384 &  0.927 &  1.465 & 3.2 $\pm$0.5 &  0.5 $\pm$ 0.0 & 0.96 $\pm$ 0.03 &3  \\
  24 &  02  51 59.7& +60 06 39.3 &  17.693 &  1.172 &  1.739 & 1.5 $\pm$0.6 &  1.7 $\pm$ 0.2 & 0.58 $\pm$ 0.02 &6  \\
  25 &  02  51 52.1& +60 07 10.0 &  16.677 &  1.157 &  1.720 & 1.7 $\pm$0.4 &  0.7 $\pm$ 0.1 & 0.59 $\pm$ 0.03 &   \\
  26 &  02  52 01.3& +60 06 15.3 &  18.891 &      - &  2.261 & 3.0 $\pm$0.7 &  1.7 $\pm$ 0.1 & 0.34 $\pm$ 0.01 &   \\
  27 &  02  51 59.9& +60 05 32.0 &  18.437 &      - &  1.924 & 3.3 $\pm$1.1 &  2.1 $\pm$ 0.5 & 0.46 $\pm$ 0.03 &   \\
&            &             &         &        &        &       &       &           \\ 
{\bf BRC 11}&            &             &         &        &        &       &    & \\  
  28 &  02  51 32.8& +60 03 54.3 &  15.967 &      - &  1.431 & 3.9 $\pm$0.4 &  0.9 $\pm$ 0.1 & 0.97 $\pm$ 0.03 &1  \\
  29 &  02  51 25.6& +60 06 04.8 &  14.372 &  0.353 &  0.860 & 4.0 $\pm$0.5 &  2.7 $\pm$ 0.3 & 2.2  $\pm$ 0.08 &\\
&            &             &         &        &        &       &       &       \\
{\bf BRC 11E}&&&&&&&& \\
  30 &  02  52 13.6& +60 03 26.2 &  20.008 &      - &  2.468 &1.0 $\pm$0.9 & 2.9 $\pm$ 0.3 & 0.27 $\pm$ 0.01 &1 \\
  31 &  02  52 14.2& +60 03 11.7 &  18.291 &  0.347 &  2.165 &0.8 $\pm$0.5 & 1.2  $\pm$ 0.1& 0.36 $\pm$ 0.01 &   \\
&            &             &         &        &        &       &              \\
{\bf BRC 13}&            &             &         &        &        &             &\\     
  32 &  03  00 51.1& +60 39 36.3 &  15.917 &  0.887 &      - &2.6 $\pm$ 0.6 & 8.0 $\pm$ 1.3 &1.45 $\pm$ 0.04  & 6 \\
  33 &  03  00 51.6& +60 39 48.9 &  19.684 &      - &  3.144 &2.00 $\pm$ 0.6 & 0.1 $\pm$ 0.0 &0.19 $\pm$ 0.01  & 7 \\
  34 &  03  00 52.7& +60 39 31.6 &  18.923 &  1.371 &  2.239 &0.7 $\pm$ 0.6 & 1.7 $\pm$ 0.1 &0.34 $\pm$ 0.01  &10 \\
  35 &  03  00 53.6& +60 40 24.9 &  13.770 &  0.492 &  0.569 &5.9 $\pm$ 0.6 & 8.6 $\pm$ 0.8 &1.72 $\pm$ 0.04  &11 \\
  36 &  03  00 55.4& +60 39 42.7 &  15.210 &      - &  0.845 &5.6 $\pm$ 0.9 & 8.0 $\pm$ 1.4 &1.41 $\pm$ 0.05  &12 \\
  37 &  03  00 56.0& +60 40 26.3 &  18.169 &      - &  2.508 &2.5 $\pm$ 0.7 & 0.1 $\pm$ 0.0 &0.29 $\pm$ 0     &13 \\
  38 &  03  00 44.8& +60 40 09.1 &  19.923 &  1.974 &  2.640 &   0$\pm$ 0.5   &  2.2 $\pm$ 0.1 & 0.36 $\pm$ 0.00         & 2   \\
  39 &  03  00 45.3& +60 40 39.5 &  17.059 &  1.329 &  1.695 &1.7 $\pm$ 0.5 & 1.0 $\pm$ 0.1 &0.60 $\pm$ 0.02  & 3 \\
&            &             &         &        &        &       &              \\
{\bf BRC 14}&            &             &         &        &        &       &       \\     
  40 &  03  01 24.0& +60 30 42.2 &  17.480 &      - &  1.947 &3.9 $\pm$0.1 & 0.9 $\pm$0.0 &0.45 $\pm$0.02 &29 \\
  41 &  03  01 24.7& +60 30 09.6 &  15.586 &      - &  1.379 &6.4 $\pm$0.1 & 0.7 $\pm$0.1 &0.98 $\pm$0.08 &30 \\
  42 &  03  01 25.6& +60 29 39.0 &  15.597 &      - &  1.258 &4.1 $\pm$0.1 & 1.1 $\pm$0.1&1.20 $\pm$0.01 &31 \\
  43 &  03  01 26.4& +60 30 53.9 &  15.126 &  1.068 &  1.317 &3.2 $\pm$0.1 & 0.5 $\pm$0.0 &1.10 $\pm$0.0  &32 \\
  44 &  03  01 27.2& +60 30 56.9 &  18.031 &      - &  1.771 &2.9 $\pm$0.3 & 2.3 $\pm$0.2 &0.56 $\pm$0.02 &33 \\
  45 &  03  01 27.4& +60 30 39.7 &  16.498 &  0.899 &  1.295 &4.2 $\pm$0.1 & 2.7 $\pm$0.3 &1.11 $\pm$0.04 &34 \\
  46 &  03  01 29.3& +60 31 13.6 &  15.511 &  1.001 &  1.366 &2.8 $\pm$0.1 & 0.7 $\pm$0.1 &0.99 $\pm$0.01 &35 \\
  47 &  03  01 34.0& +60 27 45.6 &  17.503 &  1.428 &  1.931 &2.9 $\pm$0.1 & 0.9 $\pm$ 0   &0.45 $\pm$0.01 &39 \\
  48 &  03  01 34.4& +60 30 08.5 &  14.977 &      - &  1.290 &5.5 $\pm$0.1 & 0.5 $\pm$0.0 &1.19 $\pm$0.03 &40 \\
  49 &  03  01 36.4& +60 29 06.1 &  16.706 &      - &  1.749 &4.8 $\pm$0.1 & 0.7 $\pm$0.0 &0.55 $\pm$0.03 &41 \\
  50 &  03  01 37.0& +60 31 00.2 &  17.326 &      - &  2.031 &3.1 $\pm$0.1 & 0.7 $\pm$0.0 &0.39 $\pm$0.01 &42 \\
  51 &  03  01 37.1& +60 29 41.2 &  20.355 &      - &  3.128 &   0$\pm$0.2 & 0.3 $\pm$0.0  &1.80 $\pm$0.00 &43 \\
  52 &  03  01 43.3& +60 28 51.5 &  14.893 &      - &  1.047 &7.2 $\pm$0.1 & 1.4 $\pm$0.3 &1.79 $\pm$0.11 &46 \\
\hline
\end{tabular}
\end{table*}
\begin{table*}
{\bf Table 4 cont.}\\
\begin{tabular}{|p{.7in}|p{0.5in}|p{0.6in}|p{.5in}|p{.5in}|p{.4in}|p{.6in}|p{.6in}|p{.6in}|p{.7in}|}
\hline
  53 &  03  01 50.0& +60 28 50.5 &  14.444 &      - &  0.773 &7.3 $\pm$ 0.1 & 5.6 $\pm$ 1.7 &1.90 $\pm$ 0.18   &47 \\
  54 &  03  01 04.2& +60 31 25.3 &  16.820 &      - &  1.400 &3.8  $\pm$ 0.1 &  2.4 $\pm$ 0.2&0.94 $\pm$ 0.03 & 1 \\
  55 &  03  01 06.2& +60 30 17.6 &  17.572 &  0.763 &  2.043 &3.1 $\pm$ 0.1 &  0.9 $\pm$ 0.1&0.29 $\pm$ 0.01 &3 \\
  56 &  03  01 06.6& +60 30 36.0 &  19.819 &      - &  2.760 &2.5 $\pm$ 0.3 &  1.5 $\pm$ 0.3&0.26 $\pm$ 0.01 &4 \\
  57 &  03  01 07.7& +60 29 21.8 &  16.119 &  1.233 &  1.530 &2.2 $\pm$ 0.1 &  0.8 $\pm$ 0.0 &0.75 $\pm$ 0.00 &5 \\
  58 &  03  01 11.5& +60 30 46.3 &  18.474 &      - &  1.981 &2.4 $\pm$ 0.3 &  1.8 $\pm$ 0.2&0.42 $\pm$ 0.04 &6 \\
  59 &  03  01 13.4& +60 29 31.9 &  17.696 &      - &  1.871 &4.1 $\pm$ 0.1 &  1.2 $\pm$ 0.1&0.48 $\pm$ 0.02 &8 \\
  60 &  03  01 16.1& +60 29 47.1 &  17.738 &      - &  2.075 &3.4 $\pm$ 0.1 &  0.9 $\pm$ 0.0&0.38 $\pm$ 0.01 &10 \\
  61 &  03  01 17.0& +60 29 23.2 &  16.532 &  1.359 &  1.451 &3.4 $\pm$ 0.1 &  1.3 $\pm$ 0.1&0.85 $\pm$ 0.02 &12 \\
  62 &  03  01 20.3& +60 30 02.3 &  17.826 &      - &  1.723 &2.5 $\pm$ 0.1 &  2.1 $\pm$ 0.1&0.59 $\pm$ 0.01 &18 \\
  63 &  03  01 20.6& +60 29 31.7 &  17.630 &  0.904 &  1.592 &3.1 $\pm$ 0.1 &  3.1 $\pm$ 0.3&0.72 $\pm$ 0.02 &20 \\
  64 &  03  01 21.2& +60 29 44.3 &  17.052 &      - &  1.464 &3.3 $\pm$ 0.1 &  2.4 $\pm$ 0.2&0.85 $\pm$ 0.03 &23 \\
  65 &  03  01 21.2& +60 30 10.5 &  18.789 &      - &  2.372 &2.2 $\pm$ 0.1 &  1.3 $\pm$ 0.0&0.33 $\pm$ 0.01 &24 \\
  66 &  03  01 32.0& +60 29 36.3 &  21.235 &      - &  2.460 &0.7 $\pm$ 0.3 &  9.0 $\pm$ 1.2&0.23 $\pm$ 0&    \\    
  67 &  03  01 21.9& +60 29 29.5 &  19.493 &      - &  2.515 &1.1 $\pm$ 0.9 &  1.7 $\pm$ 0.2&0.28 $\pm$ 0.01&\\    
  68 &  03  01 51.4& +60 27 22.7 &  20.701 &      - &  3.087 &1.6 $\pm$ 0.9 &  0.9 $\pm$ 0.7&0.17 $\pm$ 0.01&  \\    
  69 &  03  01 19.4& +60 29 38.9 &  21.266 &      - &  2.055 &0.7 $\pm$ 0.2 &  $>$30&   &     \\    
  70 &  03  00 47.1& +60 28 53.6 &  19.343 &      - &  2.273 &1.0 $\pm$ 0.7 &  2.3 $\pm$ 0.2&0.33 $\pm$ 0.01  &\\    
  71 &  03  01 20.3& +60 29 49.3 &  14.746 &  1.183 &  1.490 &   0$\pm$ 0.4 &  0.3 $\pm$ 0.1&0.89 $\pm$ 0.04  & \\    
  72 &  03  01 23.5& +60 31 50.6 &  19.226 &      - &  2.143 &1.6 $\pm$ 1.2 &  2.6 $\pm$ 0.5&0.36 $\pm$ 0.02  &\\    
  73 &  03  01 14.1& +60 29 37.4 &  21.553 &      - &  2.357 &0   $\pm$ 0.1 &     $>$15           &                &\\    
  74 &  03  01 01.1& +60 30 45.2 &  19.026 &      - &  1.269 &2.0 $\pm$ 0.1 &  $>$30&   &     \\    
  75 &  03  00 58.0& +60 30 13.4 &  18.655 &      - &  2.253 &1.1 $\pm$ 0.6 &  1.4 $\pm$ 0.1&0.35 $\pm$ 0.01 &  \\
  76 &  03  01 00.9& +60 33 26.7 &  20.402 &      - &  2.989 &0.3 $\pm$ 1.1 &  1.8 $\pm$ 0.4&0.21 $\pm$ 0.01 &  \\
  77 &  03  01 02.9& +60 31 22.4 &  20.978 &      - &  2.961 &0.1 $\pm$ 1.3 &  2.8 $\pm$ 0.3&0.17 $\pm$ 0.02 &   \\
  78 &  03  00 57.9& +60 31 21.7 &  20.406 &      - &  2.961 &0.4 $\pm$ 0.0 &1.8  &  0.19         &\\
  79 &  03  00 51.8& +60 32 10.8 &  19.960 &      - &  2.510 &0.8 $\pm$ 1.3 &  2.6 $\pm$ 0.4 &0.27 $\pm$ 0.01      &  \\
  80 &  03  01 05.2& +60 31 55.4 &  15.523 &  1.334 &  1.727 &0.8 $\pm$ 0.3 &  0.1 $\pm$ 0.0 &0.61 $\pm$ 0.01       & \\
&            &             &         &        &        &       &       &       \\

{\bf BRC 27}&            &             &         &        &        &       &       &  \\   
  81 &  07  03 52.8& -11 23 13.2 &  15.278 &  0.860 &  1.326 &2.2  $\pm$0.6 & 2.0 $\pm$0.5  &1.05 $\pm$0.09 & 2  \\
  82 &  07  03 53.8& -11 24 28.4 &  18.164 &      - &  2.557 &1.9  $\pm$0.6 & 1.4 $\pm$0.1  &0.29 $\pm$0.01 & 4  \\
  83 &  07  03 57.1& -11 24 32.8 &  16.618 &  0.920 &  1.711 &2.5  $\pm$0.4 & 1.9 $\pm$0.2  &0.60 $\pm$0.01 & 7  \\
  84 &  07  04 02.9& -11 23 37.3 &  15.426 &  0.642 &  1.330 &3.6  $\pm$0.6 & 2.3 $\pm$0.5  &1.03 $\pm$0.08 & 8  \\
  85 &  07  04 03.1& -11 23 50.6 &  15.726 &      - &  1.097 &4.5  $\pm$0.7 & 11.2$\pm$1.1  &1.19 $\pm$0.01 &10  \\
  86 &  07  04 04.3& -11 23 55.7 &  17.151 &  0.637 &  1.962 &2.5  $\pm$0.6 & 1.5 $\pm$0.2  &0.44 $\pm$0.01 &12  \\
  87 &  07  04 04.8& -11 23 39.8 &  15.620 &  0.920 &  1.329 &2.7  $\pm$0.5 &  3.0$\pm$0.5  &1.06 $\pm$0.55 &14  \\
  88 &  07  04 05.3& -11 23 13.2 &  16.378 &  0.660 &  1.523 &2.7  $\pm$0.9 &  3.4 $\pm$0.8 &0.80 $\pm$0.05 &15  \\
  89 &  07  04 06.0& -11 23 58.9 &  16.791 &  1.188 &  1.758 &1.4  $\pm$0.4 &  1.9 $\pm$0.2 &0.56 $\pm$0.02 &16  \\
  90 &  07  04 06.0& -11 23 15.7 &  17.568 &      - &  1.800 &2.5  $\pm$0.7&  4.5 $\pm$0.8 &0.54 $\pm$0.03 &17  \\
  91 &  07  04 06.5& -11 23 36.2 &  19.134 &      - &  3.199 &1.5  $\pm$0.6 &  0.2$\pm$ 0.0&0.18 $\pm$0.01 &18  \\
  92 &  07  04 06.5& -11 23 16.4 &  15.700 &  0.933 &  1.439 &2.4 $\pm$0.7 &  1.9 $\pm$0.3  &0.88$\pm$0.04 &19  \\
  93 &  07  03 52.6& -11 26 16.8 &  15.064 &  0.907 &  1.076 & 1.8  $\pm$0.4 &  5.3 $\pm$0.4  & 1.43 $\pm$0.02& 1 \\
  94 &  07  03 55.0& -11 25 14.5 &  16.887 &  1.067 &  1.906 & 1.9  $\pm$0.5 &  1.4 $\pm$0.1 & 0.47 $\pm$0.01 & 5 \\
  95 &  07  03 56.4& -11 25 41.5 &  20.435 &      - &  3.039 &    0 $\pm$0.9 &  3.2 $\pm$1.4 & 0.14 $\pm$0.01 & 6 \\
  96 &  07  04 04.1& -11 26 35.5 &  20.515 &      - &  3.247 &    0 $\pm$0.8 &  0.9 &0.11  $\pm$0.01         &11 \\
  97 &  07  04 08.2& -11 23 54.6 &  15.644 &  1.136 &  1.488 & 0.3  $\pm$0.5 &  1.5 $\pm$0.1 &0.81  $\pm$0.02 &22 \\
  98 &  07  04 08.2& -11 23 09.6 &  18.795 &  1.066 &  2.343 & 1.5  $\pm$1.1 &  3.2  $\pm$0.5 &0.30  $\pm$0.01&23 \\
  99 &  07  04 09.4& -11 24 38.1 &  21.053 &      - &  3.792 &    0 $\pm$0.6 &  0.3 $\pm$0.1 & 0.10 $\pm$0.00 &24 \\
 100 &  07  04 09.8& -11 23 16.4 &  14.759 &  1.039 &  1.449 & 0.4  $\pm$0.3 &  0.6 $\pm$0.1 & 0.85 $\pm$0.02 &25 \\
 101 &  07  04 12.0& -11 24 23.0 &  19.751 &      - &  3.261 & 0.6  $\pm$0.7 &  0.3  $\pm$0.0 &0.14$\pm$0.00  &27 \\
 102 &  07  04 13.0& -11 24 03.2 &  15.189 &  0.744 &  0.976 & 2.4  $\pm$0.7 &  9.4  $\pm$2.5 &1.34 $\pm$0.07 &28 \\
 103 &  07  04 13.4& -11 24 55.8 &  14.604 &  1.043 &  1.432 & 0.9  $\pm$0.3 &  0.6  $\pm$0.1 &0.89$\pm$0.02  &29 \\
 104 &  07  04 14.2& -11 23 17.2 &  17.430 &  1.365 &  1.985 & 0.2 $\pm$0.3 &  1.9 $\pm$0.1&0.42  $\pm$0.01  &31 \\
 105 &  07  04 14.2& -11 23 37.3 &  20.043 &     -  &  3.034 & 0.8 $\pm$0.8 &  2.9  $\pm$0.3 &0.18 $\pm$0.01  &32 \\
 106 &  07  04 08.4& -11 20 05.3 &  17.122 &  1.258 &  1.909 & 1.7  $\pm$0.4 &  1.7  $\pm$0.1 &0.46 $\pm$0.01  &   \\
 107 &  07  04 03.1& -11 23 27.6 &  13.749 &      - &  1.209 & 5.2  $\pm$0.5 &  0.6  $\pm$0.1 & 1.38$\pm$0.06  &   \\
 108 &  07  03 54.7& -11 20 11.0 &  20.425 &      - &  2.709 & 1.0  $\pm$1.0 &  6.3  $\pm$0.9& 0.2 $\pm$0.01   &   \\
 109 &  07  03 52.3& -11 21 01.1 &  20.808 &      - &  3.117 & 1.4  $\pm$0.9 &  4.0  $\pm$0.4& 0.11 $\pm$0.01  &   \\
 110 &  07  04 12.2& -11 20 20.8 &  15.657 &      - &  0.567 & 4.7  $\pm$&  $>$30& &    \\
 111 &  07  04 05.8& -11 20 03.8 &  16.059 &      - &  1.428 & 4.4  $\pm$0.6 &  3.5  $\pm$1.2 &0.92  $\pm$0.09  &    \\
 112 &  07  04 16.8& -11 24 32.4 &  16.685 &  0.669 &  1.545 & 0.3  $\pm$0.4 &  4.7  $\pm$0.5 &0.79 $\pm$0.02  &  \\
 113 &  07  04 15.1& -11 26 22.6 &  15.313 &  0.868 &  1.362 & 1.9  $\pm$0.4 &  1.7 $\pm$0.2 &0.97  $\pm$0.05  & \\
 114 &  07  04 19.9& -11 22 22.4 &  17.050 &  0.979 &  1.695 & 1.2  $\pm$0.3 &  3.7  $\pm$0.4 &0.63 $\pm$0.02  & \\
 115 &  07  04 15.1& -11 23 39.8 &  18.869 &      - &  2.023 & 1.8  $\pm$1.0 &  8.9  $\pm$2.4 &2.02 $\pm$0.02  & \\
&            &             &         &        &        &       &                 &      & \\
\hline                                                             
\end{tabular}                                                    
\end{table*}                                                                    
\begin{table*}
{\bf Table 4 cont.}\\
\begin{tabular}{|p{.7in}|p{0.5in}|p{0.6in}|p{.5in}|p{.5in}|p{.4in}|p{.6in}|p{.6in}|p{.6in}|p{.7in}|}
\hline
{\bf BRC 38}&            &             &         &        &         &       & &     \\
 116 &  21  40 26.2& +58 14 24.7 &  16.917 &  1.035 &  1.168 & 3.3  $\pm$0.8  &  $>$30 &  &1   \\
 117 &  21  40 28.1& +58 15 14.4 &  16.375 &   -    &  1.460 & 3.8  $\pm$0.5  & 11.3 $\pm$1.9 &0.87 $\pm$0.03  &3   \\
 118 &  21  40 31.7& +58 17 55.3 &  16.082 &  1.119 &  1.637 &  4.2 $\pm$0.4  &  3.1  $\pm$0.4 &0.67 $\pm$0.02 &4   \\
 119 &  21  40 37.0& +58 14 38.0 &  15.288 &  1.382 &  1.704 &  1.4 $\pm$0.4  &  0.9  $\pm$0.0  &0.59 $\pm$0.02&6   \\
 120 &  21  40 37.0& +58 15 03.2 &  17.644 &  1.086 &  2.130 &  2.5 $\pm$0.5  &  3.0  $\pm$0.3  &0.36 $\pm$0.01&7   \\
 121 &  21  40 41.3& +58 15 11.5 &  14.673 &  0.917 &  1.374 &  3.4 $\pm$0.4  &  1.5  $\pm$0.2  &0.95 $\pm$0.04&9   \\
 122 &  21  40 41.5& +58 14 25.8 &  17.398 &  0.913 &  1.738 &  3.2 $\pm$0.4  & 12.3  $\pm$1.8  &0.61 $\pm$0.02&10   \\
 123 &  21  40 44.9& +58 15 03.6 &  16.921 &  -     &  1.653 &  4.3 $\pm$0.5  &  9.5  $\pm$2.1  &0.7 $\pm$0.04 &11   \\
 124 &  21  40 48.0& +58 15 37.8 &  17.209 &  1.005 &  2.441 &  3.3 $\pm$0.4  &  1.3  $\pm$0.1  &0.3           &12   \\
 125 &  21  40 49.0& +58 17 09.6 &  17.374 &  -     &  2.035 &  4.2 $\pm$0.5  &  2.9  $\pm$0.4  &0.39 $\pm$0.02&15   \\
 126 &  21  40 27.4& +58 14 21.5 &  16.709 &  0.625 &  1.530 &  3.0 $\pm$0.5  & 11.9  $\pm$2.5 &0.8 $\pm$0.02   &2  \\
 127 &  21  40 36.5& +58 13 46.2 &  16.379 &  1.331 &  1.902 &  2.8 $\pm$0.4  &  1.5  $\pm$0.1 &0.47 $\pm$0.01 &5  \\
 128 &  21  40 42.7& +58 19 37.6 &  17.030 &      - &  2.129 &  4.1 $\pm$0.4  &  1.7  $\pm$0.1 &0.37 $\pm$0.01      &\\
 129 &  21  41 12.0& +58 20 33.7 &  20.125 &      - &  2.148 &  1.6 $\pm$1.2 & $>$30 & &   \\
 130 &  21  40 45.1& +58 19 50.2 &  16.277 &      - &  1.421 &  6.1 $\pm$0.4  &  10.2   &0.84 $\pm$0.03    &\\
 131 &  21  39 49.2& +58 14 37.0 &  20.312 &  1.930 &  3.308 &  0.8 $\pm$0.4  &  3.1  $\pm$0.3  &0.13 $\pm$0.01  &   \\
 132 &  21  39 56.4& +58 13 47.7 &  17.783 &  1.584 &  2.274 &  1.3 $\pm$0.4  &  2.5  $\pm$0.2  &0.33 $\pm$0.01  &   \\
 133 &  21  40 21.8& +58 14 45.6 &  20.447 &  1.749 &   3.312&  0.7 $\pm$0.5  &  3.6  $\pm$0.5  &0.13 $\pm$0.01  &   \\
\hline
\end{tabular}
\end{table*}

The ages range from 0.1 to a few Myr (with some exceptions) which are comparable 
with the lifetime of T-Tauri stars (TTSs). The masses of these YSOs, range from 
$\sim$0.1-2.0 M$_\odot$, further indicate that they are probable TTSs and their
siblings.

Here we would like to point out that the estimation of the ages of the PMS stars by
comparing the observations with the theoretical isochrones is prone to two kinds of
errors; random errors in observations and systematic errors due to the variation 
between the predictions of different theoretical evolutionary tracks (see e.g. 
Hillenbrand 2005). The effect of random errors in determination of A$_V$, age and 
mass was estimated by propagating the random errors to the observed estimation by
assuming normal error distribution and using the Monte-Carlo simulations. The use 
of different PMS evolutionary model gives different age and age spread in a cluster
(e.g. Sung et al. 2000). Here in the present study we have used model by Siess
et al. (2000) only for all the BRCs, therefore our age and mass estimations are not 
affected by the systematic errors. However, the use of different sets of PMS 
evolutionary tracks will introduce systematic shift in age determination.  
The presence of binaries may be the another source of error in the age determination.
The presence of binary will brighten a star, consequently the CMD will yield a lower age
estimate. In the case of equal mass binary we expect an error of $\sim$ 50 - 60\% in
age estimation of the PMS stars. However, it is difficult to estimate the influence
of binaries on mean age estimation as the fraction of binaries is not known.
Here we would like to point out that we are interested mainly in the  {\it relative} 
ages of the aggregate members, in particular, the spatial differences with respect to 
the bright rim.                                                        
\begin{table*}
\caption{Average age of the YSOs associated with the inside/outside regions of the BRCs.}
\begin{tabular}{|p{.6in}|p{1.0in}|p{.8in}|p{2.2in}|p{2.2in}|}
\hline
BRC    & Region  & No. of stars & Mean age $\pm$ std dev (Myr)& Mean Av $\pm$ std dev (mag)\\
\hline
\end{tabular}
{\bf Only H$\alpha$ stars}\\
\begin{tabular}{|p{.6in}|p{1.0in}|p{.8in}|p{2.2in}|p{2.2in}|}
& & & &\\
BRC 2  & On/Inside BR & 11 & 1.0 $\pm$ 1.0 & 3.1 $\pm$ 1.4 \\
       & Outside BR&-  & - &- \\
BRC 11 & On/Inside BR &  4 & 1.5 $\pm$ 1.1 & 2.4 $\pm$  1.4\\
       & Outside BR&  5 & 2.1 $\pm$ 1.4 & 2.1 $\pm$  1.4\\
BRC 13 & On/Inside  BR&  3 & 0.6 $\pm$ 0.9 & 1.7 $\pm$  0.9\\
       & Outside BR&  2 & 1.6 $\pm$ 0.9 & 1.7 \\
BRC 14 & On/Inside  BR& 13 & 1.0 $\pm$ 0.7 & 3.9 $\pm$  1.8\\
       & Outside BR& 12 & 1.6 $\pm$ 0.7 & 3.0 $\pm$  0.6\\
BRC 27 & On/Inside  BR& 11 & 2.2 $\pm$ 1.1 & 2.3 $\pm$  0.6 \\
       & Outside BR& 12 & 2.2 $\pm$ 2.5 & 0.7 $\pm$  0.7\\
BRC 38 & On/Inside  BR&  6 & 2.1 $\pm$ 1.0 & 3.2 $\pm$  1.1 \\
       & Outside BR&  1 & 1.5 & 2.8\\
& & & & \\
\end{tabular}
{\bf H$\alpha$ and NIR excess stars} \\
\begin{tabular}{|p{.6in}|p{1.0in}|p{.8in}|p{2.2in}|p{2.2in}|}
 & & & & \\
BRC 2  & On/Inside BR & 13 & 1.0 $\pm$ 1.0 & 3.0 $\pm$ 1.4\\
       & Outside BR& - &-  &-\\
BRC 11 & On/Inside BR &  8 & 1.5 $\pm$ 0.8 & 2.3 $\pm$  1.2\\
       & Outside BR&  6 & 2.1 $\pm$ 1.3 & 2.4 $\pm$  1.5\\
BRC 13 & On/Inside  BR&  3 & 0.6 $\pm$ 0.9 & 1.8 $\pm$  0.9\\
       & Outside BR&  2 & 1.6 $\pm$ 0.8 & 1.7 \\
BRC 14 & On/Inside  BR& 15 & 1.1 $\pm$ 0.7 & 3.6 $\pm$  1.9\\
       & Outside BR& 21 & 1.7 $\pm$ 0.8 & 2.0 $\pm$  1.3\\
BRC 27 & On/Inside  BR& 15 & 2.3 $\pm$ 1.2 & 2.7 $\pm$  1.2 \\
       & Outside BR& 14 & 1.9 $\pm$ 1.4 & 0.7 $\pm$  0.7\\
BRC 38 & On/Inside  BR&  7 & 2.1 $\pm$ 0.9 & 3.3 $\pm$  1.0 \\
       & Outside BR&  4 & 2.7 $\pm$ 0.9 & 1.4 $\pm$  1.0\\
\hline
\end{tabular}
\end{table*}
\section{Star Formation Scenario in BRC Regions} 
 Propagating star formation, where energetic activity of massive stars compresses 
the surrounding gas and triggers the formation of new generation of stars
at the peripheries of H II regions (see e.g. Elmegreen 1998), is quite 
common in the Galaxy. Some different triggering mechanisms may work in such regions.
Briefly, the process which has been frequently supported by the observations is RDI, 
which takes place in small remnant clouds such as BRCs. The signature of star formation 
due to RDI is the presence of bright rims and embedded IR sources just inside the dense 
head. The collect-and-collapse model is another mechanism proposed by 
Elmegreen \& Lada (1977). The signature of this process are the presence of a
collected, dense layer adjacent to the ionization front and the presence of massive 
condensations  in it (e.g. Deharveng et al. 2003).
\subsection{ Small-Scale Sequential Star Formation }
As for the $S{^4}F$ hypothesis on the RDI star formation, there has been only 
qualitative evidence such as an asymmetric distribution of probable TTSs (Ogura 
et al. 2002) and of properties of NIR excess stars (Matsuyanagi et al. 2006). Very 
recently Paper I (Ogura et al. 2007) have quantitatively verified the $S{^4}F$ 
hypothesis by using {\it BVI$_{c}$} photometry of four BRCs. In the present study 
we follow the approach as given in Paper I . 
We have divided the YSOs (H$\alpha$ stars and NIR excess stars) associated with 
BRCs into two groups: those lying on/inside and outside of the rims (see Fig. A1). 
Mean ages and mean $A_V$ values have been calculated for these regions. 
Some of the stars in Table 4 show ages older than 5 Myr. Since 
the ages of the associated ionizing sources of BRCs studied here have maximum 
ages of 4-5 Myr, therefore the TTSs having ages greater than this can not be expected 
as products of triggered star formation. We suspect that they may have formed 
spontaneously in the original molecular cloud prior to the formation of the HII region (see 
Sect. 6.3). Some of them may be background stars; larger distances make them look older 
in the CM diagram. So while calculating the mean ages we have not included those stars. 
The results are given in Table 5, which shows that in almost all the BRCs the YSOs lying on/inside 
the rim are younger than those located outside it, whereas the mean {\it A$_{V}$} is 
higher on/inside the bright rim than outside it. The only exception for the mean age is BRC 27.

 The above results are exactly the same as those obtained 
in Paper I. Therefore, the present results further confirm the $S{^4}F$ hypothesis. 
As in Paper I, we again find a big scatter in the stellar ages
 for each region of all BRCs in spite of a clear trend of the mean ages. 
Possible reasons for the scatter include photometric errors, errors in 
extinction correction, light variation of young stars, their proper motions,  
binarity of the stars, etc. Photometric 
errors and light variation as big as 0.5 mag would affect stellar ages by $\sim$ 0.25 dex, 
so they do not seem to be the major reason for the scatter. As to the extinction correction, 
it probably does not affect the results much again, because in the $V_0, (V-I_c)_0$ CMD 
the isochrones are nearly parallel to the reddening vector.
The adopted evolutionary models and distances of the BRCs causes systematic shifts in ages
of the stars, but will not introduce scatters. 
As discussed in Paper I, we speculate that the proper motions of the newly born stars may  
probably the main cause of the scatter.  

Since stars inside the rim are often deeply embedded, mid-infrared (MIR) observations through the 
{\it Spitzer Space Telescope} can provide a deeper insight into the embedded YSOs. YSOs occupy 
distinct regions in the IRAC colour plane; this makes MIR colour-
colour diagram a very useful tool for the classification of YSOs. Whitney et al.
(2003) and Allen et al. (2004) presented independent model predictions for 
IRAC colours of various classes of YSOs. Fig. 3 presents [5.8]-[8.0] versus 
[3.6]-[4.5] colour-colour diagrams for the sources lying in the BRCs 2, 27 and 13/14 
regions. The sources within the box represents the location of Class II objects 
(Megeath et al. 2004, Allen et al. 2004). The sources located around [5.8]-[8.0]=0 
and [3.6]-[4.5]=0 are foreground/background stars, as well as diskless PMS 
stars (Class III objects). Sources with [3.6]-[4.5] $\ge$0.8 and/or [5.8]-[8.0]  $\ge$ 
1.1 have colours similar to those derived from models of protostellar objects 
with in-falling dusty envelopes (Allen et al. 2004). These are Class 0/I sources.

\begin{figure}
\includegraphics[scale = .36, trim = 10 10 10 10, clip]{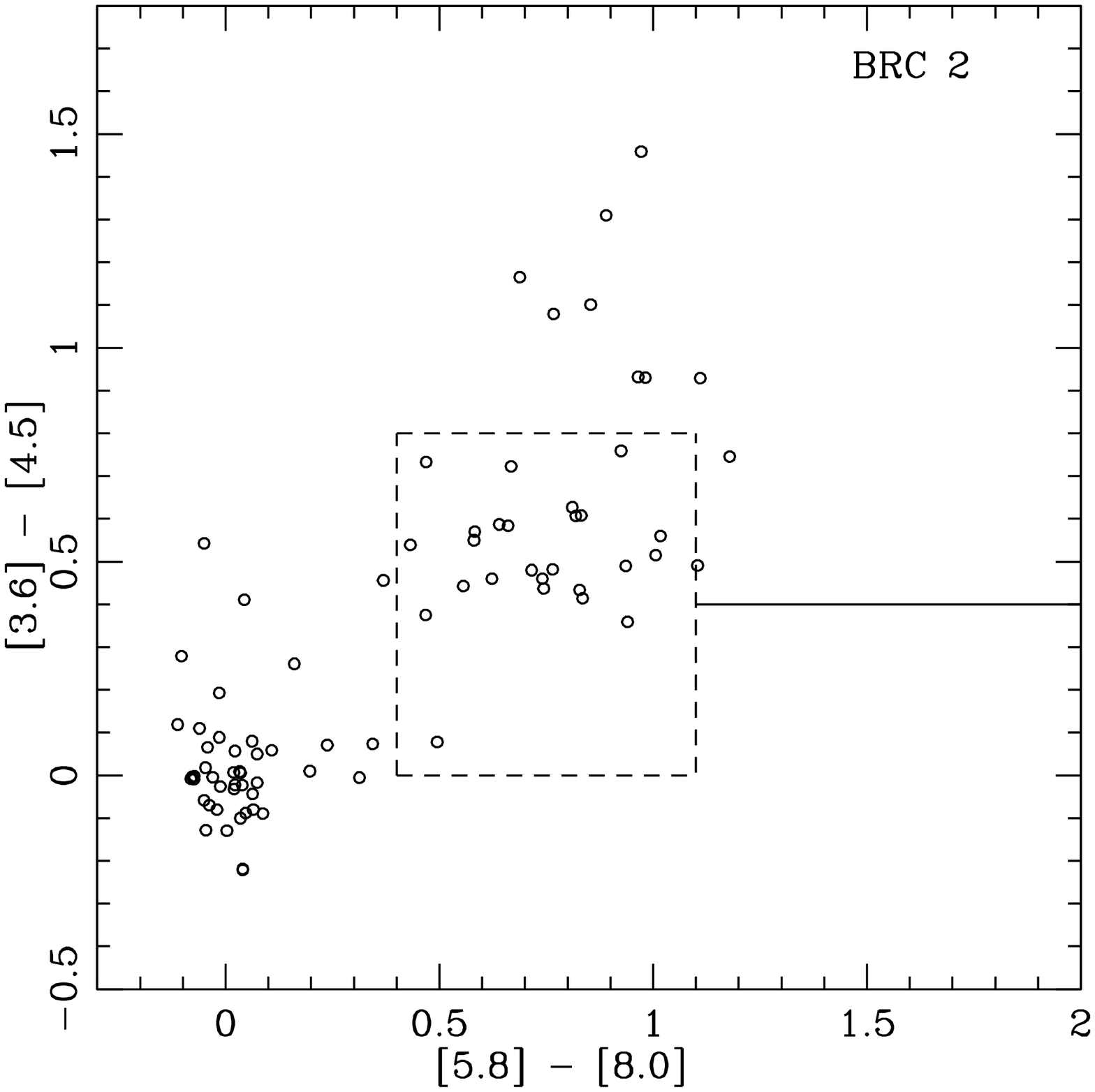}
\includegraphics[scale = .36, trim = 10 10 10 10, clip]{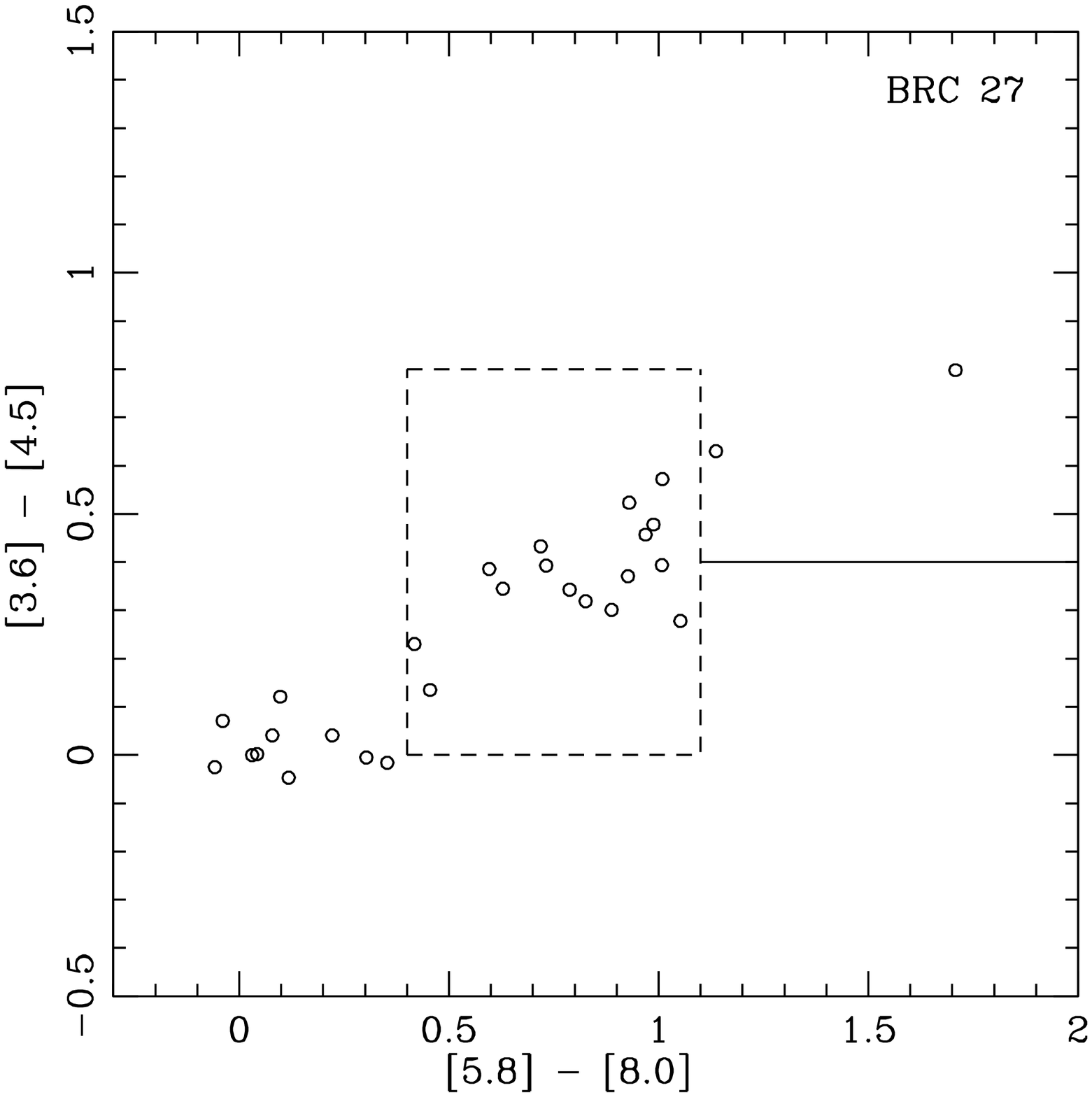}
\hspace{-.01cm}
\includegraphics[scale = .36, trim = 10 10 10 10, clip]{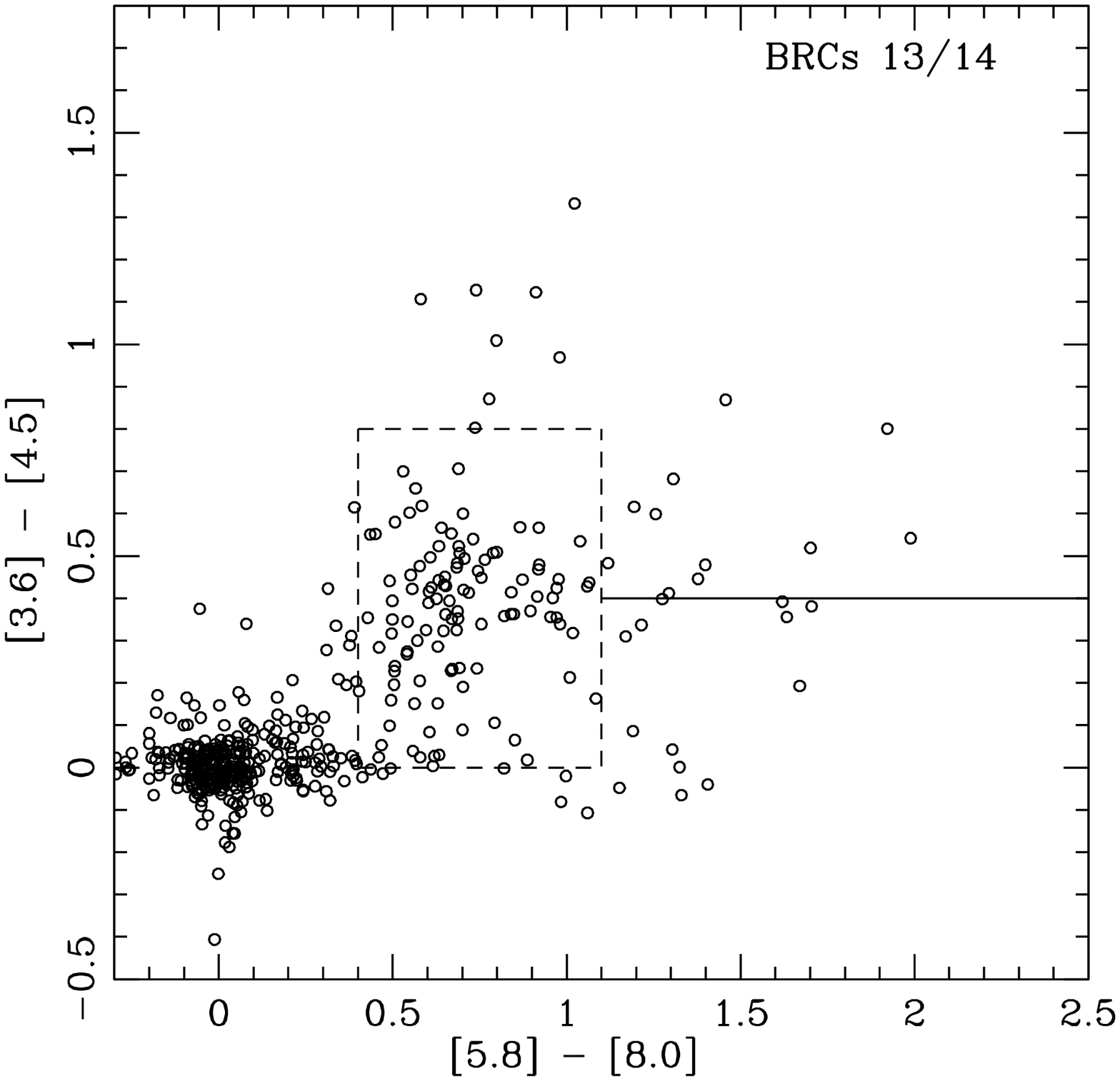}
\noindent{\footnotesize {\footnotesize \hspace{6.0cm} \bf Figure 3.} IRAC colour-colour diagrams for YSOs in BRCs 2, 27 and 13/14. The sources 
lying within the box are Class II sources. The sources located around 
[5.8]-[8.0] $\sim$  0 and [3.6]-[4.5] $\sim$ 0 are the field/ Class III stars. 
Sources with [3.6]-[4.5] $\ge$ 0.8 and/or [5.8]-[8.0]$\ge$ 0.8 represent 
Class 0/I sources. The horizontal continuous line shows the adopted division between 
Class I and Class I/II sources (see Megeath et al. 2004).}
\label{fig3}
\end{figure}
\begin{table*}
\caption{IRAC photometric magnitudes of the disk bearing candidates in BRCs 2, 27 and 13/14. The complete table is available in electronic form only.}
\begin{tabular}{|rrrrrrrrrrr|}
\hline
{\bf RA (J2000)}           &{\bf DEC (J2000)}      &{\bf [3.6]} & {\bf e[3.6]}&{\bf [4.5]} &{\bf e[4.5]}&{\bf [5.8]} & {\bf e[4.5] } &{\bf [8.0]} &{\bf e[8.0]}&{\bf IRAC type}\\
\hline
{\bf BRC 2}& &&&&&&&&&\\
 00 04 14.69  & +68 32 49.8 & 11.899&  0.033 &  10.97 & 0.03 &  10.095  &   0.052  &  8.985 &   0.028 & 0/I\\
 00 03 57.27  & +68 33 24.4 & 12.231&  0.038 &  11.74 &0.042 &  11.100  &   0.087  &  9.996 &   0.075 &0/I\\
 00 04 03.83  & +68 32 49.6 & 13.316&  0.064 &  12.57 &0.062 &  11.749  &   0.123  &  10.57 &   0.117 &0/I\\
\hline
\end{tabular}
\end{table*}

On the basis of the initial results from the {\it Spitzer } young cluster survey, 
Megeath et al. (2004) found  a cluster of young stars 
near the edge of BRC 2 along with a group of Class I sources at the northern apex 
of the cluster.  Table 6 summarizes the IRAC magnitudes of the disk bearing candidates 
of BRCs 2, 27 and 13/14, which is available in electronic form only. We reproduce the spatial distribution of the Class I and Class II 
sources in the BRCs 2 and 27 regions in Fig. 4. The upper panel for BRC 2  shows that the majority of the  
Class I sources are preferentially located away from the ionization sources 
(which lies downward in Fig. 4) as compared to the Class II sources. If we divide the 
BRC into two regions at $ Dec. \geq 60^o 34^\prime.5$, 
the fraction of Class 0/I sources in the northern region (which is  away from the 
ionizing source) is found to be 0.55 (
6 Class 0/I and 5 Class II sources), which is significantly higher than that 
(0.16, 3 Class 0/I and 16 Class II sources) 
in the southern region (towards the ionizing source). This distribution 
further manifests a small scale age sequence in the BRC 2 region.

In the cases of BRCs 13 and 14, Allen et al. (2005) reported that the Class I 
protostars are tightly clustered on the edge of the molecular clouds, coincident 
with the interface of the ionized and molecular gas, whereas the Class II stars 
are more widely distributed. The distribution of YSOs detected 
using the IRAC data is reproduced in Fig. 5, where again Class 0/I sources
are found concentrated inside the BRCs, which is in accordance with  
 the $S{^4}F$ hypothesis. In the IC 1396N = BRC 38 
region, Getman et al. (2007) found an elongated spatial distribution of YSOs with the 
youngest stars (Class 0/I) deeply embedded inside the cloud and relatively older stars 
aligned toward the exciting star, which again supports propagation of 
small-scale triggered star formation in that region.
\begin{figure}
\includegraphics[scale = .4, trim = 0 60 15 55, clip]{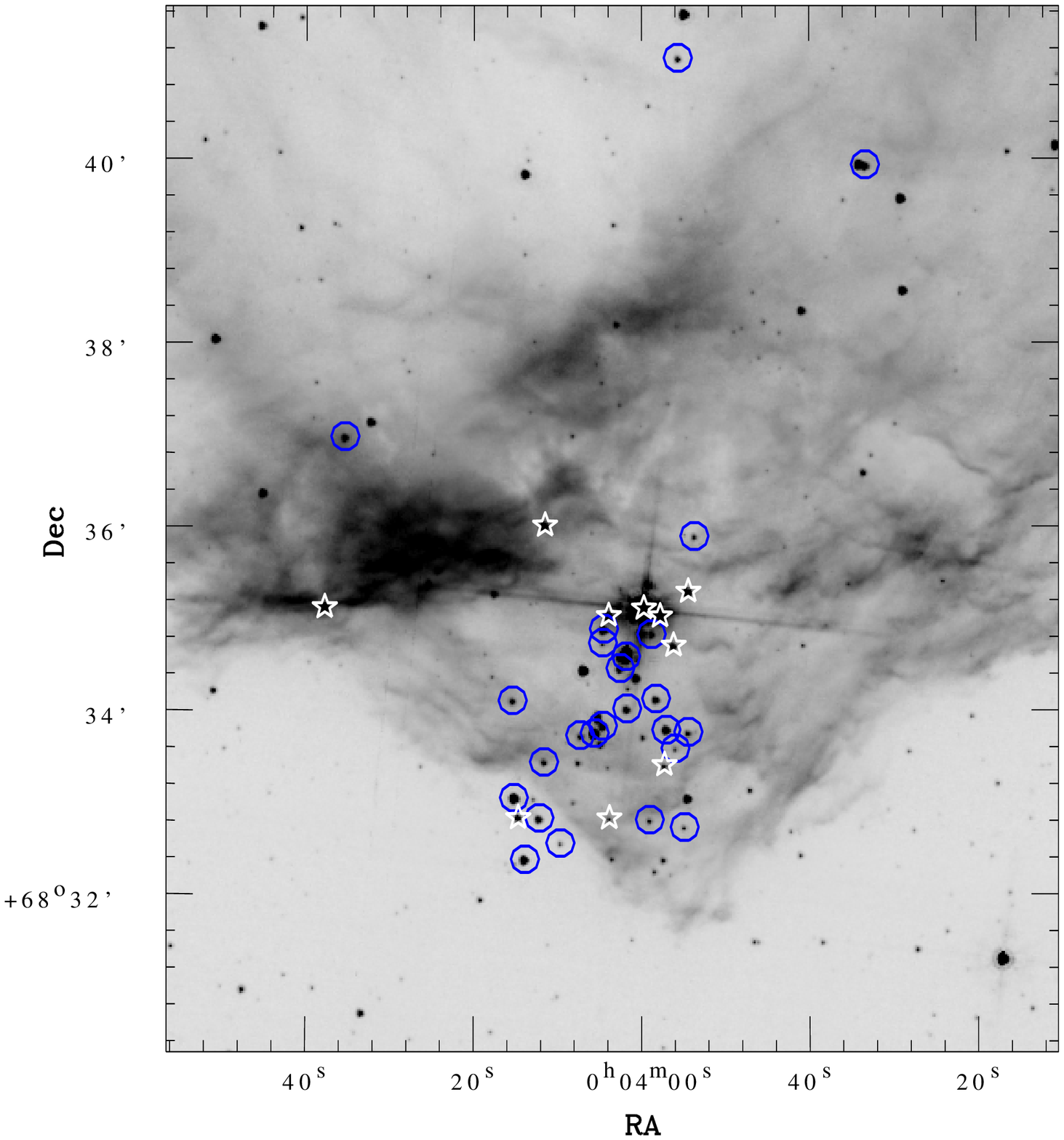}
\includegraphics[scale = .4, trim = 0 60 15 55, clip]{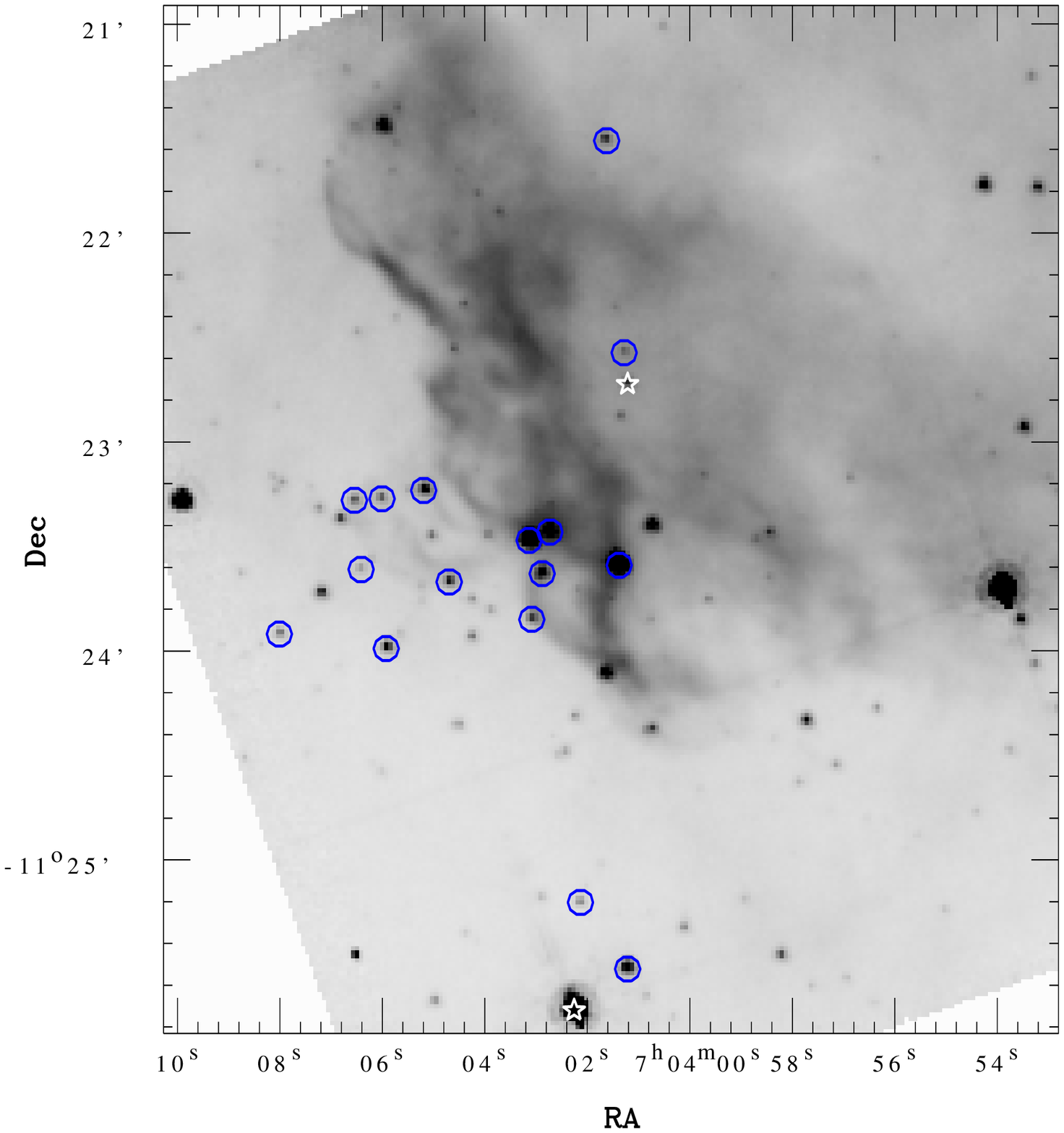}
\noindent{\footnotesize {\footnotesize \hspace{6.0cm} \bf Figure 4.} Spatial distribution of Class 0/I sources (star symbols) and Class II 
sources (open circles) in the BRC 2 (upper panel) and BRC 27 (lower panel) regions. }
\label{fig4}
\end{figure}
\begin{figure}
\includegraphics[scale = .5, trim = 0 60 15 20, clip]{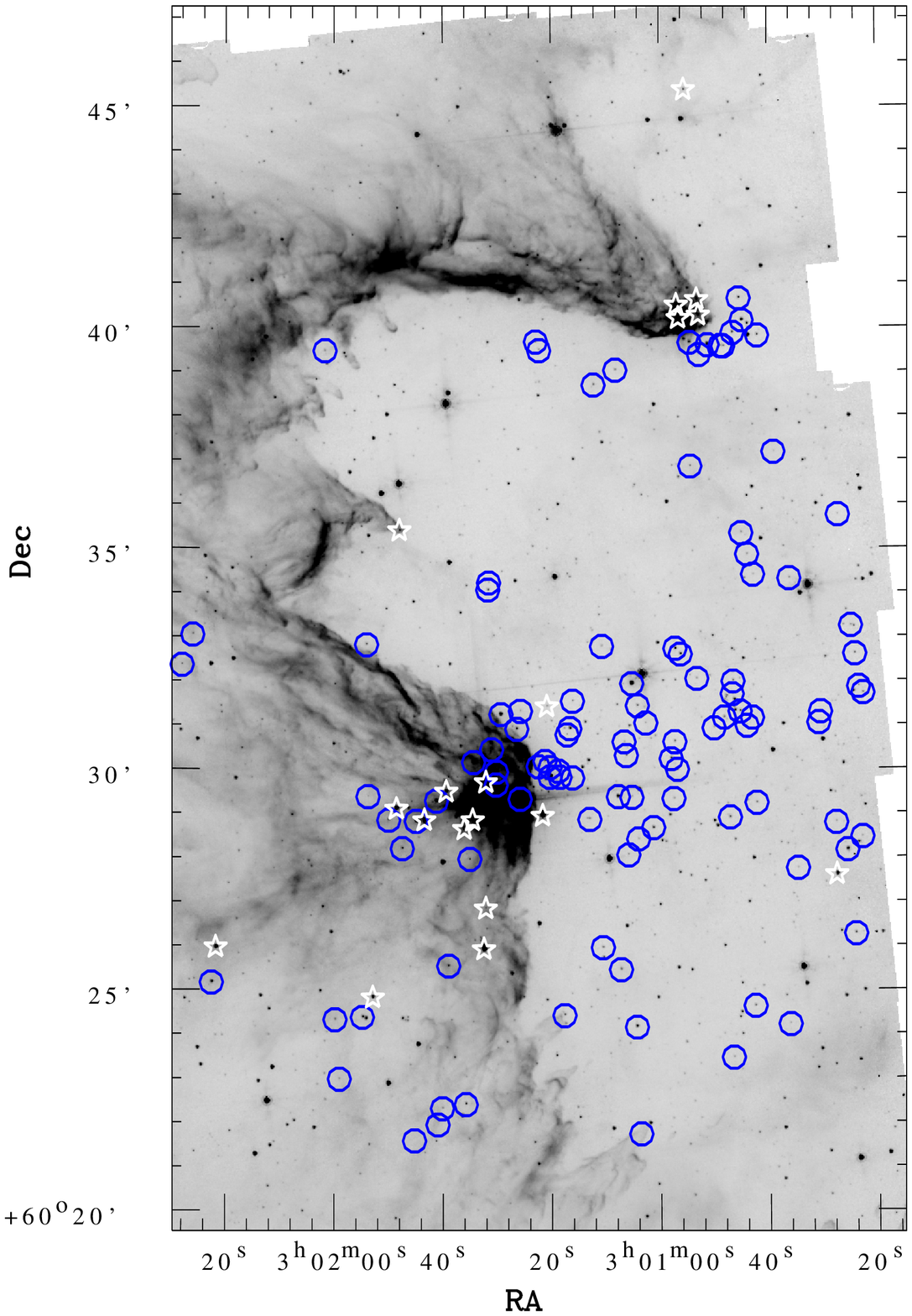}
\noindent{\footnotesize {\footnotesize \hspace{6.0cm} \bf Figure 5.} Spatial distribution of Class 0/I sources (star symbols) and Class II sources 
(open circles) in the BRCs 13 and 14 region identified in the {\it Spitzer/IRAC} data. }
\label{fig5}
\end{figure}
\subsection{Indication of Global Triggered Star Formation}                                        
 BRCs are considered to be a sort of remnants originated from dense part  
(cores) in an inhomogeneous giant molecular cloud. So, if the original cloud was 
big, the resultant BRC could have undergone a series of RDI events, leaving an 
elongated distribution of young stars; the distribution of such YSOs and its 
morphological details could be used to probe the star formation history in the 
OB association. With this expectation we have searched for NIR excess stars by 
using 2MASS PSC in the whole HII regions where the studied BRCs are located.
Figs. A2 to A5 show spatial distribution of NIR excess stars in the 
IC 1848W, IC 1848E, CMaR1 and IC 1396 regions which contain BRCs 11NE, 13/14, 
27 and 38 respectively. These figures are available in electronic form only.
Figures 6 to 8 show radial variation of  $\Delta(H-K)$ and $A_V$, for the 
stars located within the strip shown in Figs. A2 to A4. The NIR data along with 
$\Delta(H-K)$ and $A_V$ values are given in Table 7, which is available in 
electronic form only.

\begin{table*}
\caption{{\it J, H} and {\it K} magnitudes of the sources used in the analysis (cf. sec. 6.2). The complete table is available in electronic form only.}
\tiny
\centering
\begin{tabular}{|llrrrrrrrrr|}
\hline
   {\bf   RA }    & {\bf  DEC }    &{\bf 2MASS Name}    &{\bf J $\pm$ eJ} & {\bf H $\pm$ eH }& {\bf K $\pm$ eK } & {\bf Q flag}&{\bf C flag}& {\bf A$_V$} & {\bf $\Delta(H-K)$}&{\bf 2MASS/Matsuyanagi}\\
     {\bf  (J2000) } & {\bf (J2000)} &&{\bf (mag)}&{\bf (mag)}&{\bf (mag)}&& &{\bf (mag)}&{\bf (mag)}&{\bf et al. 2006(M06)}\\
\hline
{\bf  IC 1848W}&             &             &         &        &        &       &       &      &&\\
02 51 12.63 & +60 24 00.1 & 02511262+6024000  &13.719 $\pm$ 0.050  & 12.912 $\pm$ 0.051  & 12.361 $\pm$ 0.035  & AAA &000  &   0.00    & 0.05 &2MASS  \\ 
02 51 24.86 & +60 21 40.2 & 02512485+6021402  &14.305 $\pm$ 0.038  & 13.545 $\pm$ 0.043  & 13.017 $\pm$ 0.031  & AAA &000  &   0.00    & 0.05 &2MASS  \\ 
02 51 12.27 & +60 25 51.3 & 02511226+6025512  &15.955 $\pm$ 0.086  & 15.189 $\pm$ 0.103  & 14.630 $\pm$ 0.099  & AAA &c00  &   0.00    & 0.08 &2MASS  \\ 
\hline
\end{tabular}
\end{table*}
Fig. A2 shows that the NIR excess stars are aligned loosely towards the direction 
of BRC 11NE from the cluster IC 1848W which contains the ionizing sources 
(HD 17505, O6 V; HD 17520, O9V) of the HII region. 
A very recent study (while the present study was in the reviewing process) based on 
{\it Spitzer} observations by Koenig et al. (2008) also shows a nice alignment of
Class II stars towards the direction of the BRC 11NE region from the ionizing source(s)
(see their Fig. 10).
Figs. 6a and 6b show radial variation of $\Delta(H-K)$ and $A_V$, for the 
stars in the BRC 11NE region located within the strip shown in Fig. A2, as a 
function of radial distance from HD 17505. 
$\Delta(H-K)$ is defined as the horizontal displacement from the middle 
reddening vector (see Fig. 1).
The distribution of the NIR excess $\Delta(H-K)$ values shows an 
increasing trend as we move towards the BRC 11NE region. 
For the whole sample shown in Fig. 6a, the Kendall's tau test gives a 
positive correlation at a confidence level of about 85\%. The two extreme points
at radial distance $\sim28'$ have small $\Delta(H-K)$ values with small A$_V$ (0.24 
and 0.84) values. We presume that these sources are not embedded inside the rim
and lying on the outer region of the cloud. The two stars at  radial distance $\sim 5'$ and
$\sim 9'$ shows relatively higher value of  $\Delta(H-K)$ in comparison to
nearby stars. Exclusion of these four points gives a probability of $\sim$
97\% for a positive correlation between radial distance and $\Delta(H-K)$. 
Table 8 summarizes the results of the correlation analysis using the Kendall's
tau test.
\begin{figure}
\includegraphics[scale = .35]{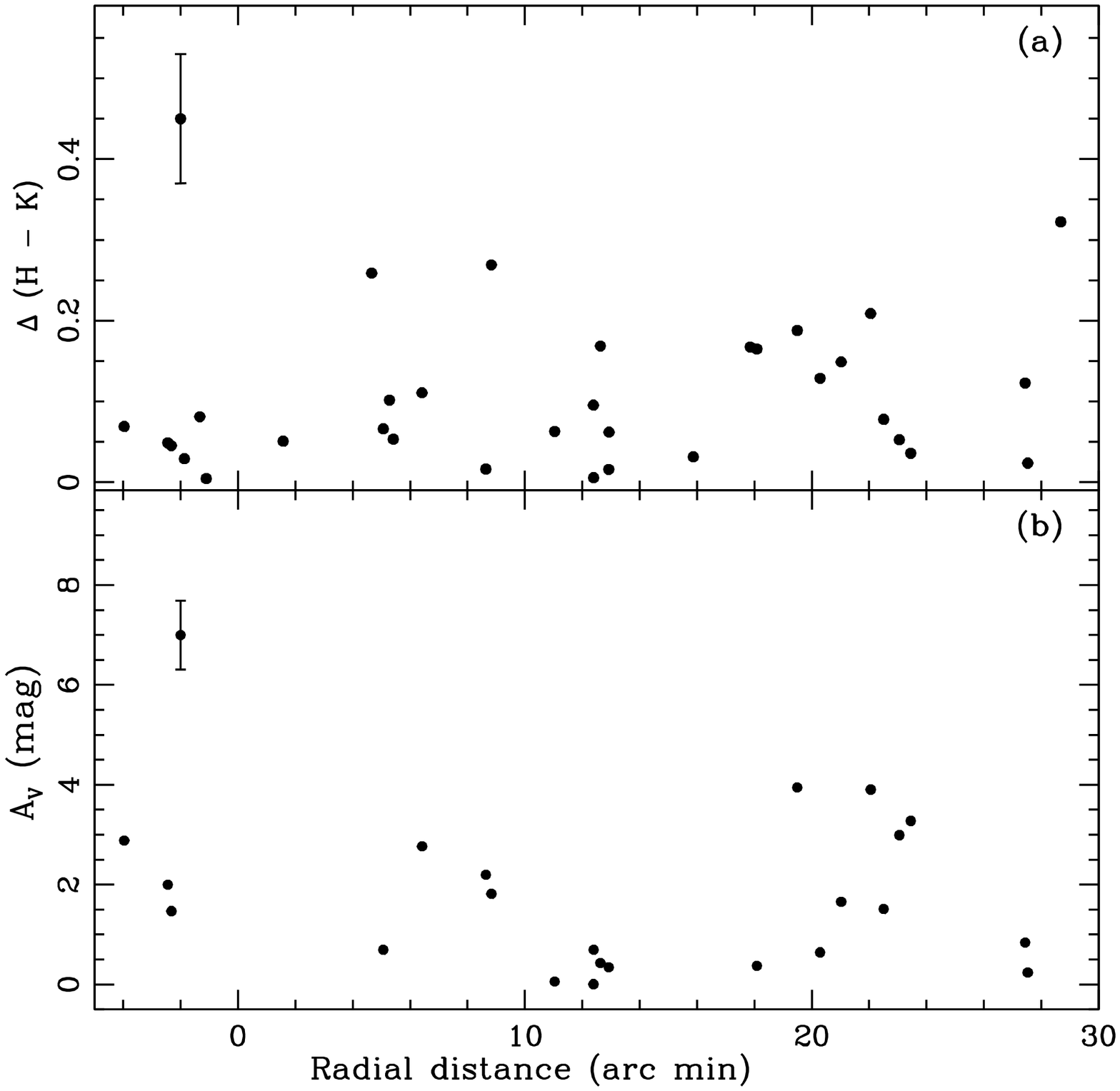}
\noindent{\footnotesize {\footnotesize \hspace{6.0cm} \bf Figure 6.} Variation of (a) NIR excess $\Delta$({\it H-K}) and (b) $A_V$ for the stars 
within the strip shown in Fig. A2 as a function of distance from HD 17505 toward BRC 11 region. 
Average error bar is shown at the upper-left corner of the plot.}
\label{fig6}
\end{figure}


\begin{figure}
\includegraphics[scale = .35]{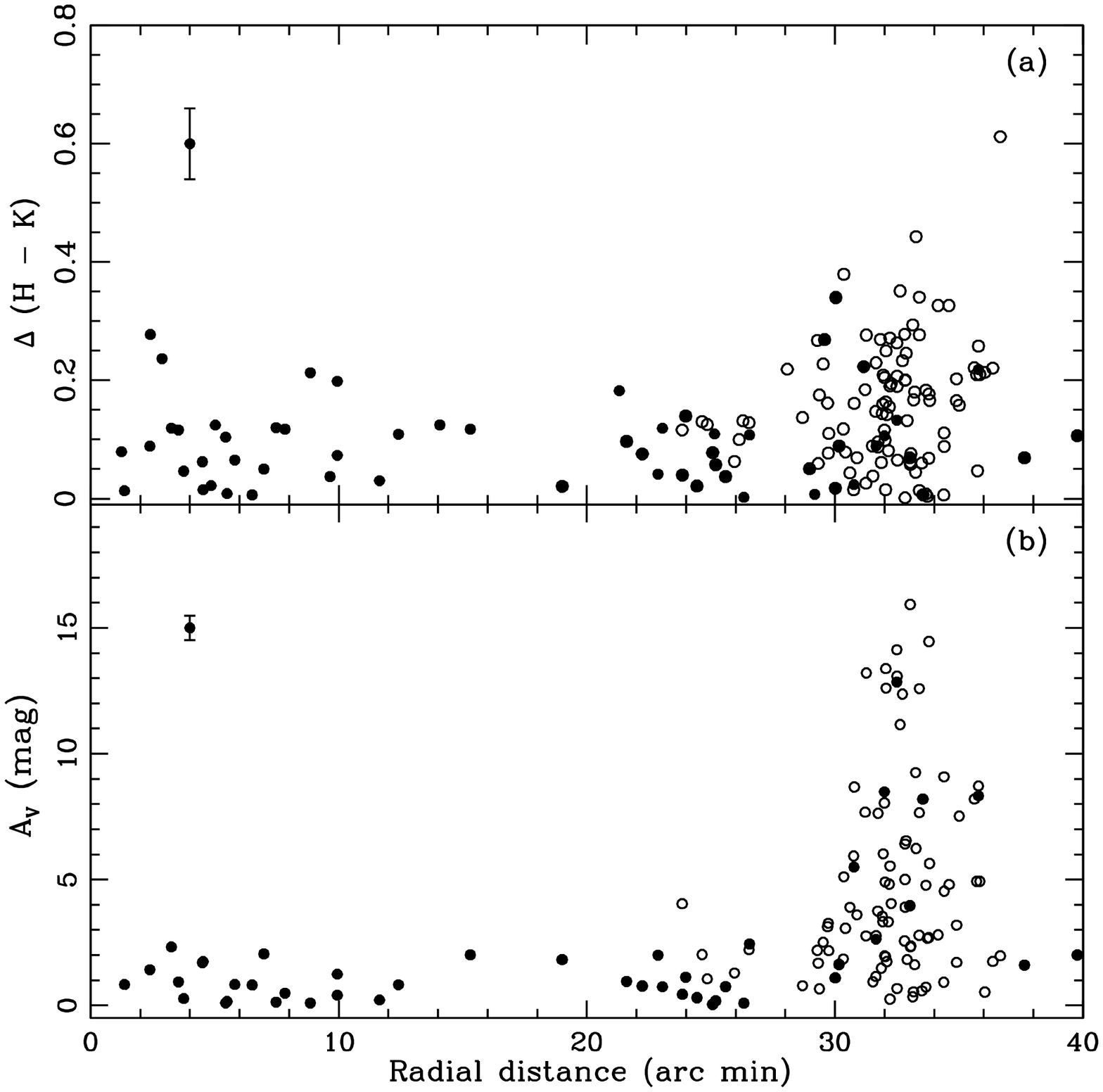}
\noindent{\footnotesize {\footnotesize \hspace{6.0cm} \bf Figure 7.} Variation of (a) NIR excess $\Delta$({\it H-K}) and (b) $A_V$ as a function 
of the distance from HD18326 toward BRC 14. Filled and open circles represents the data 
taken from the 2MASS catalogue and Matsuyanagi et al. (2006), respectively. 
Average error bar is shown at the upper-left corner of the plot.}
\label{fig7}
\end{figure}
\begin{figure}
\includegraphics[scale = .35]{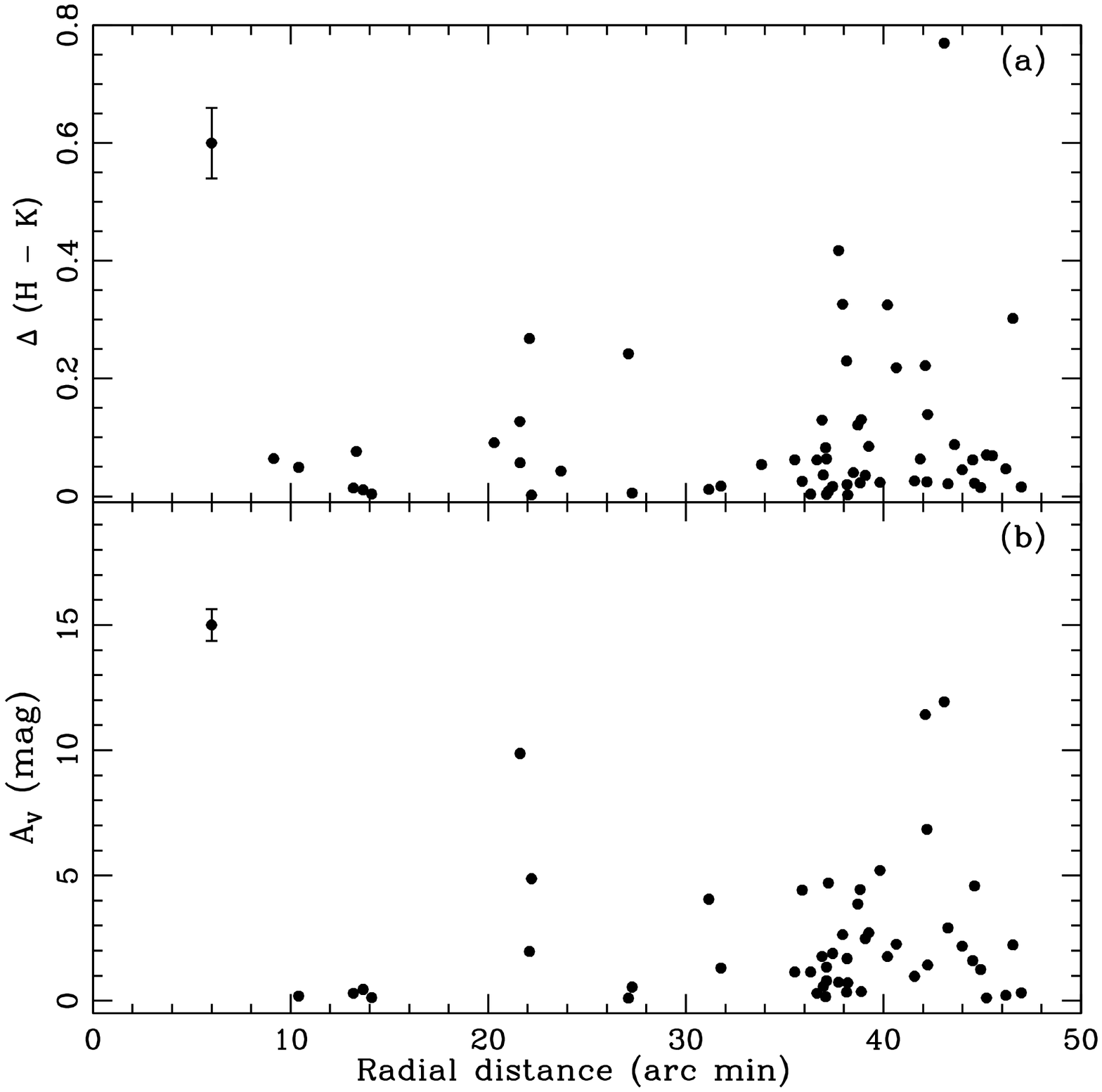}
\noindent{\footnotesize {\footnotesize \hspace{6.0cm} \bf Figure 8.} Variation of (a) NIR excess $\Delta$({\it H-K}) and (b) $A_V$ for the stars
 within the strip shown in Fig. A4 as a function of distance 
from the probable ionizing source (HD 53974) of the CMaR1 region. 
Average error bar is shown at the upper-left corner of the plot.}
\label{fig8}
\end{figure}


On the basis of the pressure of the ionized boundary layer (IBL) and that of the
molecular cloud, Thompson et al. (2004) have concluded that the cloud is
in pressure balance with the exterior ionized gas and photo ionization 
induced shocks are propagating in the cloud. They also concluded that overall
morphology of the cloud is similar to that predicted by RDI models 
(Bertoldi 1989, Lefloch \& Lazareff 1994).  
They have also estimated the duration over 
which the BRC 11NE region might have been exposed to the UV flux. Assuming that the 
rims are located at a distance of $\sim$22 pc from the ionizing sources, an 
ionization front expanding into a medium of homogeneous density at a speed of 11.4 km/s 
will take about 1.5 Myr to reach the rims. The mean age of the YSOs 
(H$\alpha$ stars and NIR excess stars) associated with BRC 11NE 
(both inside and outside the bright rim)
is found from Table 4 to be 1.7$\pm$1.0 Myr. Thus the sum of these two values
yields a time scale of $\sim$ 3.2 Myr, which is comparable to the 
MS lifetime ($\sim$ 4.0 Myr) of HD 17505 (Lang 1992, Schaller et al. 1992).
The above facts seems to support the triggered star formation scenario in the IC 1848W region.

Figure A3 shows that the distribution of the NIR excess stars in the IC 1848E region. We see 
they are aligned beautifully from the vicinity of the O7 star HD 18326 to the direction of BRC 14. 
A more impressive alignment of the Class II sources can be seen in Fig. 7 of Koenig et al. 
(2008). This spatial distribution of NIR excess stars resembles that in NGC 1893, 
where a similar nice distribution of NIR excess stars is noticed from the centre 
of the cluster containing several OB stars to the direction of the cometary globules 
Sim 129 and 130 (see Fig. 22 of Sharma et al. 2007). In the case of NGC 1893 evidence 
for triggered star formation due to RDI is also found. 
In Fig. 7a (upper panel) we plot the amount of NIR excess $\Delta(H-K)$ for the 
stars shown in Fig. A3 as a function of radial distance from the center of the cluster. 
Fig. 7a manifests an increase in NIR excess near BRC 14. A similar trend is noticed 
for the spatial distribution of $A_V$ (Fig. 7b). 
Kendall's tau test yields a positive correlation for the radial variation of $\Delta(H-K)$
and $A_V$ at a confidence level of better than 99.9\%.
As discussed in Matsuyanagi et al. (2006), these features indicate that stars 
located near BRC 14 should be younger than the rest of the stars. 

In Fig. A3 a loose clustering is also clearly visible around HD 18326. To our 
knowledge this clustering has not been designated so far as a known cluster\footnote{In a very recent study based on {\bf Spitzer} observations, Koenig 
et al. (2008) have also identified this cluster.}. 
$J/(J-H)$ CM diagram of the cluster region yields an age of $\sim$2 Myr. This 
cluster will be studied in detail in a forthcoming paper. On the other hand, the 
mean age of the YSOs associated with BRCs 13 and 14 (again, both inside and outside 
of the rims) is derived from Table 4 to be 1.0$\pm$0.9 Myr 
and 1.5$\pm$0.9 Myr, respectively, which are younger than the age of the cluster. 
Recently Nakano et al. (2008) reached the same conclusion, obtaining the ages of 
4 Myr and 1 Myr for a groups of H$\alpha$ emission stars around HD 18326 and that 
near eastern edge of the HII region, respectively. This again indicates that the 
star formation in the BRCs 13/14 region is triggered by the O star in the cluster 
region. Thus all the above mentioned  evidences clearly support a series of RDI 
processes which took place in the past starting from the vicinity of the O star. 

The spatial distribution of the NIR excess stars in the BRC 27 region is shown in Fig. A4.
Assuming that B0.5IV (HD 53974; marked as `2') and B1V (HD 54025; marked as `1') stars
are the ionizing sources for the BRC 27 region, the $\Delta(H-K)$ and $A_V$ 
distribution for the sources lying within the strip marked in Fig. A4 as a 
function of radial distance from  HD 54025 is shown in Fig. 8, which indicates 
relatively higher NIR excess and  $A_V$ near the BRC 27 region. 
The Kendall's tau test for the entire sample indicates a positive correlation between 
radial distance and $\Delta(H-K)$ and $A_V$ at a confidence level of 
$\sim$80\% and $\sim$95\%, respectively. The sources having radial distance 
$>43^\prime$ show small value of $A_V$ as well as $\Delta(H-K)$ as compared to
the sources lying around $40^\prime - 41^\prime$. We presume that these sources
are not embedded inside the rim and are lying on the outer periphery of the cloud. 
Exclusion of these points gives a probability of $\sim$98\% or better and 
99.9\% for a positive correlation between radial distance and $\Delta(H-K)$; 
and $A_V$, respectively. 
If the B1V/ B0.5 IV star(s) is (are) actually the ionizing source(s) 
for the region, the maximum MS life-time of the star(s) is $\sim$10 Myr 
(Lang 1992, Schaller et al. 1992), whereas the mean age of the YSOs associated with 
BRC 27 is estimated as 2.1$\pm$1.3 Myr, which is not in contradiction with that star formation in the BRC 27 
region may be initiated by the UV-radiation from these star(s).

Sicilia-Aguilar et al. (2004) have shown that in the case of the Tr 37/ IC 1396 Globule region,
CTTSs are found to be aligned towards the direction of IC 1396 Globule from the ionizing 
source, HD 206267 (O6). Sicilia-Aguilar et al. (2005) found that 
most of the younger ($\sim$1 Myr) members appear to lie near or within the IC 1396 
Globule. They concluded that it can be indicative of the triggered star formation.
 Fig. A5 shows distribution of NIR  excess stars in the Tr 37/IC 1396 
Globule/BRC 38 region, where they seem to align loosely towards the direction of IC 1396 
Globule and BRC 38. Their radial distribution of NIR excess $\Delta$({\it H-K}) and 
$A_V$ does not show any trend, however. By using the ages of the YSOs near IC 1396 
Globule given by Sicilia-Aguilar et al. (2005) we obtained their mean age of 
$\sim$1.8$\pm$1.1 Myr, whereas for the YSOs near BRC 38 the mean age is estimated from 
Table 4 to be $\sim$2.2$\pm$0.9 Myr. The upper main-sequence turn off age of Tr 37 is 
found to be $\sim$3 Myr (Contreras et al. 2002). Thus the aligned distribution of YSOs
from the ionizing source HD 206267 towards IC 1396 Globule and BRC 38 and their younger age as 
compared to the central cluster Tr 37 suggest a triggered star formation scenario in the region.
 
We conclude that the global distribution of YSOs, their radial distribution of the 
amount of NIR excess $\Delta$(H-K) as well as of $A_V$ in each HII region studied 
here clearly show evidence that a series of RDI processes proceeded in the past from 
near the central O star(s) towards the peripheries of the HII region.
\begin{table}
\caption{Correlation between radial distance and $\Delta$(H-K), A$_V$.
The probability P(0) indicates that no correlation is found with the generalized
non-parametric Kendall's tau statistics.}
\begin{tabular}{|llll|}
\hline
Radial distance from   &P (0)  & P (0)  & comment\\
the ionizing source  &$\Delta$(H-K)&A$_V$ & \\
(arc min)&&&\\
\hline
&{\bf BRC 11}&&\\
5 - 30   & 0.150 & - &  \\
5 - 30   & 0.026 & - & Excluding Outliers \\
&&&(see text)\\
            &&&\\
&{\bf BRC 14}                   &                  &                   \\
0 - 40   & $<0.00$ & $<$ 0.00 &  \\
&&&\\
&{\bf BRC 27}                   &                  &                   \\
0 - 48   & 0.230 & 0.04 &  \\
0 - 43   &0.025  &0.001& (see text)\\
\hline
\end{tabular}
\end{table}

\subsection{Star Formation inside `A'-Type BRCs}

 The {\it Spitzer} IRAC data on BRC 2, BRC 13 and BRC 14 manifest
that the Class 0/I sources are concentrated inside the rim. The SCUBA imaging 
survey of submillimeter continuum emission from BRCs by Morgan et al.
(2008) has shown that the embedded cores are likely to contain
Class 0 protostars. On the basis of combination of the observed submillimeter
flux excess and high dust temperature they concluded that star formation
may be on-going within the BRCs. They have further concluded that the
majority of the sources have $L_{bol} > 10 L_\odot$, indicating that the sources
are intermediate to high-mass stars. Some of the higher luminosity sources 
(e.g. in BRCs 13 and14) may be proto-clusters. The {\it Spitzer} IRAC data manifest 
that in fact these two BRCs host a proto-cluster (cf. Fig. 6).

Morgan et al. (2008) did not find evidence for interaction of the external 
ionization field with the star formation inside `A' type BRCs (for the morphological 
types of BRCs we refer to SFO91) and concluded that the star formation in these 
clouds is not subjected to the RDI process. The present work includes four BRCs 
of the `A' type, namely, BRCs 2, 14, 27 and 38 (as for BRC 38, see Sect. 8) and 
provides strong evidence for star formation due to RDI occurring in BRCs, however. 
As we have seen in Sect. 6.1, BRCs 2, 14 and 38 show such age gradients that stars 
located on/inside the rim are younger than those located outside it, i.e., toward 
the ionizing source, evidencing the most recent RDI phenomenon. In addition, our 
results in Sect. 6.2 as well as recent study based on {{\it Spitzer} observations 
by Koenig et al. (2008)} manifest a nice, global alignment of NIR excess 
stars in IC 1848E from the O7 star HD 18326 to BRCs 13 and 14. The spatial 
distribution of H$\alpha$ emission stars found by Nakano et al. (2008) also 
revealed a similar alignment. Thus the ages of the YSOs and their spatial distribution 
in the region clearly support a series of RDI processes which have 
been taking place in the past until very recently. These results do not support 
the notion of Morgan et al. (2008) that star formation in/around `A'-type BRCs 
is not subjected to the RDI triggering process.

\section{ Evolution of H$\alpha$ EW and disk of T-Tauri stars}

H$\alpha$ emission and IR excess are important signatures of young PMS stars. 
These signatures in CTTSs indicate the existence of a well-developed circumstellar 
disk actively interacting with the central star. Strong H$\alpha$ emission (equivalent 
width EW $>$ 10{\AA}) in CTTSs is attributed to the magnetospheric accretion of the 
innermost disk matter onto the central star (Hartmann et al. 1994; Edwards et al. 
1994; Muzerolle et al 2001 and references therein). On the other hand the weak H$\alpha$ 
emission (EW $<$ 10\AA) in weak-line TTSs (WTTSs), which lack disks (or, at least inner disks), 
is believed to originate from their chromospheric activity (e.g., Walter 1988;  Martin 1998). 
In 1990s a large number of WTTSs were found in and over wide areas around T associations by 
X-ray surveys with ROSAT, which aroused active studies on the nature of the so-called dispersed 
WTTSs. For a detailed discussions on this topic we refer to Caillault et al. (1998). 
As for the relation of the WTTS to the CTTS, the "standard model" (Kenyon 
and Hartmann 1995) postulates that the latter evolves to the former by losing the circumstellar 
disk (or, at least its inner part). Actually analysis of the age distribution derived from the 
HR diagram of, e.g., the Taurus region indicated that the WTTSs are systematically older than 
the CTTSs, but the statistical significance was low (Kenyon and Hartmann 1995; Hartmann 2001; 
Armitage et al. 2003). 

On the other hand, there also have been many observations which claimed 
that the CTTS and the WTTS are coeval and have indistinguishable stellar properties (e.g., Walter 
et al. 1988; Lawson et al. 1996; Gras-Velazquez and Ray 2005). From the analyses of the HR diagram 
of the CTTSs and WTTSs in Chamaeleon I, Lawson et al. (1996) concluded that some stars may be born 
even almost diskless or lose the disk at very early stages (age $<$ 1 Myr). However, in order to 
explain the co-existence and approximate coevality of CTTSs and WTTSs in a star forming region, 
it is usually postulated that YSOs display a wide range of disk masses and their accretion 
activity and/or the dispersal of the disk takes place in a correspondingly wide range of 
time-scales (Bertout et al. 2007; Furlan et al. 2006). Based on L-band surveys of clusters of 
various ages, Haisch et al. (2001) reached the quantitative conclusion that the disk fraction 
is initially very high ($\ge$ 80\%) and that one half the stars lose their disks in $\sim$ 3 Myr 
and almost all in $\sim$ 6 Myr. Armitage et al. (2003) obtained similar results that around 
30\% of stars lose their disks within 1 Myr, while the remainder have disk lifetimes that 
are typically in the 1 - 10 Myr range. Recently, Bertout et al. (2007),  by using new 
parallaxes for CTTS and WTTS in the Taurus-Auriga T association, concluded that their 
observed age and mass distribution can be explained by assuming that a CTTS evolves 
into a WTTS when the disk is fully accreted by the star. 

In the present work we have derived the ages of 93 H$\alpha$ emission stars, hence we 
can study the evolution of the H$\alpha$ emission activity in TTSs. The advantage of 
our sample in addressing this issue is that the stars are spatially, i.e., three-dimensionally, 
very close to each other, so there should be no problem of the distance  difference, contrary to 
the extended T associations. The H$\alpha$ EWs are taken from Ogura et al. (2002); however 
the values reported as EWs in their Table 5 are values in pixels. To convert these 
values into {\AA} we multiply the reported values by a factor of 3.8 (see Ikeda et al. 2008). 

\begin{figure}
\includegraphics[scale = .4, trim = 10 10 10 10, clip]{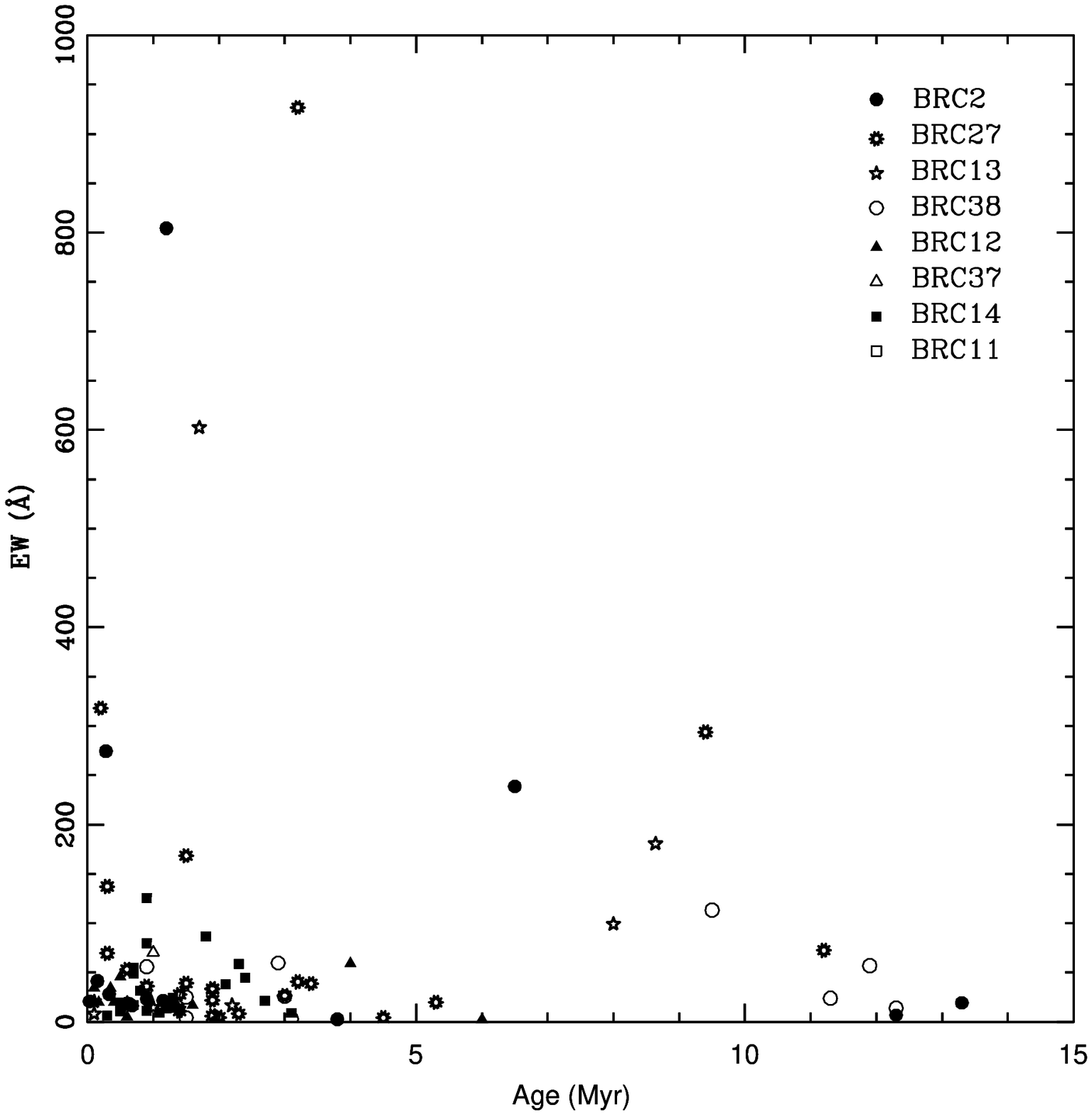}
\noindent{\footnotesize {\footnotesize \hspace{6.0cm} \bf Figure 9.} EWs of H$\alpha$ emission stars in our sample as a function of stellar ages.  }

\label{fig9}
\end{figure}

In Fig. 9 we plot the EWs of H$\alpha$ emission stars as a function of age to explore 
possible evolutionary trends. Although, the dispersion around younger side is quite large, still  in 
general there seems to be a decreasing trend in EW with the age. Here it is worthwhile to mention 
that a rather similar trend in the EWs of H$\alpha$ emission line of HAe/Be stars is reported by Manoj et al. (2006).
The distribution of EWs in Fig. 9 indicates that the accretion activity in the TTSs associated with BRCs drops substantially by 5 Myr.
In Fig. 9, there seems to be a small group of H$\alpha$ emission stars having far larger ages ($\ge$ 5 Myr) and a relatively 
elevated level of EWs. The masses of these stars lie in the range
$0.6\ge M/M_\odot \ge1.9$, whereas the majority of the YSOs having age $\le$5 Myr have
masses in the range $0.1\ge M/M_\odot \ge1.2$. 
If we take their ages at their face values, they presumably are not products  
of triggering. Since the ages of the ionizing sources of BRCs studied here 
have maximum age of 4-5 Myr, stars having ages greater than $\sim$5 Myr can 
not be expected as results of triggered star formation, but must have formed 
spontaneously prior to the formation of the HII region. 
The stars with ages $\ge$ 5 Myr seems to born with large 
disk masses and spent a substantial part, say, half of their ages unexposed to 
UV radiation from O stars, the long lifetime of their accretion activity may be 
understood. Johnstone et al. (2004) have 
reported that the far-UV  radiation from nearby massive star(s) may cause 
photo-evaporation of YSO disks resulting in short ($\sim$ 10$^6$ yr) disk lifetimes. 
However, Fig. A1, where these stars are marked with crosses, shows that they 
are located both inside and outside the bright rims mixed with H$\alpha$ stars of 
younger ages. So their origin remains a mystery. But in the case of BRC 38, which 
contributes four to this group of altogether eleven stars, Getman et al. (2007) 
recognized, apart from young stars associated with the BRC, an older population of 
PMS stars dispersed in IC 1396. We suspect the above four stars may belong to 
this population and formed in the original molecular cloud prior to the formation of 
HD 206267. In Fig. A1 they look concentrated along the bright rim, but note 
that the H$\alpha$ survey by Ogura et al. (2002) is limited down to +58 13 35, 
which is only a few arcmin south of the bright rim.
Here it is worthwhile to mention that in the case of cluster Tr 37 (age 1-5 Myr), 
Sicilia-Aguliar et al. (2005) have found a few stars having age $>$ 5 Myr. They pointed
out that in some clusters intermediate-mass stars seem older than low-mass stars and 
this effect seems to be related to a problem defining the birth line for intermediate-mass
stars (Hartmann 2003).  
\begin{figure}
\includegraphics[scale = .43, trim = 10 10 10 10, clip]{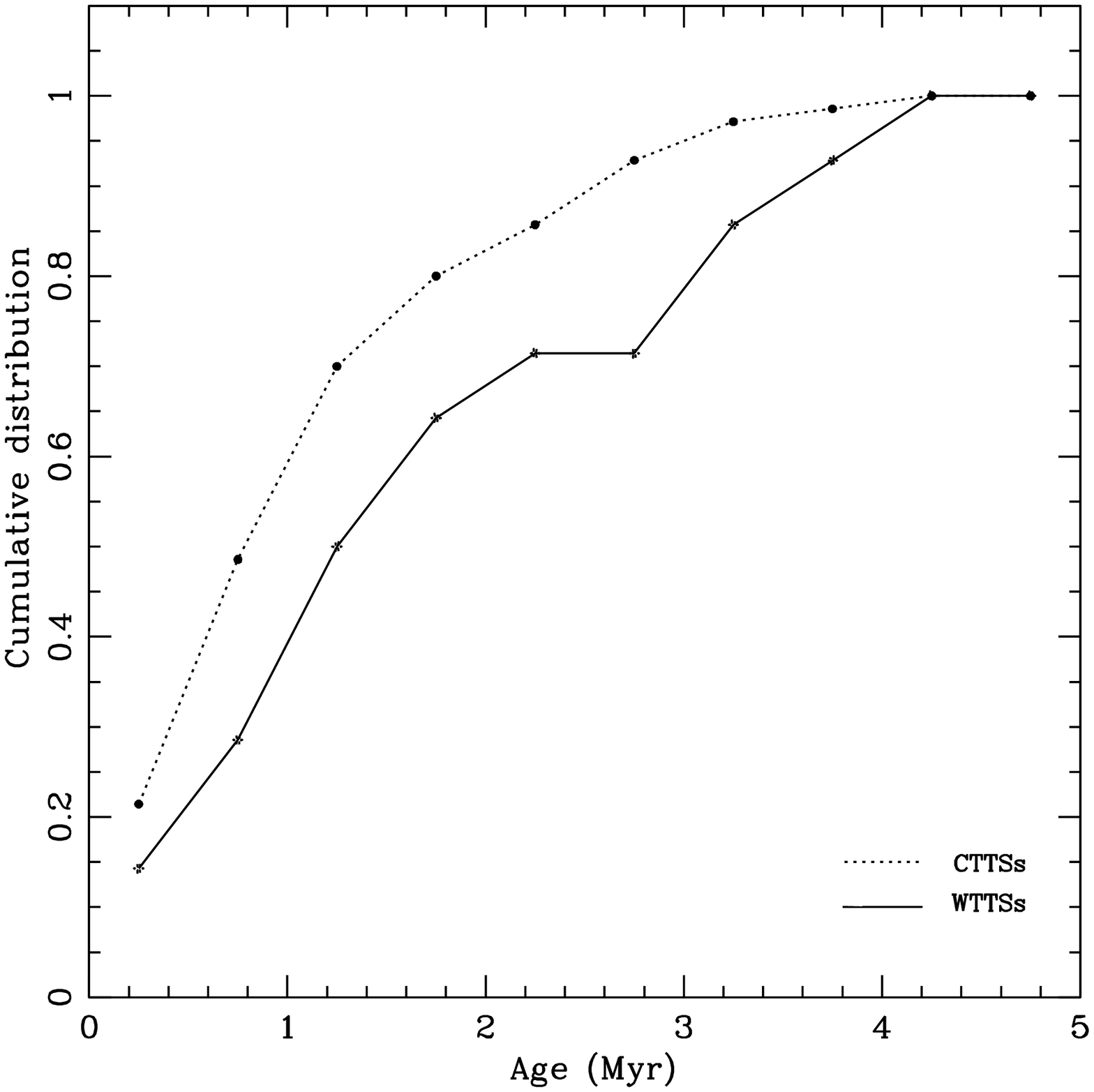}
\noindent{\footnotesize {\footnotesize \hspace{6.0cm} \bf Figure 10.} Cumulative distributions of CTTSs and WTTSs in our sample as a function of stellar age.}
\label{fig10}
\end{figure}

\begin{figure}
\includegraphics[scale = .43, trim = 10 10 10 10, clip]{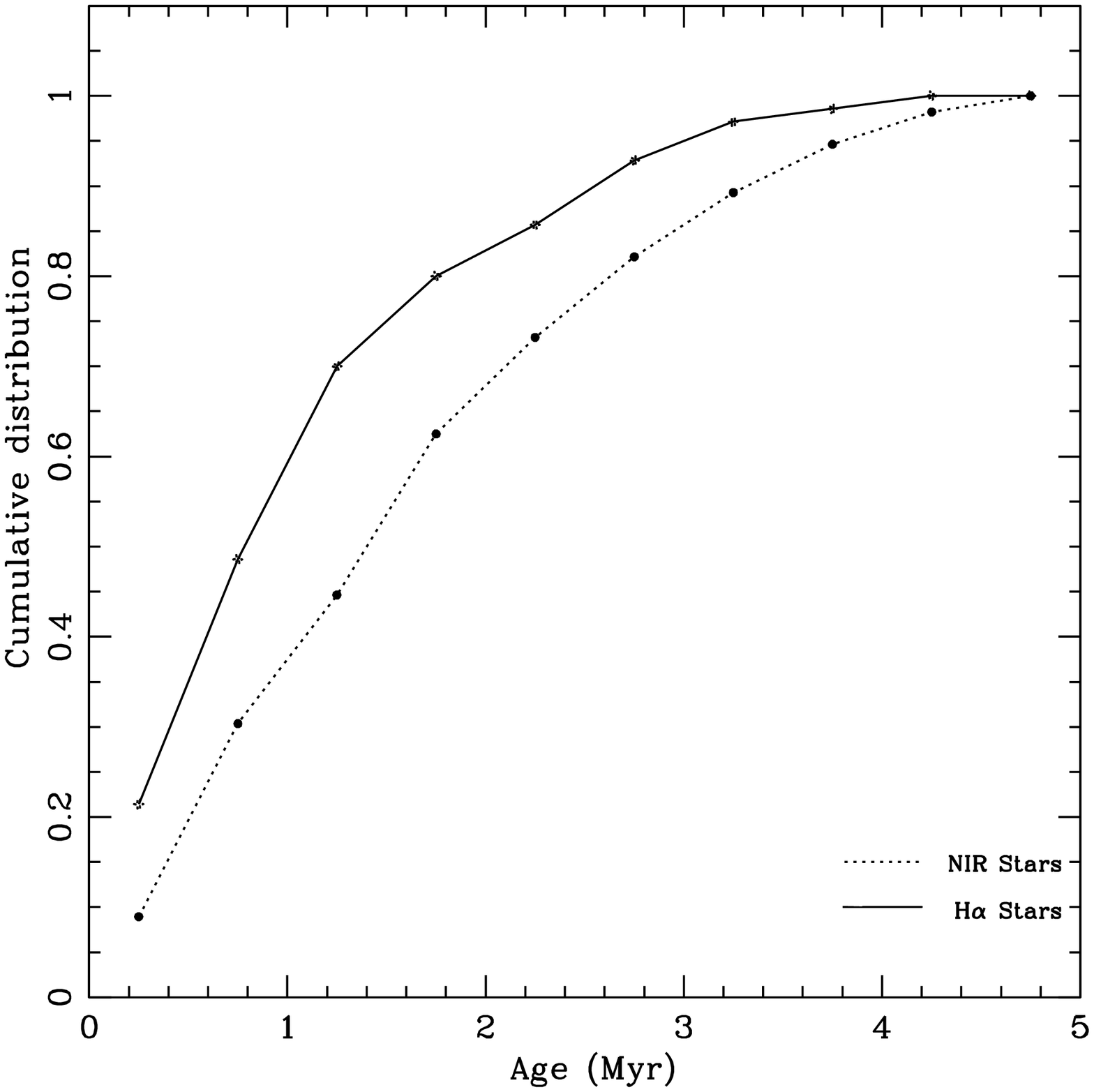}
\noindent{\footnotesize {\footnotesize \hspace{6.0cm} \bf Figure 11.} Cumulative distributions of H$\alpha$ emission and 
NIR excess stars in our sample as a function of stellar age.}
\label{fig11}
\end{figure}

Figure 10 shows the cumulative distribution of CTTSs (EW $\geq$ 10{\AA}) and WTTSs  (EW $<$ 10\AA) 
(for stars having age $\le$ 5 Myr) as a function of age. Fig. 10 manifests that CTTSs are relatively 
younger than WTTSs. A Kolmogorov-Smirnov test confirms the statement that the cumulative distributions 
of CTTSs and WTTSs are different at a mean confidence level of $\sim$70\% with minimum and maximum  
confidence level (obtained using the Monte Carlo simulations) of $\sim$55\% and $\sim$90\%, respectively. 
This result is in agreement with that of Bertout et al. (2007) for the  Taurus-Auriga T-association, that 
WTTSs are older than CTTSs and CTTSs evolve into WTTSs. 
In Fig. 11 we plot cumulative age distribution of H$\alpha$ emission stars (EW $\geq$ 10\AA) and of NIR 
excess stars. Fig. 11, at a mean confidence level of $\sim$98\% (with a minimum and maximum confidence 
level of $\sim$92\% and $\sim$99.4\%) indicates that YSOs exhibit NIR excess for 
a relatively longer time as compared to  accretion activity.
Although our sample is small and the age span is very short, the obtained CTTS fraction (from Tables 3 
and 4) in BRCs seems to follow the trend of TTSs in the Taurus region  as given by Armitage et al. (2003). 

\section{Mass Function of BRC aggregates}

The initial mass function (IMF) is an important tool to study the star 
formation process. Morgan et al. (2008), using SCUBA observations, have 
estimated the masses of 47 dense cores within the heads of 44 BRCs. They 
concluded that the slope of the mass function of these cores is significantly 
shallower than that of the Salpeter mass function. They also concluded that it depends 
on the morphological type of BRCs (for the morphological description of BRCs 
we refer to SFO91): `A' type BRCs appear to follow the mass spectrum of the
clumps in the Orion B molecular cloud, whereas the BRCs of 
the `B' and `C' types have a significantly shallower mass function. 

It would be worthwhile to compare the mass function of protostars 
given by Morgan et al. (2008) with that of BRC aggregates. In Fig. 12 we plot 
cumulative mass function (CMF) of the YSOs in 7 BRCs, namely BRCs 2, 11NE, 12, 
13, 14, 27 and 38, in the mass range of 0.2$\le$ M/M$_\odot$ $\le$1.2. 
Here we have supplemented the present data with the data of BRC 12, taken from Paper I,
because among the present sample of BRCs there are fewer number of BRCs of type `B'
than those of type `A'. 
The CMF of the dense cores by Morgan et al. (2008) is also plotted for comparison.
 \begin{figure}
\includegraphics[scale = .43, trim = 5 5 5 5, clip]{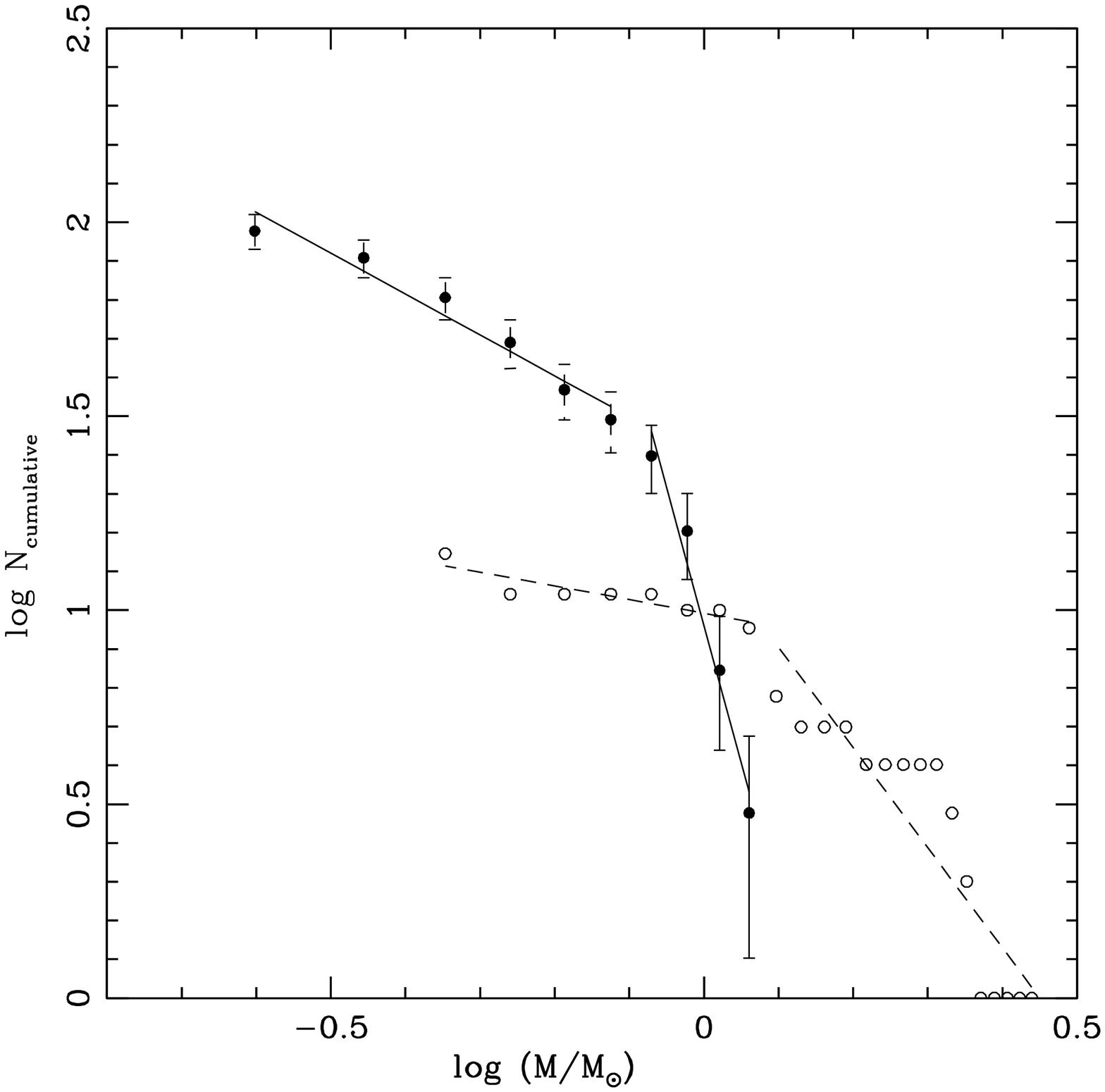}

\noindent{\footnotesize {\footnotesize \bf Figure 12.} Cumulative mass function (CMF) of YSOs in the 7 BRCs (filled circles). 
Error bars represent $\pm$$\sqrt{N}$ errors. Open circles represent the CMF for the cores by Morgan et al. (2008).}
\label{fig12}
\end{figure}

\begin{figure}
\includegraphics[scale = .43, trim = 5 5 5 5, clip]{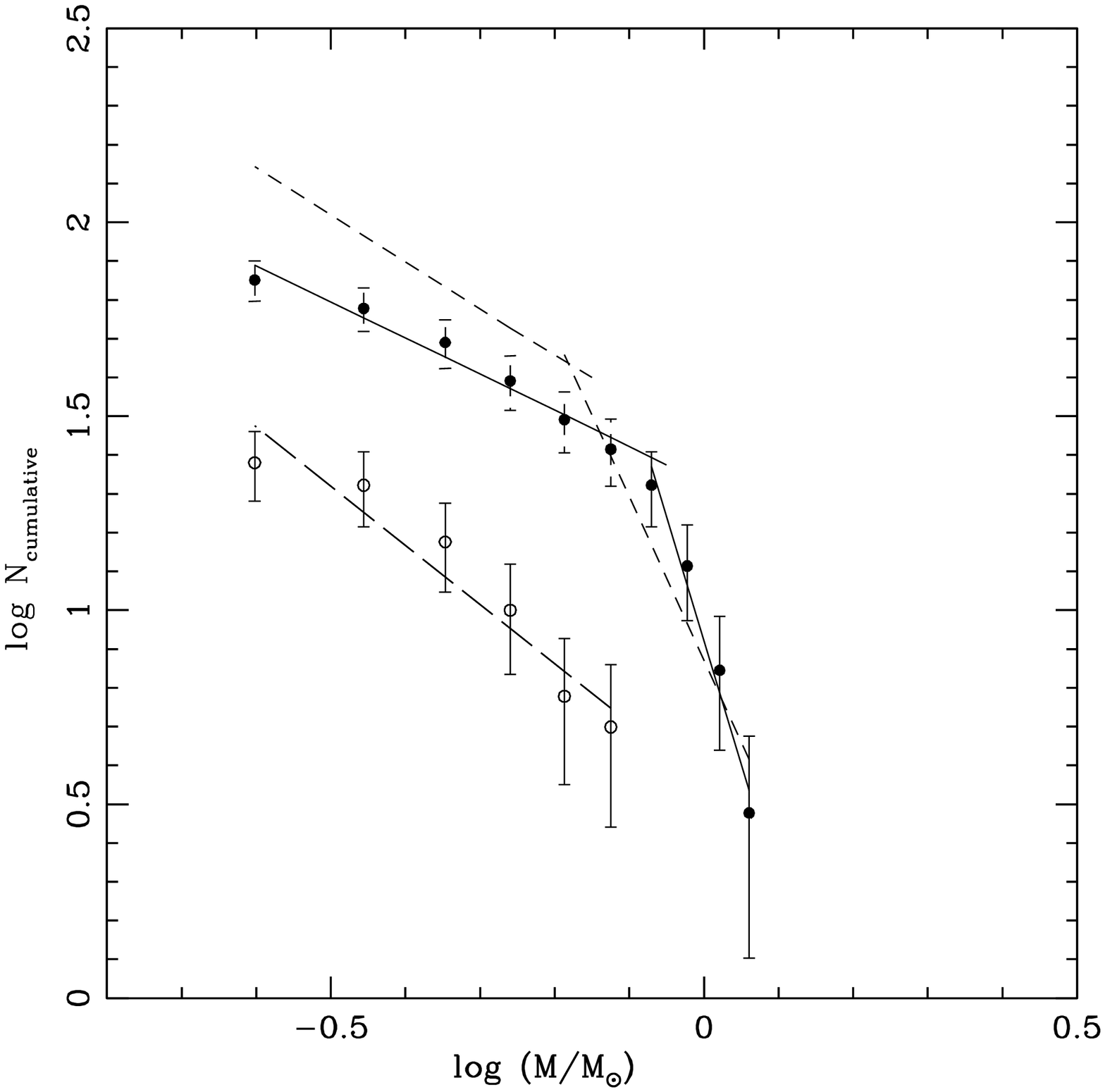}

\noindent{\footnotesize {\footnotesize \bf Figure 13.} Cumulative mass function (CMF) of the `A' type (filled circles) and `B/C' type BRCs 
(open circles). Error bars represent $\pm$$\sqrt{N}$ errors. The CMF for the standard MF is shown by short dashed lines (see the text).}
\label{fig13}
\end{figure}
It is interesting to notice that both CMFs show a roughly similar shape with a 
break in power law. Obviously a detailed comparison 
manifests differences. In the case of the YSOs we find a break in the slope of the CMF at 
$\sim$0.8 M$_\odot$. In the mass range 0.8$\le$ M/M$_\odot$ $\le$1.2 the slope of 
the CMF is -7.1$\pm$0.9 and it becomes shallower (-1.0$\pm$0.1) for masses 
0.2$\le$ M/M$_\odot$ $\le$0.8, whereas the CMF of the cores can be represented 
by a power law with a shallower slope of -0.4$\pm$0.1 in the mass range 
0.4$\le$ M/M$_\odot$ $\le$1.2. The core CMF becomes steeper for masses 
$\ge$1.2 M$_\odot$ (slope=-2.6$\pm$0.3). Morgan et al. (2008) have reported 
that their sample is complete down to 0.5 M$_\odot$. Our sample toward lower 
mass may be affected by incompleteness, however the correction due to 
incompleteness will further steepen the CMF slope of the YSOs. The shallower CMF 
slopes in the case of dense cores than those for YSOs indicates that the star 
formation in the next sequence/ generation favours formation of relatively massive stars.
 
If the star formation within the BRCs depends on morphology of the clouds, as 
suggested by Morgan et al. (2008), it would be interesting to study the CMF 
of YSOs by separating the target BRCs on the basis of the morphology of BRCs. 
Here we assign type A to BRC 38 rather than type B given in SFO91. BRC 11NE, 
which is not included in SFO91, is classified as type B. In Fig. 13 we plot the CMFs 
of the YSOs in 4 `A' type BRCs, namely BRCs 2, 14, 27 and 38, and of those
in 3 `B/C' type BRCs, namely BRCs 11NE, 12 and 13. In the YSO mass
range 0.2$\le$ M/M$_\odot$ $\le$0.8 the slope of the CMF for  the `B/C' type BRCs is found 
to be -1.5$\pm$0.2 which is steeper than that (-0.9$\pm$0.1) obtained 
for `A' type BRCs. This is in contradiction with the results 
reported by Morgan et al. (2008). They reported a shallower mass function slope 
for `B/C' type BRCs in comparison to that of `A' type BRCs (see their Figure 11); 
however a close inspection of their figure 11 manifests that in the mass range 
0.5$\le$ M/M$_\odot$ $\le$3.0, the MF slope of the cores of `A' type 
BRCs is definitely shallower than that for `B/C' type BRCs. This suggests 
that `A' type rims, in the mass range 0.4$\le$ M/M$_\odot$ $\le$1.2, appear 
to follow a mass function that is more biased toward formation of 
relatively massive stars in comparison to that in case of `B' and `C' type BRCs.

In Fig. 13 we have also plotted the CMF generated for a sample aggregate having an average
Galactic initial mass function, i.e., $\Gamma$ = -1.35 for 0.6$\le$ M/M$_\odot$ $\le$1.2, and 
$\Gamma$ = -0.3 for 0.2$\le$ M/M$_\odot$ $\le$0.6 (Kroupa 2001, 2002). 
The slope of the CMF in the mass range 0.2$\le$ M/M$_\odot$ $\le$0.6 comes out to be 
$\sim$1.1$\pm$0.1, which is close to the slope of the CMF (-0.9$\pm$0.1) of the YSOs 
(0.2$\le$ M/M$_\odot$ $\le$0.8) in the `A' type BRCs. Whereas, the CMF slope for 
YSOs in the `B/C' type BRCs is significantly steeper (-1.5$\pm$0.2) than the standard 
MF. This suggests that in the mass range 0.2$\le$ M/M$_\odot$ $\le$0.8 the YSOs in 
`A' type BRCs follow the standard form of MF, whereas aggregates in `B/C' type 
BRCs is more biased towards relatively less massive objects. 
We have also estimated the effect of errors on estimation of mass function. The 
results are given in  Table 9 which indicate an insignificant effect on the mass function slopes.

\begin{table}
\small
\caption{Mass function of BRC aggregates. The maximum and minimum value of the slopes are estimated 
by propagating the random errors using the Monte Carlo simulations.}
\begin{tabular}{|p{0.95in}|p{.65in}|p{0.65in}|p{0.65in}|}
\hline
Mass Range  &Mean value  &Maximum value & Minimum value\\
(M$_\odot$) &of the slope& of the slope & of the slope\\
\hline
{\bf  All BRCs}&&&\\
0.2 - 0.8   & -0.97 $\pm$ 0.14 & -0.99 $\pm$ 0.15 & -0.95 $\pm$ 0.15 \\
0.8 - 1.2   & -7.08 $\pm$ 0.89 & -8.17 $\pm$ 0.86 & -6.40 $\pm$ 0.62 \\
            &&&\\
{\bf A-type BRCs} &                  &                  &                   \\
0.2 - 0.8   & -0.92 $\pm$ 0.09 & -0.96 $\pm$ 0.10 & -0.87 $\pm$ 0.11 \\
0.8 - 1.2   & -6.40 $\pm$ 0.78 & -7.60 $\pm$ 0.74 & -5.60 $\pm$ 0.55 \\
&&&\\
{\bf B/C-type BRCs} &                  &                  &                   \\
0.2 - 0.8   & -1.53 $\pm$ 0.20 & -1.63 $\pm$ 0.20 & -1.20 $\pm$ 0.17 \\
\hline
\end{tabular}
\end{table}

\section{Conclusions}
On the basis of the present optical and NIR analysis of six BRC aggregates we reached 
the following conclusions.

We estimated the ages of individual stars associated with BRCs from the
reddening-corrected $V_0, (V-I_c)_0$ CM diagrams. By comparing the average
ages of the stars on/inside and outside the bright rim, we again found quantitative
age gradients in almost all the studied BRCs (the only exception being BRC 27), 
although the number of the sample stars are small and their age scatters are large. 
The results are quite similar to the results reported in Paper I. In addition the 
youngest objects, obtained from $Spitzer$ MIR data, are found to be deeply embedded 
inside the BRCs, supporting the above conclusion. These results further confirm $S^4F$
hypothesis.

The distribution of NIR-excess stars in the studied HII regions indicates
that they are aligned from the ionizing source to the BRC direction. The age 
indicators, viz., infrared excess ($\Delta$($H-K$)) and $A_V$ as well as the age 
itself of the YSOs manifest an age gradient toward the ionizing source.  This 
global distribution indicates that a series of triggered star formation took place in the 
past from near the central O star(s) towards the peripheries of the HII region. 

It is found that the EW of H$\alpha$ emission in TTSs associated with the BRCs
decreases with age. We found some  H$\alpha$ emission 
stars that are significantly older than those TTSs associated with the BRCs. They
apparently must have formed spontaneously before the main star formation event which gave
birth to the massive stars in the region; however their origin is not clear. We found
that in general WTTSs are older than CTTSs. It is also found that the
fraction of CTTSs among the TTSs associated with the BRCs is found to decrease
with age, as found in Taurus region by Armitage et al. (2003). These facts are 
in accordance with the conclusion by Bertout et al. (2007) that CTTSs evolve into WTTSs.

The CMF of `A' type BRCs seems to follow a mass function similar
to that found in young open clusters, whereas `B/C' type BRCs have a significant
steeper CMF, indicating that BRCs of the latter type tend to form relatively more
low mass YSOs of the mass range 0.2$\le$ M/M$_\odot$ $\le$0.8.

\section{ACKNOWLEDGMENTS}
We are thankful to the anonymous referee for the critical comments which 
improved the scientific contents and presentation of the paper.
We are thankful to the TAC and staff of HCT for the time allotment and for 
their support during the observations, respectively. This 
publication makes use of data from the Two Micron All Sky Survey (a 
joint project of the University of Massachusetts and the Infrared Processing 
and Analysis Center/ California Institute of Technology, funded by the 
National Aeronautics and Space Administration and the National Science 
Foundation), archival data obtained with the {\it Spitzer Space Telescope}
(operated by the Jet Propulsion Laboratory, California Institute 
of Technology, under contract with the NASA). This study is a part 
of the DST (India) sponsored project and NC is thankful to DST for the support. NC also 
acknowledges the financial support provided by TIFR during her visit to TIFR. AKP and KO
 acknowledge the financial support received from DST (India) and JSPS (Japan).


\section{Appendix}
\appendix
\section{}
\begin{figure*}
\centering
\includegraphics[scale = .39, trim = 0 10 10 180, clip]{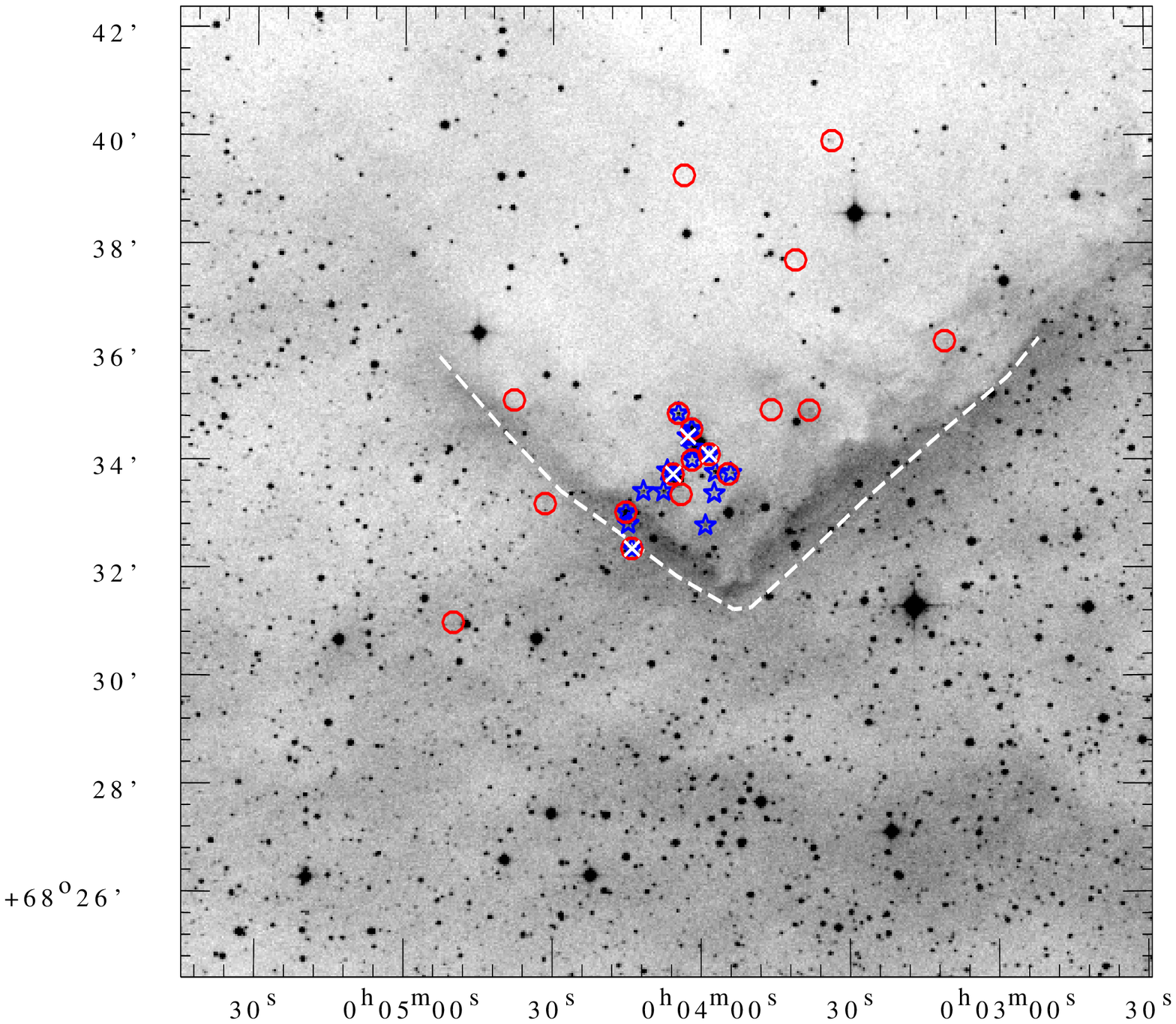}
\includegraphics[scale = .39, trim = 0 10 10 180, clip]{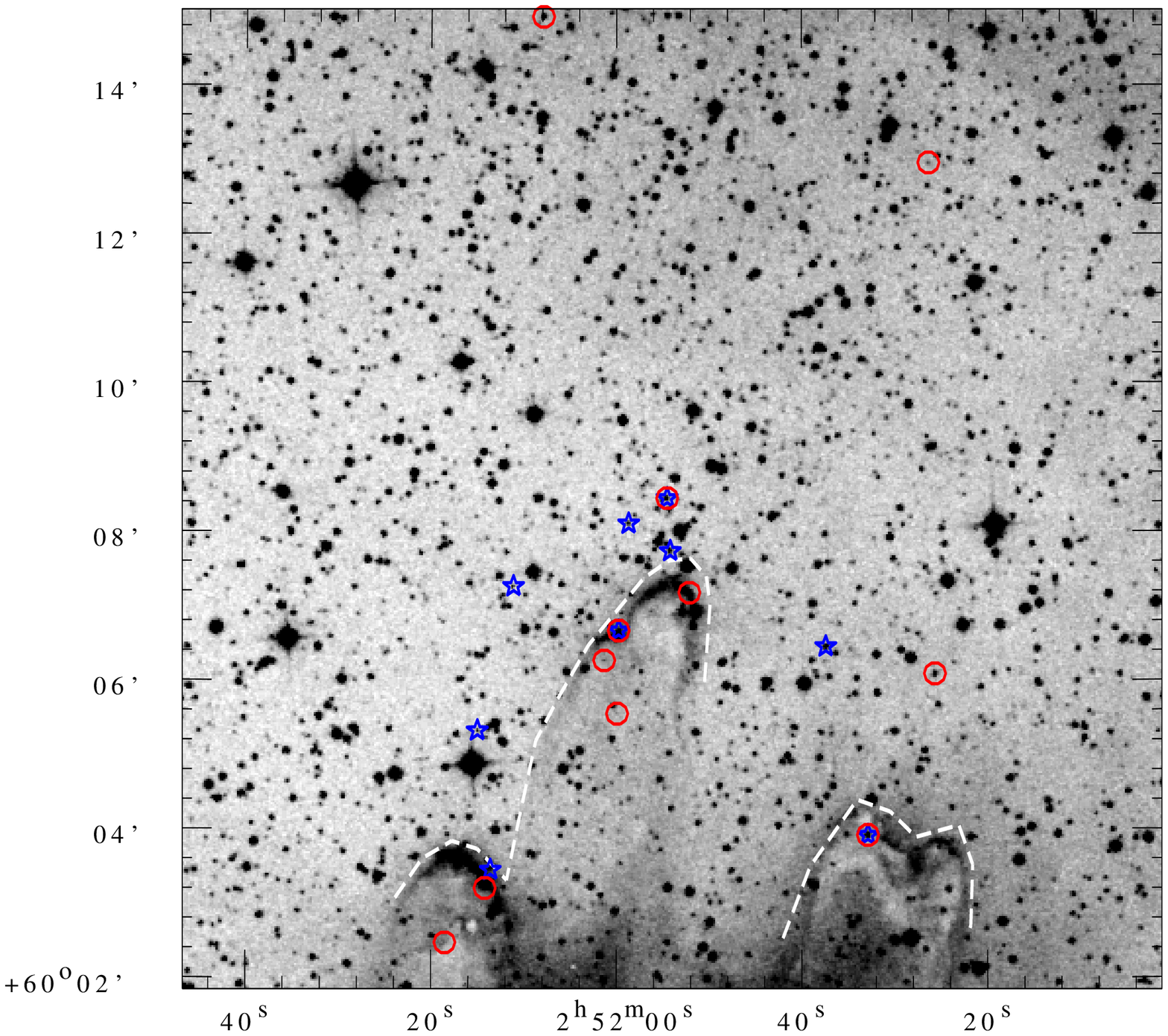}
\includegraphics[scale = .39, trim = 0 10 10 180, clip]{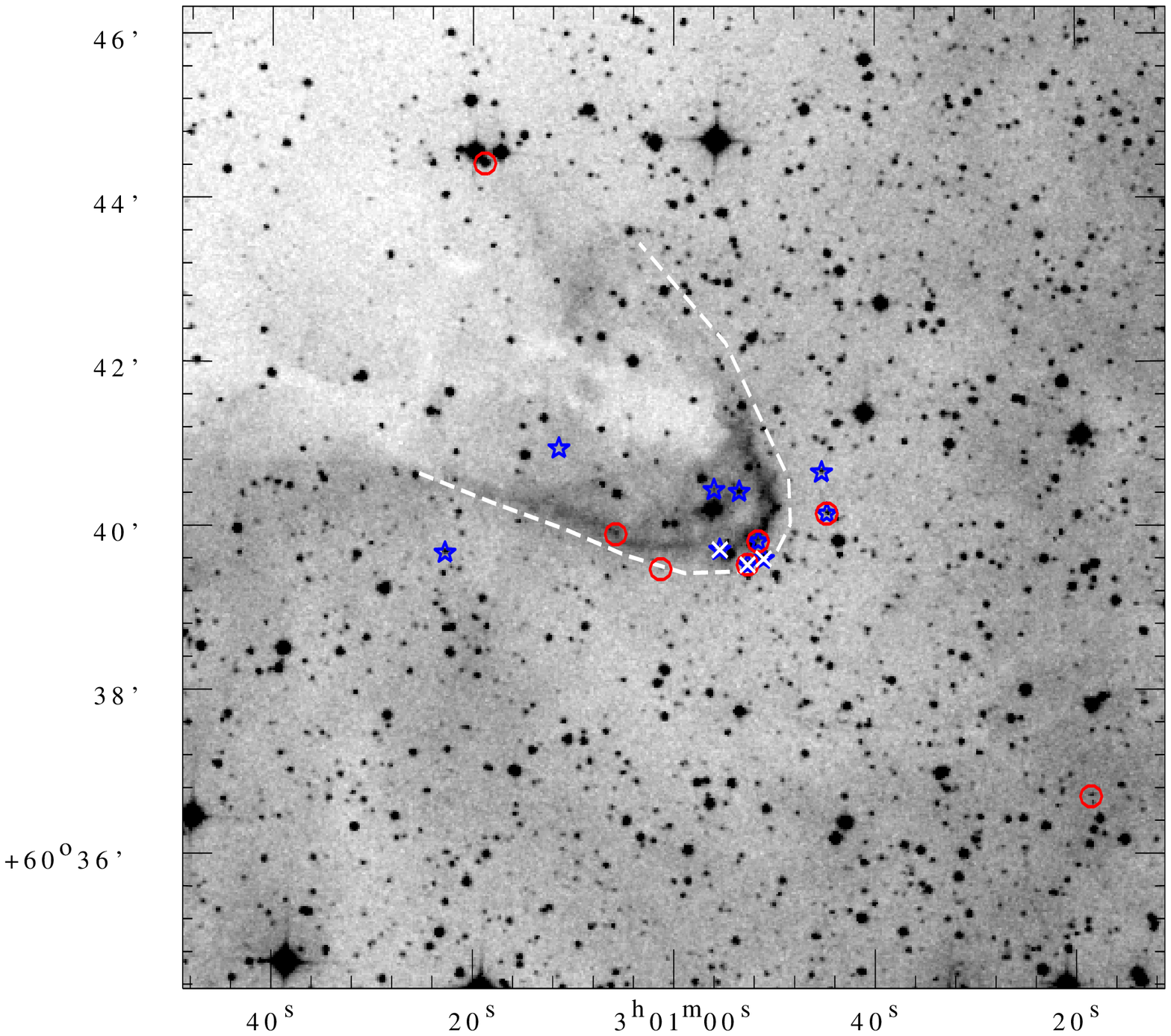}
\includegraphics[scale = .39, trim = 0 10 10 180, clip]{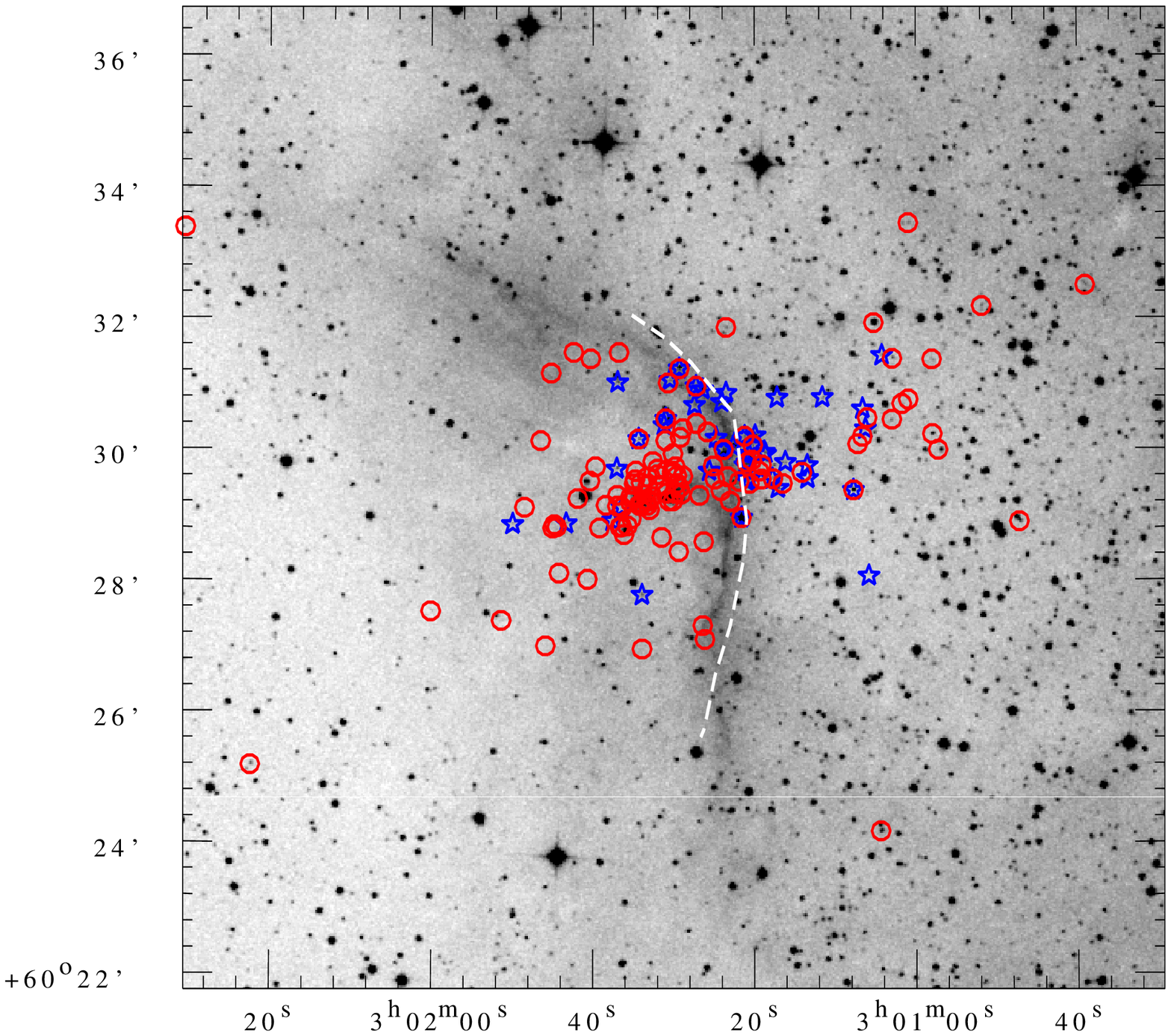}
\includegraphics[scale = .39, trim = 0 10 10 180, clip]{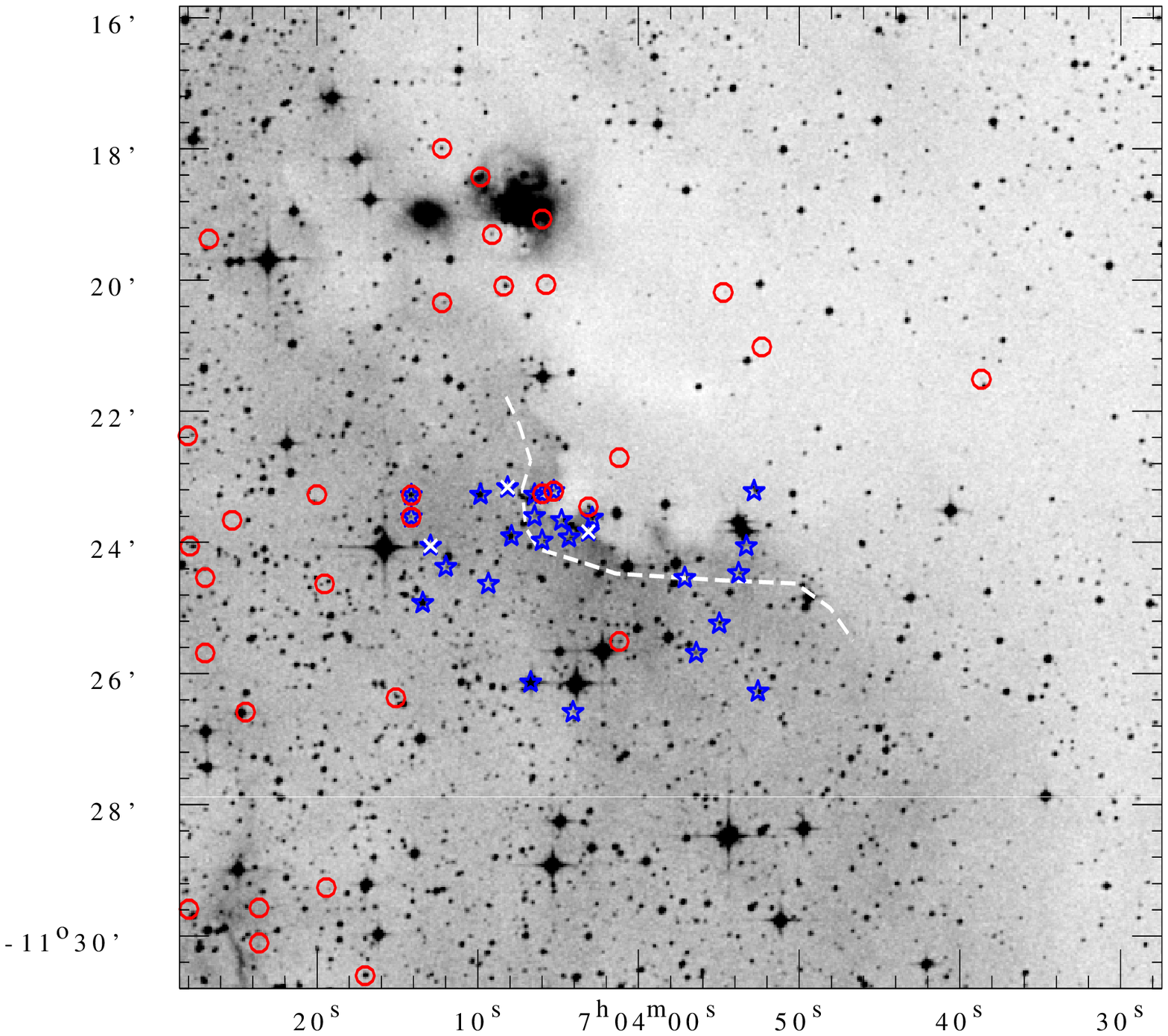}
\includegraphics[scale = .39, trim = 0 10 10 180, clip]{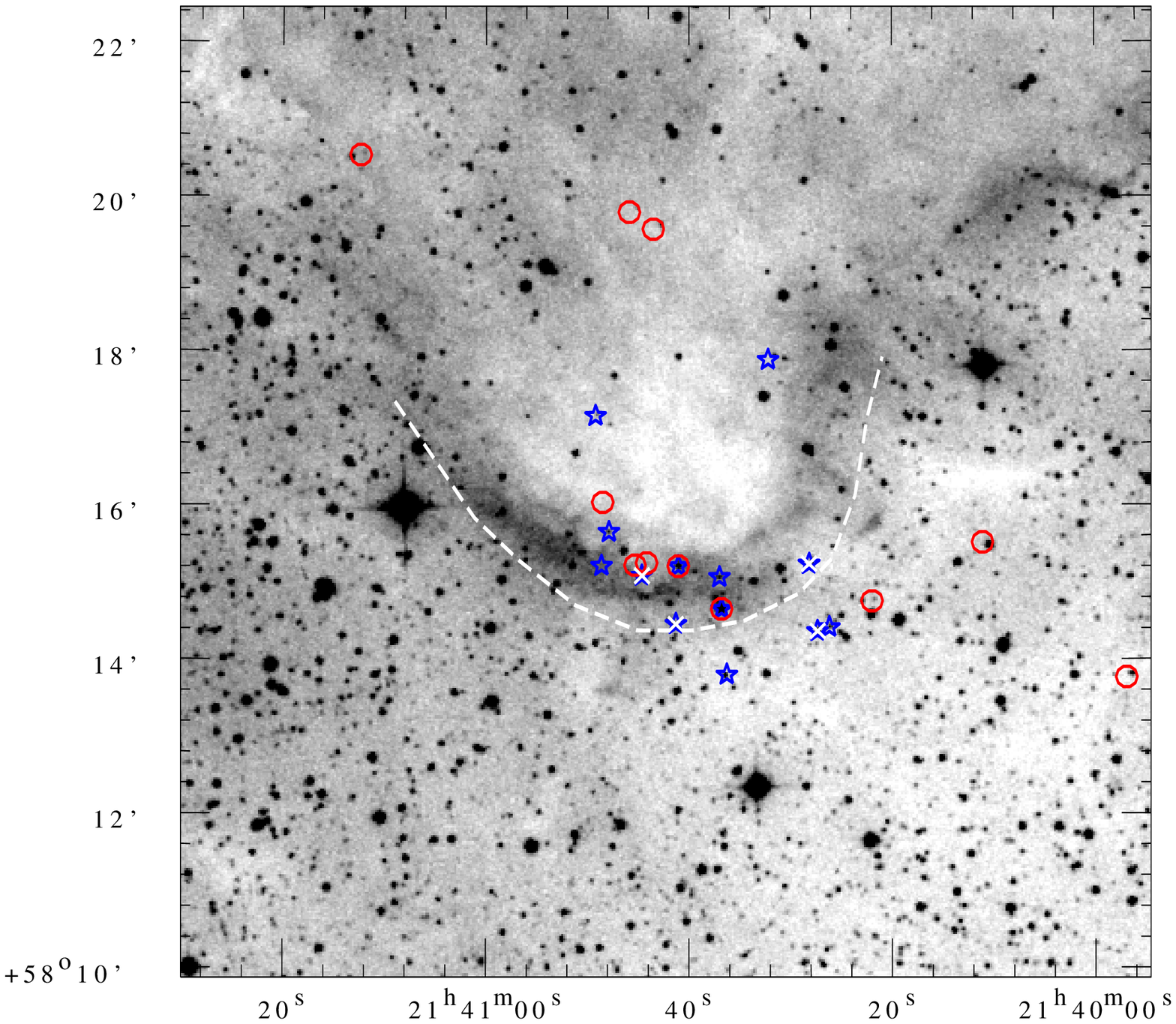}

\caption{ Spatial distribution of H$\alpha$ emission (star symbol) 
and NIR excess (open circle) stars overlaid on DSS2 R-band image of BRC 2 $(upper-left$ $panel)$; BRC 11NE 
$(upper-right$ $panel)$; BRC13 $(middle-left$ $panel)$; BRC 14 $(middle-right$ $panel)$; BRC27 $(lower-left$ $panel)$; BRC 38 $(lower-right$ $panel)$. Dashed line demarcates the regions on/ inside and outside 
bright-rims. White crosses represent H$\alpha$ emission stars having 
ages larger than 5 Myr (see sec. 7 and Fig. 9).}
\label{figA}
\end{figure*}

\clearpage
\begin{figure}
\centering
\includegraphics[scale = .47, trim = 0 60 10 170, clip]{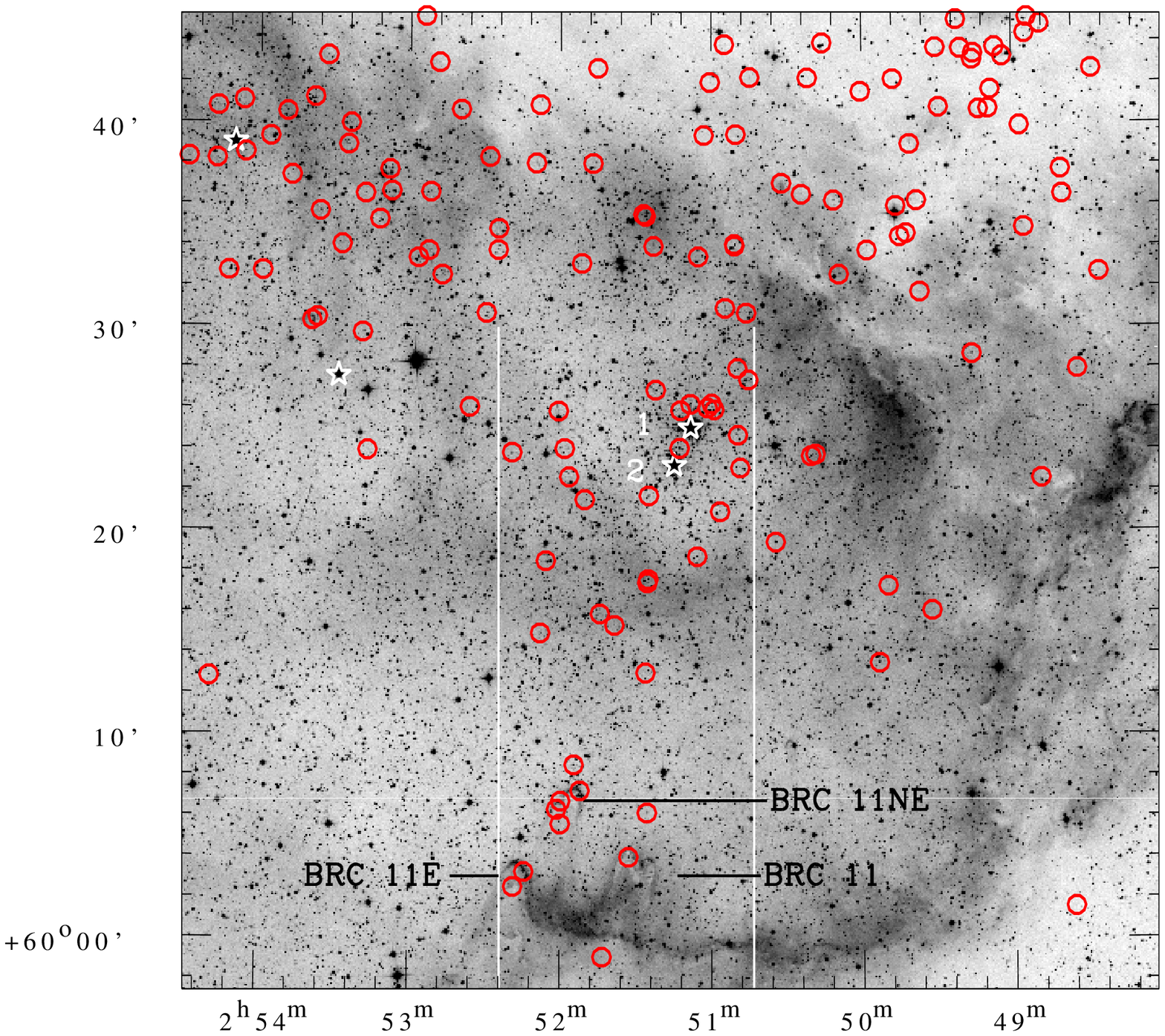}
\caption{ Spatial distribution of NIR excess stars (open circles) overlaid on DSS2 R-band image of
 the IC 1848W region. Star symbols indicate O type stars. The stars marked as `1' and 
`2' are HD 17505 (O6 V) and  HD 17520 (O9 V) respectively. The variation of 
NIR excess $\Delta$(H-K) and $A_V$ for the stars 
within the strip as a function of distance from HD 17505 is shown in Fig. 6. }
\label{fig7}
\end{figure}
\begin{figure}
\centering
\includegraphics[scale = .47, trim = 0 60 10 240, clip]{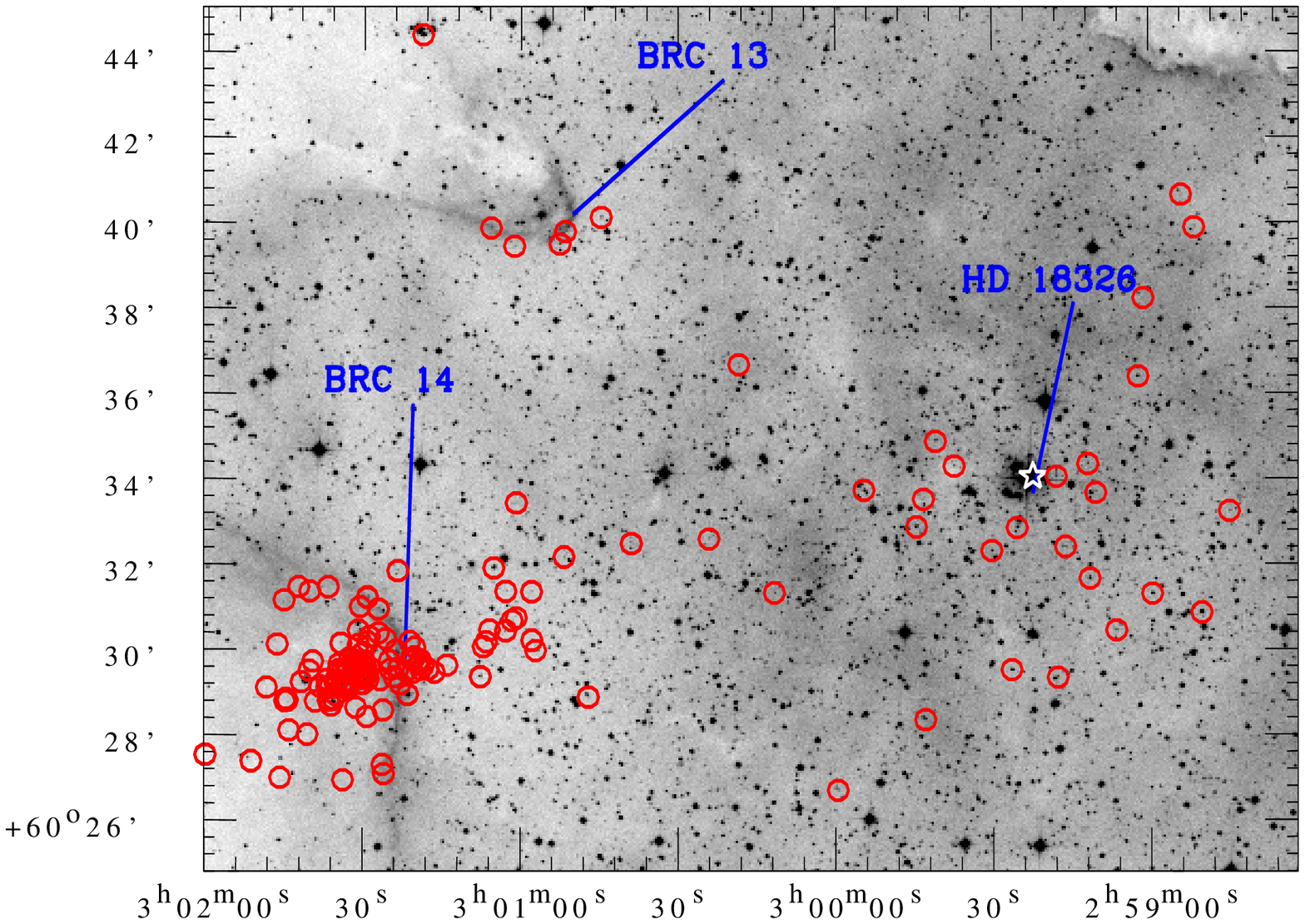}
\caption{ Same as Fig. A2 but for IC 1848E region.}
\label{fig8}
\end{figure}
\begin{figure}
\centering
\includegraphics[scale = .47, trim = 0 50 0 170, clip]{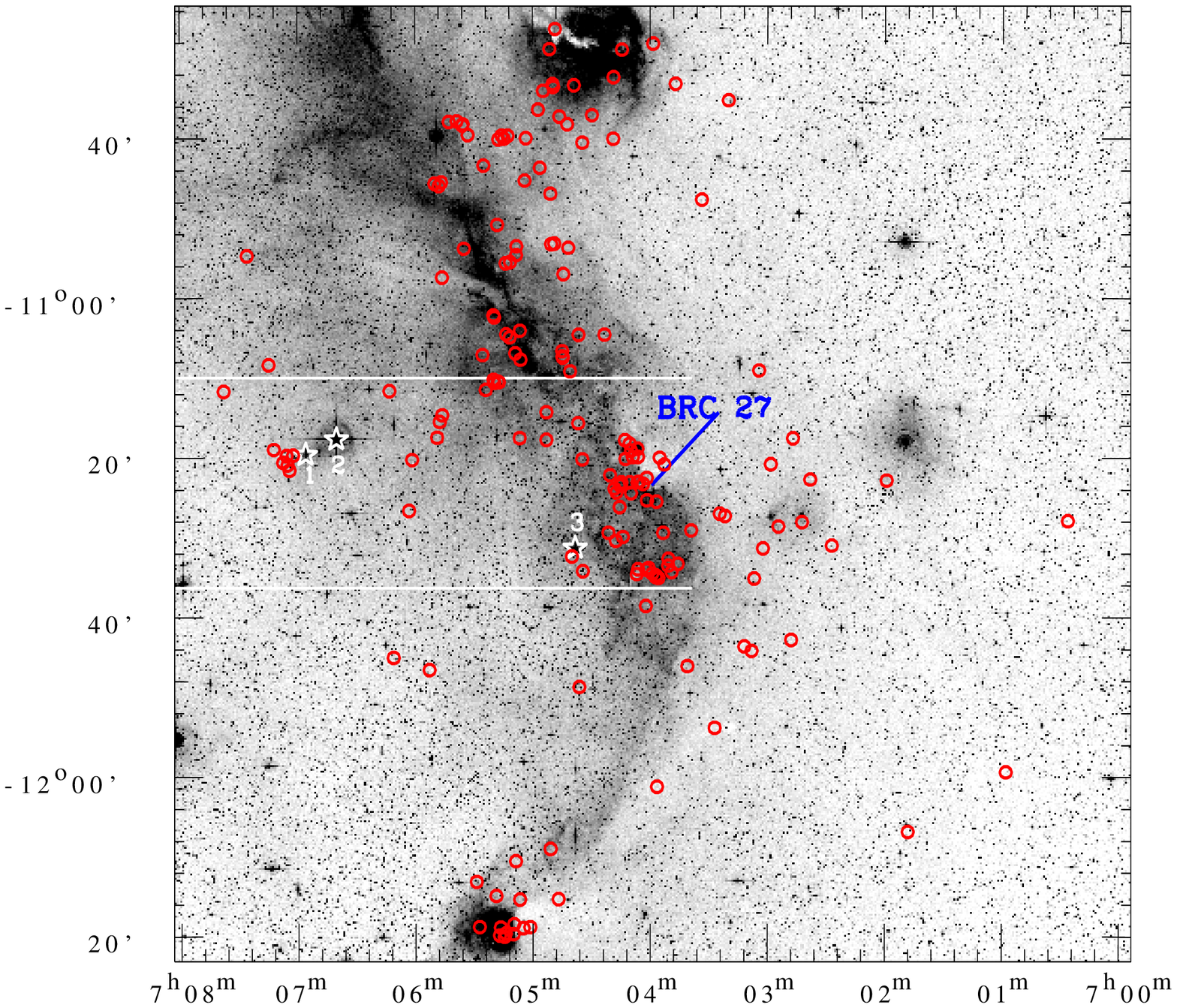}
\caption{ Same as Fig. A2 but for CMaR1 region. Stars marked as `1', `2' and `3' 
are HD 54025 (B1 V); HD 53974 (B0.5 IV) and HD 53456 (B0 V) respectively.}
\label{fig9}
\end{figure}
\begin{figure}
\centering
\includegraphics[scale = .47, trim = 0 50 10 180, clip]{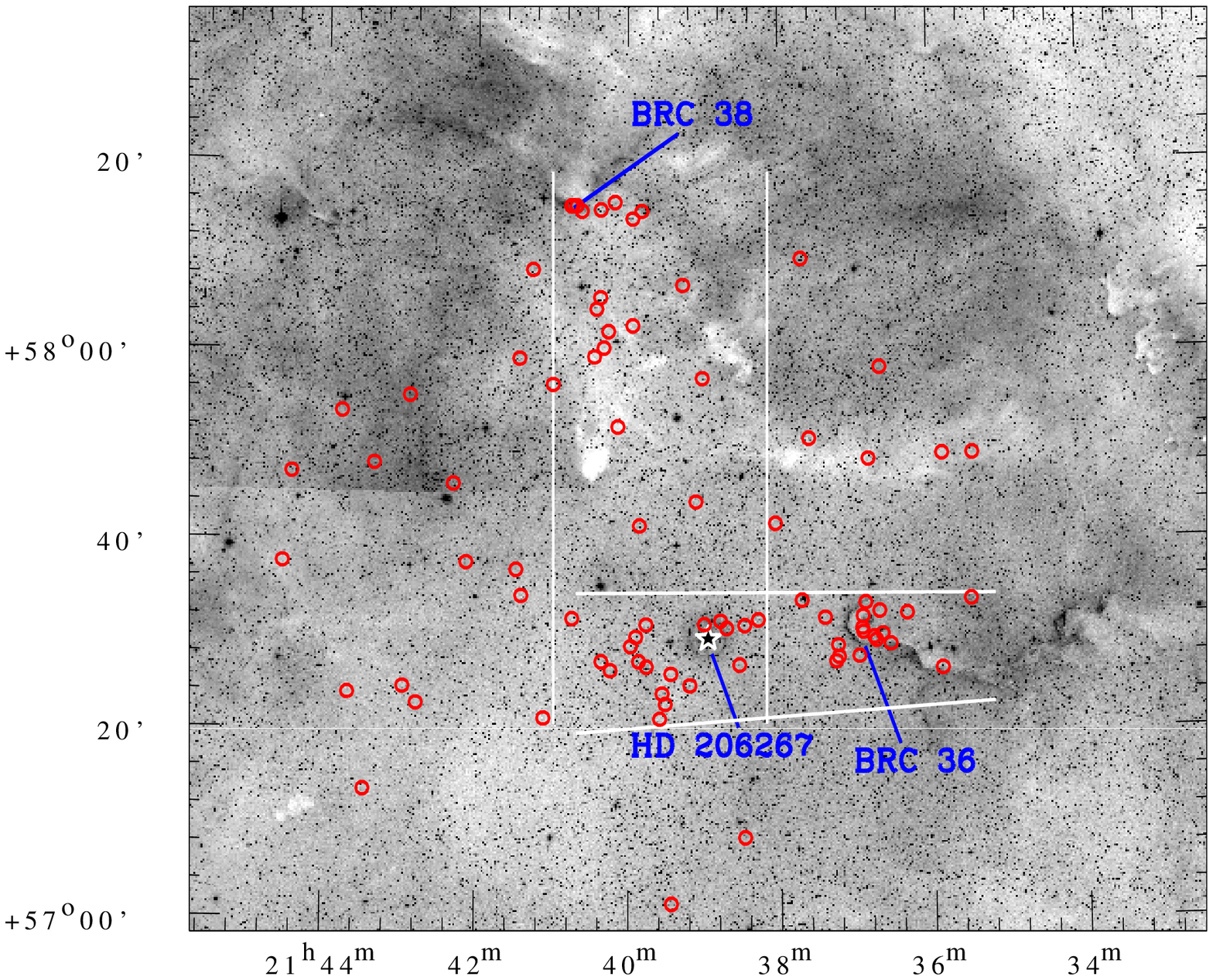}
\caption{ Same as Fig. A2 but for IC 1396.}
\label{fig10}
\end{figure}
\clearpage

\begin{table*}
{{\bf Table 6.} IRAC photometric magnitudes of the disk bearing candidates in BRCs 2, 27 and 13/14.}
\begin{tabular}{|rrrrrrrrrrr|}
\hline
{\bf RA (J2000)}           &{\bf DEC (J2000)}      &{\bf [3.6]} & {\bf e[3.6]}&{\bf [4.5]} &{\bf e[4.5]}&{\bf [5.8]} & {\bf e[4.5] } &{\bf [8.0]} &{\bf e[8.0]}&{\bf IRAC type}\\
\hline
{\bf BRC 2}& &&&&&&&&&\\
 00 04 14.69  & +68 32 49.8 & 11.899&  0.033 &  10.97 & 0.03 &  10.095  &   0.052  &  8.985 &   0.028 & 0/I\\
 00 03 57.27  & +68 33 24.4 & 12.231&  0.038 &  11.74 &0.042 &  11.100  &   0.087  &  9.996 &   0.075 &0/I\\
 00 04 03.83  & +68 32 49.6 & 13.316&  0.064 &  12.57 &0.062 &  11.749  &   0.123  &  10.57 &   0.117 &0/I\\
 00 03 59.71  & +68 35 06.9 &  7.237&  0.004 &  5.778 &0.004 &   4.558  &   0.004  &  3.586 &   0.002 &0/I \\
 00 04 37.79  & +68 35 07.4 & 10.274&  0.015 &  9.342 &0.014 &   8.595  &   0.027  &  7.630 &   0.026 &0/I\\
 00 03 57.78  & +68 35 01.6 & 10.344&  0.021 &  9.414 &0.021 &   8.707  &   0.034  &  7.725 &   0.035 &0/I\\
\hline												   
\end{tabular}	
\end {table*}	
\begin{table*}	
{\bf Table 6 cont.}\\
\begin{tabular}{|rrrrrrrrrrr|}
\hline
 03 01 32.14  & +60 26 48.9 & 13.510&  0.073 & 13.064 &0.079&  12.478   & 0.215&   11.100&    0.234&0/I\\
 03 00 55.39  & +60 45 21.8 & 13.864&  0.081 & 13.322 &0.088&  11.816  &  0.115&   9.827 &   0.042&0/I\\
 03 01 48.00  & +60 35 22.9 & 13.194&  0.088 & 12.325 &0.059&  11.115  &  0.131&   9.658 &   0.144&0/I\\
 03 00 53.09  & +60 40 36.5 & 13.176&  0.093 & 12.375 &0.065&  11.151  &  0.139&   9.229 &   0.095&0/I\\
 03 01 32.47  & +60 25 53.7 & 10.677&  0.019 &  9.344 &0.015&   8.289  &  0.023&   7.266 &   0.013&0/I\\
 03 01 48.54  & +60 29 05.1 & 10.980&  0.021 & 10.109 &0.020&   9.219  &  0.035&   8.442 &   0.021&0/I \\
 03 01 43.43  & +60 28 50.2 & 11.130&  0.025 & 10.327 &0.025&   9.835  &  0.046&   9.098 &   0.041&0/I \\
 03 01 32.10  & +60 29 40.5 & 11.271&  0.035 & 10.262 &0.030&   9.418  &  0.067&   8.620 &   0.071&0/I \\
\hline												   
\end{tabular}											   
\end {table*}
\begin{table*}
{\bf Table 6 cont.} \\
\begin{tabular}{|rrrrrrrrrrr|}
\hline
 03 02 26.07  & +60 33 02.8 & 12.837&  0.050 & 12.819 &0.070&  12.613  &  0.179&  11.726 &   0.192&II \\
 03 01 31.04  & +60 30 26.1 & 12.750&  0.052 & 12.322 &0.056&  11.634  &  0.129&  10.575 &   0.154&II \\
 03 01 06.26  & +60 30 16.8 & 12.962&  0.052 & 12.536 &0.061&  12.177  &  0.135&  11.566 &   0.084&II \\
 03 01 49.97  & +60 28 49.3 & 12.902&  0.053 & 12.367 &0.057&  11.797  &  0.117&  10.758 &   0.092&II \\
 03 01 07.27  & +60 25 25.7 & 13.011&  0.054 & 12.520 &0.061&  11.887  &  0.119&  11.122 &   0.076&II \\
 03 01 08.16  & +60 39 00.5 & 13.070&  0.055 & 12.627 &0.063&  12.271  &  0.142&  11.639 &   0.106&II \\
 03 02 28.02  & +60 32 21.2 & 13.074&  0.055 & 13.035 &0.077&  12.920  &  0.200&  12.362 &   0.193&II \\
 03 00 46.58  & +60 31 58.4 & 13.122&  0.056 & 12.615 &0.063&  12.410  &  0.152&  11.718 &   0.095&II \\
\hline                                                                                         
\end{tabular}  	                                                                                
\end{table*}   	                                                                                 


\begin{table*}
\begin{sideways}
\begin{minipage}{245mm}
{{\bf Table 7.} {\it J, H} and {\it K} magnitudes of the sources used in the analysis (cf. sec. 6.2)}
\centering
\tiny
\begin{tabular}{|llrrrrrrrrr|}
\hline
   {\bf   RA }    & {\bf  DEC }    &{\bf 2MASS Name}    &{\bf J $\pm$ eJ} & {\bf H $\pm$ eH }& {\bf K $\pm$ eK } & {\bf Q flag}&{\bf C flag}& {\bf A$_V$} & {\bf $\Delta(H-K)$}&{\bf 2MASS/Matsuyanagi}\\
     {\bf  (J2000) } & {\bf (J2000)} &&{\bf (mag)}&{\bf (mag)}&{\bf (mag)}&& &{\bf (mag)}&{\bf (mag)}&{\bf et al. 2006(M06)}\\
\hline
{\bf  IC 1848W}&             &             &         &        &        &       &       &      &&\\
02 51 12.63 & +60 24 00.1 & 02511262+6024000  &13.719 $\pm$ 0.050  & 12.912 $\pm$ 0.051  & 12.361 $\pm$ 0.035  & AAA &000  &   0.00    & 0.05 &2MASS  \\ 
02 51 24.86 & +60 21 40.2 & 02512485+6021402  &14.305 $\pm$ 0.038  & 13.545 $\pm$ 0.043  & 13.017 $\pm$ 0.031  & AAA &000  &   0.00    & 0.05 &2MASS  \\ 
02 51 12.27 & +60 25 51.3 & 02511226+6025512  &15.955 $\pm$ 0.086  & 15.189 $\pm$ 0.103  & 14.630 $\pm$ 0.099  & AAA &c00  &   0.00    & 0.08 &2MASS  \\ 
02 51 08.31 & +60 26 09.9 & 02510831+6026099  &14.390 $\pm$ 0.075  & 13.636 $\pm$ 0.065  & 13.160 $\pm$ 0.062  & AAA &000  &   0.00    & 0.01 &2MASS  \\ 
02 50 56.55 & +60 20 53.1 & 02505654+6020530  &15.257 $\pm$ 0.060  & 14.296 $\pm$ 0.066  & 13.646 $\pm$ 0.043  & AAA &ccc  &   0.69    & 0.07 &2MASS  \\
\hline
\end{tabular}
\end{minipage}
\end{sideways}
\end{table*}
\begin{table*}
\tiny
\begin{sideways}
\begin{minipage}{230mm}

\begin{tabular}{|llrrrrrrrrr|}
&\multicolumn{9}{c}{\bf Table 7 cont.}& \\
\hline
03 01 31.0  & +60 30 07   &			 & 19.27 $\pm$  0.06  &   17.81 $\pm$  0.04  &   16.79 $\pm$  0.03 & &  & 	         4.82  &      0.16 &M06  \\
03 01 29.2  & +60 30 10   &			 & 16.35 $\pm$$<$0.01 &   14.64 $\pm$$<$0.01 &   13.54 $\pm$$<$0.01& &  & 	         7.64  &      0.09 &M06  \\
03 01 28.9  & +60 30 18   &			 & 19.09 $\pm$  0.05  &   18.04 $\pm$  0.04  &   17.27 $\pm$  0.04 & &  & 	         1.15  &      0.15 &M06  \\
03 01 31.0  & +60 30 27   &          		 & 16.33 $\pm$  0.01  &   15.10 $\pm$  0.01  &   14.29 $\pm$$<$0.01& &  &               3.32  &       0.08 &M06  \\
03 01 30.7  & +60 30 60   &          		 & 17.96 $\pm$  0.02  &   16.54 $\pm$  0.01  &   15.60 $\pm$  0.01 & &  &               4.91  &       0.10 &M06  \\
03 01 29.3  & +60 31 13   &          		 & 14.72 $\pm$$<$0.01 &   13.42 $\pm$$<$0.01 &   12.42 $\pm$$<$0.01& &  &               2.77  &       0.23 &M06  \\
03 01 45.9  & +60 26 59   &			 & 19.64 $\pm$  0.10  &   18.46 $\pm$  0.08  &   17.54 $\pm$  0.07 & &  & 	         1.75  &      0.22 &M06  \\
\hline
\end{tabular}
\end{minipage}

\end{sideways}

\end{table*}
\begin{table*}
\tiny
\begin{sideways}
\begin{minipage}{230mm}

\begin{tabular}{|llrrrrrrrrr|}
&\multicolumn{9}{c}{\bf Table 7 cont.}& \\
\hline

03 00 24.11 & +60 32 36.3 & 03002411+6032363 &  16.195 $\pm$ 0.107  & 15.027 $\pm$ 0.104  & 14.213 $\pm$ 0.078 & BAA &  000  &   2.01  &    0.12    &2MASS 	  \\
03 02 00.11 & +60 27 31.0 & 03020010+6027309 &  15.401 $\pm$ 0.056  & 14.244 $\pm$ 0.053  & 13.447 $\pm$ 0.042 & AAA &  000  &   2.01  &    0.11    &2MASS 	  \\
03 00 11.65 & +60 31 20.2 & 03001164+6031202 &  16.000 $\pm$ 0.103  & 14.982 $\pm$ 0.082  & 14.258 $\pm$ 0.073 & AAA &  000  &   0.82  &    0.11    &2MASS 	  \\
03 01 01.33 & +60 39 28.5 & 03010133+6039285 &  15.860 $\pm$ 0.092  & 14.942 $\pm$ 0.095  & 14.272 $\pm$ 0.093 & AAA &  000  &   0.00  &    0.11    &2MASS 	  \\
03 00 52.66 & +60 39 31.8 & 03005265+6039317 &  15.043 $\pm$ 0.050  & 14.025 $\pm$ 0.050  & 13.291 $\pm$ 0.042 & AAA &  000  &   0.74  &    0.12    &2MASS 	  \\
03 01 05.81 & +60 39 54.1 & 03010581+6039541 &  15.868 $\pm$ 0.088  & 15.040 $\pm$ 0.092  & 14.526 $\pm$ 0.092 & AAA &  000  &   0.09  &    0.002   &2MASS 	  \\
03 00 18.43 & +60 36 41.9 & 03001843+6036419 &  15.687 $\pm$ 0.078  & 14.891 $\pm$ 0.080  & 14.272 $\pm$ 0.082 & AAA &  c00  &   0.00  &    0.13    &2MASS       \\
\hline
\end{tabular}
\end{minipage}
\end{sideways}
\end{table*}
\begin{table*}
\tiny
\begin{sideways}
\begin{minipage}{230mm}

\begin{tabular}{|llrrrrrrrrr|}
&\multicolumn{9}{c}{\bf Table 7 cont.}& \\
\hline

07 03 45.54 & -11 33 28.6 & 07034554-1133286 &  15.968 $\pm$ 0.079  & 14.597 $\pm$ 0.065  & 13.488 $\pm$ 0.038 & AAA &  ccc  &   2.23  &    0.30    &2MASS 	   \\
07 04 00.41 & -11 33 59.6 & 07040041-1133596 &  12.099 $\pm$ 0.030  & 10.916 $\pm$ 0.028  & 10.190 $\pm$ 0.026 & AAA &  000  &   2.91  &    0.02    &2MASS 	   \\
07 04 01.60 & -11 34 13.3 & 07040160-1134133 &  15.430 $\pm$ 0.052  & 12.449 $\pm$ 0.030  &  9.997 $\pm$ 0.026 & AAA &  000  &   11.94 &     0.77   &2MASS 	   \\
07 04 05.85 & -11 34 10.0 & 07040584-1134099 &  16.174 $\pm$ 0.086  & 13.780 $\pm$ 0.043  & 12.195 $\pm$ 0.030 & AAA &  000  &   11.43 &     0.22   &2MASS 	   \\
07 03 58.50 & -11 34 42.7 & 07035849-1134426 &  15.671 $\pm$ 0.070  & 14.552 $\pm$ 0.087  & 13.837 $\pm$ 0.055 & AAA &  000  &   2.18  &    0.05    &2MASS 	   \\
07 03 48.69 & -11 34 30.3 & 07034869-1134303 &  15.658 $\pm$ 0.062  & 14.772 $\pm$ 0.066  & 14.182 $\pm$ 0.068 & AAA &  0cc  &   0.22  &    0.05    &2MASS 	   \\
07 03 56.67 & -11 34 55.4 & 07035666-1134553 &  12.938 $\pm$ 0.024  & 11.872 $\pm$ 0.027  & 11.169 $\pm$ 0.023 & AAA &  000  &    1.60 &     0.06   &2MASS 	   \\
\hline                                                                           
\end{tabular}
\end{minipage}
\end{sideways}
\end{table*}

\end{document}